%% file: Two-loop_Yang-Mills_diagrams_from_superstring_amplitudes.tex
 \documentclass[12pt,a4paper]{article}

\usepackage[
      colorlinks=false,
      linkcolor=darkblue,
      urlcolor=blue,
      filecolor=blue,
      citecolor=red,
linktocpage=true,
      pdfstartview=FitV,
      bookmarksopen=true
      ]{hyperref}

 \usepackage{amsmath,amssymb,graphicx,subfig,bm,euscript}

 \usepackage{cancel,mathtools,setspace}

 \def\eq#1{Eq.~(\ref{#1})}
 \newcommand{\secn}[1]{Section~\ref{#1}}
 \def\fig#1{fig.~{\ref{#1}}}
 \def\Fig#1{Fig.~{\ref{#1}}}
 \def\beq{\begin{equation}}
 \def\eeq{\end{equation}}
 \def\beqa{\begin{eqnarray}}
 \def\eeqa{\end{eqnarray}}
 \newcommand{\e}{\epsilon}
 \newcommand{\bs}{\boldsymbol}
 \def\one{\!\!{\hbox{ 1\kern-.8mm l}}}
 \newcommand{\bra}[1]{\langle{#1}|}
 \newcommand{\ket}[1]{|{#1}\rangle}
 
 \newcommand{\ex}[1]{{\rm e}^{#1}}
 \renewcommand{\d}{d}
 \def\ii{{\rm i}}
 
 \newcommand{\veps}{\ensuremath{\vec{\epsilon}}}
 \newcommand{\btau}{\ensuremath{\boldsymbol{\tau}}}
 
\newcommand{\dotminus}{-}
\newcommand{\zn}{\ex{ \ii \pi \vec{\varsigma} \cdot \vec{N}_\alpha}}

\newcommand{\kso}{\ex{\ii \pi\varsigma_1}k_1^{1/2}}
\newcommand{\kst}{\ex{\ii \pi\varsigma_2}k_2^{1/2}}
\newcommand{\ksot}{\ex{ \ii \pi (\varsigma_1 + \varsigma_2)} k_1^{1/2} k_2^{1/2}}

\newcommand{\tran}{^{\text{t}}}

\DeclareMathAlphabet\EuRoman{U}{eur}{m}{n}
\SetMathAlphabet\EuRoman{bold}{U}{eur}{b}{n}
\newcommand{\eurom}{\EuRoman}
\newcommand{\dd}{\eurom{d}}

\usepackage{graphicx}
\usepackage{color}
\usepackage{bm}
\usepackage{mathrsfs}
\usepackage{subfig}
\usepackage{tikz}
\usetikzlibrary{decorations.markings}
\usetikzlibrary{decorations.pathmorphing,decorations.markings,trees}
\usetikzlibrary{calc}

\numberwithin{equation}{section}

\tikzset{
    photon/.style={decorate, decoration={snake}},
    chargedphoton/.style={decorate, decoration={snake,markings,mark=at position .5 with {\arrow[draw=blue]{>}}}},
    particle/.style={draw=blue, postaction={decorate},
        decoration={markings,mark=at position .5 with {\arrow[draw=blue]{>}}}},
    antiparticle/.style={draw=blue, postaction={decorate},
        decoration={markings,mark=at position .5 with {\arrow[draw=blue]{<}}}},
    gluon/.style={decorate, draw=black,
        decoration={coil,amplitude=4pt, segment length=4pt}},
	ghost/.style={very thick,dotted}
     }

\newcommand{\prop}[1]{
     \begin{tikzpicture} [baseline=-2,scale=0.3]
 \draw[#1] (0,0) -- (4,0);
  \end{tikzpicture}
  }
\newcommand{\apple}[3]{
  \begin{tikzpicture} [baseline=-2,scale=0.3]
    \draw[#1] (0,-2) arc (270:90:2) ;
    \draw[#3] (0,2) arc (90:-90:2) ;
 \draw[#2] (0,2) -- (0,-2);
  \end{tikzpicture}
}
\newcommand{\figeight}[2]{
  \begin{tikzpicture} [baseline=-2,scale=0.3]
        \draw[#2] (0,0) -- ($(135:2)+(3,0)$) arc (135:-135:2)  -- (0,0);
 \draw[#1]  (0,0) -- ($(-45:2)-(3,0)$) arc (-45:-315:2)  -- (0,0);
  \end{tikzpicture}
}

\newcommand{\figeightgluons}{\figeight{gluon}{gluon}}

\newcommand{\figeighttwoNscalars}{\figeight{}{}}
\newcommand{\stringeightdiagram}{
  \begin{tikzpicture} [scale=0.25]
      	\draw ($(5,0)+(225:4)$) arc (225:495:4) .. controls (0,0.5) and (0,0.5) ..  ($(-5,0)+(45:4)$)  arc (45:315:4) .. controls (0,-0.5) and (0,-0.5) ..   ($(5,0)+(225:4)$);
	\draw (-5,0) circle (3.7);
	\draw (5,0) circle (3.7);
  \end{tikzpicture}
}
\newcommand{\stringapplediagram}{
  \begin{tikzpicture} [scale=0.25]
        \draw (0,0) circle(4);
	\draw (105:3.7) arc (105:255:3.7).. controls (-0.3,-3.6) and (-0.3,-3.6) .. (-0.15,-3)--(-0.15,3).. controls (-0.3,3.6) and (-0.3,3.6) ..(105:3.7) ;
	\draw (75:3.7) arc (75:-75:3.7).. controls (0.3,-3.6) and (0.3,-3.6) .. (0.15,-3)--(0.15,3).. controls (0.3,3.6) and (0.3,3.6) ..(75:3.7) ;
  \end{tikzpicture}
}
\newcommand{\stringhandlediagram}{
  \begin{tikzpicture} [scale=0.25]
      	\draw ($(8,0)+(225:4)$) arc (225:495:4)  .. controls (4,1.5) and (4,0.15).. (3,0.15) -- (-3,0.15).. controls (-4,.15) and (-4,1.5).. ($(-8,0)+(45:4)$)  arc (45:315:4) .. controls (-4,-1.5) and (-4,-0.15).. (-3,-0.15)-- (3,-0.15) .. controls (4,-.15) and (4,-1.5)..  ($(8,0)+(225:4)$);
	\draw (-8,0) circle (3.7);
	\draw (8,0) circle (3.7);
  \end{tikzpicture}
}

\usepackage[left=25mm,right=25mm,top=30mm,bottom=40mm]{geometry}

\allowdisplaybreaks

\begin{document}

\begin{titlepage}

\begin{flushright}
QMUL-PH-15-06\\
\vspace*{-25pt}
\end{flushright}

\begin{center}

\vspace{1cm}

{\Large \bf Two-loop Yang-Mills diagrams from superstring amplitudes}

\vspace{8mm}

Lorenzo Magnea$^{\, a,}$\footnote{email: {\tt lorenzo.magnea@unito.it}},
Sam Playle$^{\, a,}$\footnote{email: {\tt playle@to.infn.it}},
Rodolfo Russo$^{\, b,}$\footnote{email: {\tt r.russo@qmul.ac.uk}},
and Stefano Sciuto$^{\, a,}$\footnote{email: {\tt sciuto@to.infn.it}}

\vskip .5cm

$^a\,${\sl Dipartimento di Fisica, Universit\`a di Torino  \\
and INFN, Sezione di Torino}\\
{\sl Via P. Giuria 1, I-10125 Torino, Italy}\\

\vskip .3cm

$^b\,${\sl Centre for Research in String Theory \\ School of Physics and Astronomy\\
Queen Mary University of London\\
Mile End Road, London, E1 4NS,
United Kingdom}\\

\vskip 1.2cm

\begin{abstract}

\noindent

Starting from the superstring amplitude describing interactions among
D-branes with a constant world-volume field strength, we present a detailed analysis
of how the open string degeneration limits reproduce the corresponding field theory
Feynman diagrams. A key ingredient in the string construction is represented by the
twisted (Prym) super differentials, as their periods encode the information about the
background field. We provide an efficient method to calculate perturbatively the determinant
of the twisted period matrix in terms of sets of super-moduli appropriate to the degeneration
limits. Using this result we show that there is a precise one-to-one correspondence
between the degeneration of different factors in the superstring amplitudes and
one-particle irreducible Feynman diagrams capturing the gauge theory effective
action at the two-loop level.

\end{abstract}

\end{center}

\vfill

\end{titlepage}


\section{Introduction}
\label{intro}

The study of scattering amplitudes has played a central r\^ole in the development
of string theory since its very beginning. In the seventies and the eighties it was
instrumental in showing that superstring theories provide perturbative gravitational
models that, at loop level, are free of ultraviolet divergences and anomalies. The analysis
of string amplitudes was also crucial in the discovery of D-branes and in the development
of the web of dualities among different superstring theories. It is, then, not surprising
that this field continues to be under intense study. Recently, there has been renewed
interest in several aspects of string perturbation theory in the RNS formalism, with particular focus on contributions
beyond one loop: for example, higher-loop diagrams with Ramond external states were
discussed in Refs.~\cite{Witten:2015hwa,D'Hoker:2015kwa}; further, Refs.~\cite{Pius:2014iaa,
Pius:2014gza,Sen:2014pia}~focused on the off-shell extension of amplitudes, studying
various situations where this is necessary; finally, Ref.~\cite{D'Hoker:2013eea,D'Hoker:2014gfa} derived
an explicit result for the $D^6{\cal R}^4$ term in the type-IIB effective action, checking
the predictions following from S-duality and supersymmetry. Another interesting approach
to the point-like limit of closed string amplitudes as a `tropical' limit was discussed in
\cite{Tourkine:2013rda}. For recent reviews on multiloop string amplitudes, with a
more complete list of references, we refer the reader to~\cite{Witten:2012bh,D'Hoker:2014nna}.

Two themes in particular have been at the center of much important progress in our
understanding of string interactions: the study of the mathematical properties of the
world-sheet formulation of string amplitudes, and their relation to the effective actions
describing the light degrees of freedom present in the theory. In this paper, we touch
on both these aspects by studying in detail the open string degeneration limits of two-loop
amplitudes described by a world-sheet with three borders and no handles. In particular,
we expand upon the results of \cite{Magnea:2013lna}: starting from the Neveu-Schwarz
(NS) sector of the open superstring partition function in the background of a constant
magnetic field strength, we derive the Euler-Heisenberg effective action for a gauge
theory coupled to scalar fields in the `Coulomb phase'. The idea of using string theory to
investigate effective actions in constant electromagnetic fields has a long history, and
was studied at one loop in~\cite{Fradkin:1985qd,Abouelsaood:1986gd,Bachas:1992bh},
with some results for the bosonic theory at two loops given in \cite{Magnea:2004ai}. In
our analysis we find exact agreement between calculations in field theory and string
theory, in the infinite-tension limit, for the two-loop correction to the effective action.
Furthermore, we find that the correspondence holds not just for the whole amplitude,
but we can precisely identify the string origin of all individual one-particle irreducible
(1PI) Feynman diagrams contributing to the effective action. In order to do so, on the
string theory side we need to use appropriate world-sheet super-moduli, respecting
the symmetry of the Feynman graphs, while on the field theory side we need to use
a version of the non-linear gauge condition introduced by Gervais and Neveu in
\cite{Gervais:1972tr}, modified by dimensional reduction to involve the scalars also,
and given here in \eq{scalargaugecon}.

On the formal side, it is advantageous to use the formalism of \emph{super}
Riemann surfaces~\cite{Crane:1986uf,Martinec:1986bq,Giddings:1987wn,
D'Hoker:1988ta,Witten:2012ga,Witten:2012bh}, in which the complex structure
is generalized to a super-conformal structure, with local super-conformal
coordinates $(z|\theta)$. We follow this approach by constructing the two-loop
amplitude in the Schottky parametrization, since there is a close relationship
between Schottky super-moduli, in particular the `multipliers', and the sewing
parameters of plumbing fixtures. This in turn relates the bosonic world-sheet
moduli to the Schwinger parameters associated to the propagators in Feynman
graphs, which provides the ideal framework for studying the connection between
string integrands and field theory Feynman diagrams. In the bosonic case, it is
possible to describe genus $h$ Riemann surfaces as quotients of the Riemann
sphere (with a discrete set of points removed) by a discrete (Schottky) group, freely
generated by $h$ M\"obius transformations. Heuristically, quotienting the Riemann
sphere by a M\"obius transformation has the effect of cutting out a pair of circles
and gluing them to each other along their boundaries. Schottky groups arose
naturally in the early treatment of multi-loop string amplitudes~\cite{Lovelace:1970sj,
Kaku:1970ym,Alessandrini:1971cz,Olive:1971fu,Alessandrini:1971dd,Montonen:1974jj}
and remained useful \cite{DiVecchia:1986uu,Martinec:1986bq,DiVecchia:1987uf,
DiVecchia:1988jy,DiVecchia:1988cy,DiVecchia:1989id} even after alternative
methods of analysis were found. In the supersymmetric case, higher genus super
Riemann surfaces are similarly generated by quotienting the super manifold
${\bf CP}^{1|1}$ (with a discrete set of points removed) by a discrete group,
generated by $h$ `super-projective' ${\rm OSp}(1|2)$ transformations.

As is well known, the presence of a constant background field strength in the
space-time description of the amplitudes translates on the world-sheet side into the
presence of non-trivial monodromies along either the $a$ or the $b$ cycles of the
Riemann surface. It is thus not surprising that the amplitudes we are interested in
involve super $1|1$-forms (sections of the Berezinian bundle) with twisted periodicities,
also known as Prym differentials. The bosonic counterparts of these objects
was discussed, in the Schottky parametrization, in~\cite{Russo:2003tt}, and their
periods along the untwisted cycles appear in any string amplitude where the fields
have non-trivial monodromies~\cite{Aoki:2003sy,Russo:2003yk,Magnea:2004ai,
Antoniadis:2005sd}. We extend these past results in two directions: first we
generalise the twisted period matrix to the supersymmetric case; then we must
calculate the supersymmetric version of the twisted determinant to sufficiently
high order in the complete degeneration limit, so as to obtain the gauge theory
Feynman graphs with multiple gluon propagators. In order to do this, we introduce
an alternative formulation of the twisted super-determinant in terms of an integral
along a Pochhammer contour, and we show that this simplifies drastically
its perturbative evaluation in the Schottky parametrization.

\begin{figure}[h!]
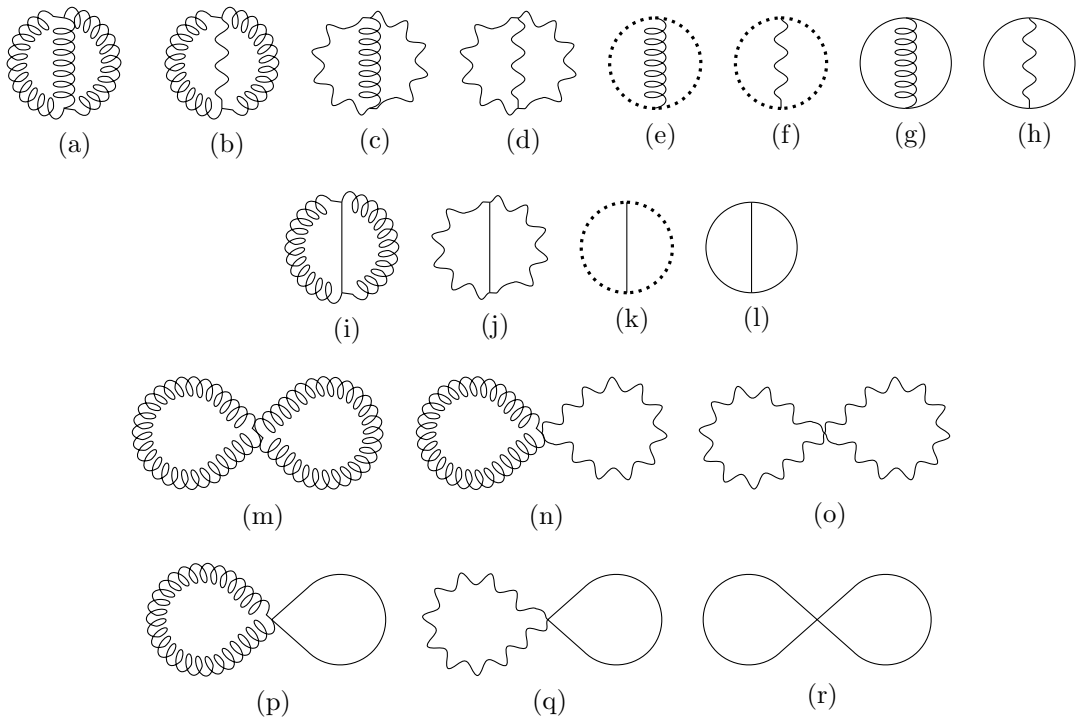

\begin{center}
\subfloat[]{\apple{gluon}{gluon}{gluon} \label{fig:mmm1} }
\subfloat[]{\apple{gluon}{photon}{gluon} \label{fig:mnm1}}
\subfloat[]{\apple{photon}{gluon}{photon} \label{fig:nmn1}}
\subfloat[]{\apple{photon}{photon}{photon} \label{fig:nnn1}}
\subfloat[]{\apple{ghost}{gluon}{ghost} \label{fig:gmg1} }
\subfloat[]{\apple{ghost}{photon}{ghost} \label{fig:gng1}}
\subfloat[]{\apple{}{gluon}{} \label{fig:sms1}}
\subfloat[]{\apple{}{photon}{} \label{fig:sns1}} \\
\subfloat[]{\apple{gluon}{}{gluon} \label{fig:msm1}}
\subfloat[]{\apple{photon}{}{photon} \label{fig:nsn1}}
\subfloat[]{\apple{ghost}{}{ghost} \label{fig:gsg1}}
\subfloat[]{\apple{}{}{} \label{fig:sss1}}  \\
\subfloat[]{\figeight{gluon}{gluon} \label{fig:mm1}}
\subfloat[]{\figeight{gluon}{photon} \label{fig:mn1}}
\subfloat[]{\figeight{photon}{photon} \label{fig:nn1}} \\
\subfloat[]{\figeight{gluon}{} \label{fig:ms1}}
\subfloat[]{\figeight{photon}{} \label{fig:ns1}}
\subfloat[]{\figeight{}{} \label{fig:ss1} }
\caption{Two-loop 1PI vacuum Feynman graphs in Yang-Mills with adjoint scalars
with VEVs. The dotted edges signify Faddeev-Popov ghosts, and the plain edges
symbolize scalars, the helical edges denote gluons polarized \emph{parallel} to
the plane of the background magnetic field and the wavy edges indicate gluons
polarized \emph{perpendicular} to the background magnetic field.}
\label{fig:1PIgraphs}
\end{center}
\end{figure}

The main result of this paper is to show how the two-loop 1PI Feynman diagrams
listed in \Fig{fig:1PIgraphs} arise from the degeneration limits of the superstring
result. The graphical notation for the field propagators is explained in detail in Appendix
\ref{Appc}; here we note in particular that we are using two different types of edges to
denote gluons, depending on whether they are polarized parallel or perpendicular to
the plane of the background field. We note also that some of the graphs (those in
Figs.~\ref{fig:msm1}--\ref{fig:sss1}) include vertices with an odd number of scalars:
these vertices arise because of the non-vanishing scalar vacuum expectation values
(to which these graphs are proportional); these diagrams appear automatically in the
string calculation, and they appear on the field-theory side as a result of having imposed
the gauge condition of Gervais and Neveu \cite{Gervais:1972tr} \emph{before} dimensional
reduction.
Our investigation is thus also a contribution to a long-standing program aimed to
use string theory to gain insights into field-theory amplitudes, which was started
in Ref.~\cite{Scherk:1971xy} in the language of dual models, and generalized to
the superstring framework in~\cite{Green:1982sw}. The practical usefulness of
string theory as an organizing principle for tree-level gauge-theory amplitudes
was first noticed and applied in~\cite{Mangano:1987xk,Mangano:1990by}.
At genus one, several results are available in the literature: they include the
derivation of the leading contribution to the Callan-Symanzik $\beta$-function
of pure Yang-Mills theory in \cite{Bern:1987tw}, as well as a general analysis of
one-loop scattering amplitudes in \cite{Bern:1990cu,Bern:1990ux,Bern:1991aq,
Bern:1991an,Bern:1992cz}. This was later used to calculate the one-loop five-gluon
amplitude in QCD for the first time in \cite{Bern:1993mq}. String theory also inspired
many developments in the world-line approach to perturbative quantum field theory
(QFT), starting with the work of Strassler in~\cite{Strassler:1992zr}, with subsequent
progress in~\cite{Schmidt:1993rk,Schmidt:1994da}, summarized in~\cite{Schubert:2001he},
and more recently, for example, in~\cite{Dai:2006vj,Bastianelli:2007pv}. Bosonic
strings were also used to compute Yang-Mills renormalization constants at one loop
in \cite{DiVecchia:1996iz}, and one-loop off-shell gluon Green's functions in
\cite{Frizzo:2000ez}. At the two-loop level much less is known: explicit QFT
amplitudes with only scalar fields were obtained from bosonic strings in
\cite{DiVecchia:1996kf} and \cite{Frizzo:1999zx,Marotta:1999re}. Two-loop
amplitudes with gluons, however, have proved difficult to study with this
technology \cite{Magnea:1997kh,Magnea:1997kv,Kors:2000bb}. Our analysis
here marks significant progress in this direction, showing that the prescriptions
discussed in~\cite{Magnea:2013lna} are indeed sufficient to derive from string
theory all the bosonic two-loop 1PI gauge-theory diagrams listed in~\Fig{fig:1PIgraphs}.

The structure of the paper is as follows. In \secn{sts} we describe the D-brane
setup in which our calculations are carried out. In \secn{dmf} we recall the
integration measure for the NS sector of open superstrings in the super Schottky
parametrization and explain how to modify it in order to accommodate our background.
In \secn{QFTlim} we expand the measure in powers the Schottky multipliers,
and then we identify the appropriate parametrizations to describe the two
degenerations of the Riemann surface which are relevant for our purposes: the
{\it symmetric} degeneration leading to the diagrams with the topology of Figs.
\ref{fig:mmm1}--\ref{fig:sss1}, and the {\it incomplete} degeneration, leading to
diagrams with only two field-theory propagators and a four-point vertex, depicted
in Figs.~\ref{fig:mm1}--\ref{fig:ss1}. An analysis of the various factors contributing
to the string amplitude, arising from different world-sheet conformal field theories,
then enables to unambiguously identify each diagram in the field-theory limit.
In \secn{YMQFT} we obtain and discuss the Lagrangian for the world-volume QFT
in the appropriate non-linear gauge, and we use it to compute example Feynman diagrams.
Finally, in \secn{cfQFT} we compare our string-theory and QFT calculations, and in
\secn{cfBos} we discuss the differences between the present calculation and the
analogous calculation using the bosonic string. In Appendix \ref{Appa} we discuss
super-projective transformations and the super Schottky group, in Appendix \ref{Appb}
we give the calculation of the twisted (Prym) super period matrix, and in Appendix
\ref{Appc} we list the values of all of the Feynman graphs in \Fig{fig:1PIgraphs} with
our choice of background fields.


\section{The string theory setup}
\label{sts}

\begin{figure}
\centering
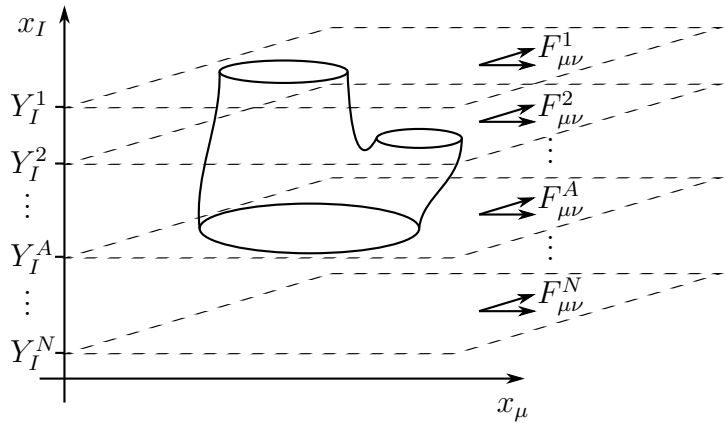
\caption{A stack of of spatially separated D-branes with constant gauge fields on
their world-volumes, connected by open strings ending on three different branes,
in a double-annulus configuration.}
\label{fig:DBranesFields}
\end{figure}

\begin{figure}
\centering
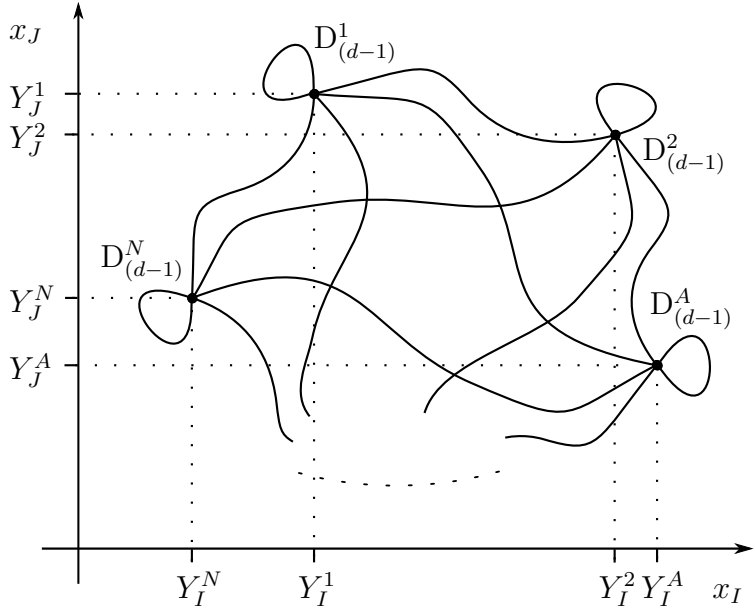
\caption{A two-dimensional section of the space transverse to the D-branes,
which therefore appear as points, connected by a web of open strings.}
\label{fig:DBranePositions}
\end{figure}

We consider a stack of $N$ parallel $d$-dimensional D-branes embedded in
a ${\cal D}$-dimensional Minkowski space-time, where, as usual, ${\cal D} =
10$ for type II theories and ${\cal D} = 26$ for bosonic string theory. When
$d < {\cal D} - 2$, and provided the string coupling $g_s$ is small, so that
$g_s N \ll 1$, this configuration can be described in terms of open strings
moving in flat space and being supported by the D-branes. We will work generically
in the `Coulomb phase' where the D-branes are spatially separated from each
other in the directions perpendicular to their world-volumes. Furthermore, on each
of the D-branes we switch on a uniform U(1) background field in the
$\{x_1, x_2\}$ plane, with a field strength tensor given by
\beq
  F_{\mu \nu}^A \, = \, B^A \, \big( \eta_{\mu 1} \eta_{2 \nu} - \eta_{\mu 2}
  \eta_{\nu 1} \big) \, ,
\label{eq:FB}
\eeq
where $B^A$ is a constant `magnetic' field on the $A$-th brane (thus $A = 1,
\ldots, N)$. The positions of the D-branes in the transverse directions will be
labelled by $Y_I^A$, with $I = d, d+1, \ldots , {\cal D} - 1$. Such a D-brane
configuration is depicted from various viewpoints in Figs. \ref{fig:DBranesFields}
and \ref{fig:DBranePositions}.
A string stretched between branes $A$ and $B$ will have squared length
\beq
Y_{A B}^2 \, = \, \sum_{I = d}^{{\cal D} - 1} \Big( Y_I^A - Y_I^B \Big)^2 \, ,
\label{length}
\eeq
and will receive a classical contribution $m_{AB}$ to its mass from the elastic
potential energy associated with the stretching of the string, given by
\beq
  m_{AB} \, = \, T \, Y_{AB} \, = \, \frac{Y_{AB}}{2 \pi \alpha' } \, ,
\label{eq:m2Y}
\eeq
where $T$ is the string tension and $\alpha'$ the related Regge slope. These
strings will also be charged under the magnetic fields $B^A$ and $B^B$, with the
sign of the charge depending on their orientation. Open strings that start and end
on the same D-brane are uncharged and their mass is independent of $Y_I^{A}$.
For generic values of $Y_I^A$, this configuration breaks the symmetry of the
world-volume theory from ${\rm U}(N)$ to ${\rm U}(1)^N$.

The theory describing open strings supported by this D-brane configuration is
free~\cite{Fradkin:1985qd,Abouelsaood:1986gd}. The constant background
magnetic fields on the D-brane world-volumes manifest themselves in the
world-sheet picture by altering the boundary conditions of string coordinates in the magnetized plane. On the double cover of the surface, this gives twisted
boundary conditions, or, in other words, non-trivial monodromies, to the zero
modes in the two magnetized space directions. To describe this setup, we
will use the conventions of Section~3 of Ref.~\cite{Magnea:2013lna}, which
we summarize below.


To begin with, let us briefly consider the spectrum of low-lying string excitations.
In the bosonic case, the world-sheet theory, in a covariant approach, comprises
${\cal D}$ embedding coordinates $X^\mu$ and the ghost system $(b,c)$. The
holomorphic components of these fields admit the mode expansions
\beq
  b(z) \, = \, \sum_{n \in \mathbf{Z}} b_n z^{- n - 2} \, ,
  \quad
  c(z) \, = \, \sum_{n \in \mathbf{Z}} c_n z^{ - n + 1} \, ,
  \quad
  \partial_z X^\mu \, =  \, - \, \ii \, \sqrt{2 \alpha'} \, \sum_{n \in \mathbf{Z}}
  \alpha_n^\mu \, z^{- n - 1} \, .
\label{eq:bcme}
\eeq
In the presence of constant abelian background fields, the theory remains free,
but string coordinates in directions parallel to the magnetized plane acquire
twisted boundary conditions and must be treated separately. Considering strings
ending on branes $A$ and $B$, it is convenient to introduce the combinations
$Z^\pm_{AB} = (X^1_{AB} \pm \ii X^2_{AB})/\sqrt{2}$. These combinations
diagonalize the boundary conditions and yield the mode expansions
\beq
  \partial_z Z^\pm_{AB} \, = \, - \, \ii \, \sqrt{2 \alpha'}  \sum_{n \in \mathbf{Z}}
  \alpha_{n \pm \epsilon_{AB}}^\pm \, z^{- n - 1 \pm \epsilon_{AB}} \, ,
\label{eq:XmeUa}
\eeq
where we defined
\beq
    \tan \left( \pi \epsilon_{AB} \right) \, \equiv \, 2 \pi \alpha' \left( B^A - B^B \right) \, .
\label{eq:eB}
\eeq
After canonical quantization, the modes introduced above satisfy standard
commutation relations, except for magnetized directions, where one finds
\beq
  \Big[ \alpha_{n + \epsilon_{A B} }^+ , \alpha_{m - \epsilon_{A B} }^- \Big] \, = \,
  \left( n + \epsilon_{A B} \right) \delta_{n + m} \, .
\label{osccomm}
\eeq
As usual in covariant quantization, not all states in the Fock space obtained by
acting with the creation modes on the $SL(2, {\bf R})$-invariant vacuum $\ket{0}$
are physical: we need to select only the states belonging to the cohomology of
the world-sheet BRST charge
\beq
  Q_B^W \, = \, \oint \frac{\d z}{2\pi \ii} \, c \left( - \frac{1}{4 \alpha'} \partial X^M
  \partial X_M + (\partial c) b \right) \, .
\label{eq:QW}
\eeq
In the bosonic theory, the lowest-lying physical state is a tachyon $\ket{\mathbf k}
\equiv c_{1} | k, 0 \rangle$, with mass-shell condition $k^2 = - m^2 = 1/\alpha'$.
The next mass level comprises $({\cal D} + 2)$ massless states, which will be the
focus of our analysis in the field theory limit: one finds two unphysical states, two
null states, and $({\cal D} - 2)$ physical polarization states appropriate for massless
gauge bosons. A crucial ingredient of our analysis is the mapping between these
string states and the space-time states in the limiting quantum field theory: as noticed
for instance in Chapter~4 of~\cite{Polchinski:1998rq}, the action of the worldsheet
BRST charge~\eqref{eq:QW} on the $({\cal D} + 2)$ massless states mirrors the
linearized action of the space-time BRST charge for the U($N$) gauge symmetry:
in particular, the states created by world-sheet ghost oscillators, $c_{-1} \ket{\mathbf k}$
and $b_{-1} \ket{\mathbf k}$, behave as the spacetime ghosts $C$ and $\overline{C}$.
Acting with the $\alpha_{-1}^M$ oscillators, on the other hand, generates $d$ states
along the D-brane, and $n_s = {\cal D} - d \,$ states associated to the $n_s$ directions transverse to the D-brane, representing
respectively the $d$ polarisations of the gauge vectors (including two unphysical ones),
and $n_s$ adjoint scalars. To be precise, the world-sheet BRST charge $Q_B^W$
acts as
\beq
  Q_B^W  b_{-1} \ket{\bf k}  = \sqrt{2\alpha'} k \cdot \alpha_{-1} \ket{\bf k}  \, ; \quad
  Q_B^W  \alpha_{-1}^M \ket{\bf k} = \sqrt{2\alpha'} k^M c_{-1} \ket{\bf k}  \, ; \quad
  Q_B^W  c_{-1} \ket{\bf k}  \, = \, 0 \, ,
\label{QbrstWS}
\eeq
while the linearised space-time BRST transformation $\delta_B$ acts as
\beq
  \delta_B \left( \, \overline{C}^{\, a} \right) \, \sim \, \partial \cdot Q^a \, ; \quad
  \delta_B \left( Q_\mu^{\, a} \right) \, \sim \, \partial_\mu C^{\, a}  \, ; \quad  \delta_B \left( Q_I^{\, a} \right) \, \sim 0 ; \quad
  \delta_B \left( C^{\, a} \right) \, \sim 0 \, ,
\label{QbrstST}
\eeq
where $a$ is an adjoint index, $Q_\mu^{\, a}$ and $Q_I^{\, a}$ stand for a gluon mode and a
scalar, depending on whether $X^M$ is parallel or perpendicular to the D-brane, and $k^M=\{k^\mu, 0\}$.

This simple relation between world-sheet and space-time states is preserved in
perturbation theory, when the string coupling is switched on and non-linear terms
in the BRST operators must be taken into account. This is expected, since, in a
perturbative analysis, fields propagating between interaction vertices are free. In
practice, we will test this statement by calculating a string diagram with the world-sheet
topology of a degenerating double-annulus, and identifying the contributions coming
from the various massless states listed above, as they propagate through the
diagram. We will show that each contribution matches the gauge theory result,
where the corresponding space-time fields propagate in the matching edge of the
relevant Feynman diagram, provided that the gauge used in field theory is the
nonlinear Gervais-Neveu gauge, introduced in \cite{Gervais:1972tr}. In this way,
we can identify individual Feynman diagrams in the target field theory directly
at the level of the string amplitude, picking a specific boundary of the string
moduli space, and identifying the string states as they propagate along the
degenerating surface.

A similar analysis holds also in the superstring case. In the RNS formalism one
needs to introduce the extra world-sheet fields $\psi^\mu$, $\beta$ and $\gamma$,
that are the partners under world-sheet supersymmetry of the $\partial X^\mu$, $b$
and $c$ fields mentioned above. The monodromies for these new fields will be the
same as those of their partners, except for a possible extra sign, which is allowed
for fields of half-integer weight, and distinguishes the Ramond from the Neveu-Schwarz
sectors. In this paper we will focus on the Neveu-Schwarz contributions: the analysis of the states at the first mass level, above the tachyonic ground
state $\ket{\mathbf k}$, parallels that of the bosonic case. The only difference is
that the relevant modes are $\psi_{-1/2}$, $\beta_{-1/2}$ and $\gamma_{-1/2}$:
in the superstring partition function, the low energy limit will be performed by
focusing on the contributions of states with half-integer weight.


\section{The superstring partition function for the NS sector}
\label{dmf}

\begin{figure}
\centering
\def\svgwidth{10cm}
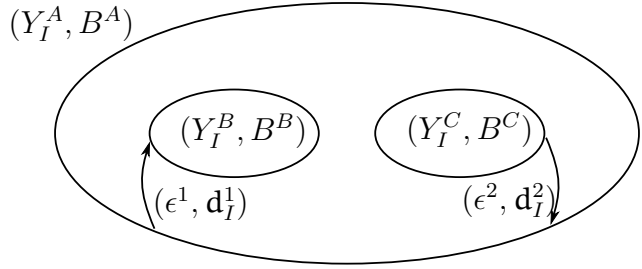
\caption{The double annulus world-sheet, with three boundaries labeled with $i = 0, 1, 2$
attached to three D-branes, with Chan-Paton factors A, B, C. The relative positions and
background field strengths of branes B and C with respect to brane A determine the
masses and the twisted boundary conditions, as described in the text. }
\label{fig:borlab}
\end{figure}

From the world-sheet point of view, the interaction among D-branes is described by
the string vacuum amplitude (the partition function) with boundaries, as  depicted in
Fig.~\ref{fig:DBranesFields}. The case of two magnetized D-branes, corresponding
to a one loop-amplitude, has been well studied~\cite{Fradkin:1985qd,Abouelsaood:1986gd,
Metsaev:1987ju,Bachas:1992bh}. Here we will focus on planar world-sheets, and most
of what we will say in this section applies to surfaces with $(h + 1)$ borders,  corresponding
to $h$-loop open superstring diagrams, but restricted to the NS sector, where the
super-Schottky formalism described in Ref.~\cite{DiVecchia:1988jy} can be used.
In particular, as discussed in \secn{sts}, we consider parallel magnetized D-branes
that can be separated in the directions transverse to their world-volumes. As a consequence,
and as depicted in Fig.~\ref{fig:borlab} for a (two-loop) surface with three boundaries,
the partition function depends on two set of variables: the relative distances among
D-branes, and the magnetic field gradients between pairs of D-branes.

To be precise, let us label the $(h + 1)$ world-sheet borders with $i = 0, 1, \ldots, h$.
Then we can label the D-brane to which the $i$-th border is attached with the integer
$A_i$, with $A_i \in \{1, \ldots, N\}$, and $A_i \leq A_{i + 1}$. To get the full amplitude,
we will have to sum over the $A_i$'s. Having fixed $A_0, \ldots, A_h$, we can take the
$A_0$-th brane as a reference and define
\beqa
  \dd_I^i & = & Y_I^{A_0} - Y_I^{A_i}  \, \qquad \left( I \, = \, d, \ldots , {\cal D} - 1 \right) \, ,
  \nonumber \\
  \tan \left( \pi \epsilon^i \right) & = & 2 \pi \alpha' \left(B^{A_0} - B^{A_i} \right) \, .
\label{deftewovec}
\eeqa
The variables $\epsilon^i$ thus form an $h$-dimensional vector, which we will denote
by $\vec{\epsilon} \, $; similarly, the variables $\dd_I^i$, which have dimension of length,
form $n_s \,$ $h$-dimensional vectors, which we will label $\vec{\dd}_I$. The classical
mass of the string stretching between the $A_0$-th brane and the $A_i$-th brane is then
given by
\beq
  m^2_i \, = \,  \frac{1}{(2 \pi \alpha')^2} \, \sum_{I = d}^{{\cal D} - 1}
  \left( \dd_I^i \right)^2 \, .
\label{massagain}
\eeq
Notice finally that, for $h = 2$, as depicted in Fig.~\ref{fig:borlab}, we make a slight
variation in this notation by flipping the sign of the second component of the
two-dimensional vectors $\vec{\epsilon}$ and $\vec{\dd}_I$, which will be useful
to take full advantage of the extra symmetry at two loops.

The string partition function in this setup can be written as follows. For our purposes,
it is useful to keep separate the contributions of the different conformal field theory
sectors, which leads to the expression
\beq
  {\bf Z}_h \big( \veps, \vec{\dd}\, \big) \, = \, {\cal N}_h^{\, (\veps \,)} \! \int d {\bs \mu}_h \,
  {\bf F}_{\rm gh} \left( {\bs \mu} \right) \, {\bf F}_{\rm scal}^{(\vec{\dd}\,)}
  \left( {\bs \mu} \right) \, {\bf F}_{\parallel}^{(\veps\,)} \left( {\bs \mu} \right) \,
  {\bf F}_{\perp} \left( {\bs \mu} \right) \, .
\label{eq:mSetup}
\eeq
Here ${\cal N}_h^{\, (\veps \,)}$ is a field-dependent normalization factor, to be
discussed in \secn{GSO}, and we denoted the contributions of the world-sheet ghost
systems $b, \, c$ and $\beta, \, \gamma$ by ${\bf F}_{\rm gh}$, that of the string fields
$X_I, \, \psi_I$ perpendicular to the D-branes by ${\bf F}_{\rm scal}^{(\vec{\dd} \,\, )}$,
while the contribution of the fields along the D-branes has been separated into sectors
parallel (${\bf F}_{\parallel}^{(\veps \,)}$) and perpendicular (${\bf F}_{\perp}$) to the
magnetized directions. Finally, ${\bs \mu}$ denotes collectively the supermoduli: here
we use the super-Schottky formalism, reviewed in Appendix~\ref{Appa}, where the
supermoduli are the sewing parameters $\ex{\ii \pi \varsigma_i} {k_i}^{1/2}$ (with
$\varsigma_i \in \{ 0, 1 \}$) and the fixed points $(u_i | \theta_i)$, $(v_i | \phi_i)$
of $h$ super-projective transformations $i = 1, \ldots, h$. Note that we explicitly associate
with each Schottky multiplier $k_i$ the phase $\varsigma_i$ associated with the NS spin
structure around the $b_i$ homology cycle. In this parametrization the measure
$d {\bs \mu}_h$ reads~\cite{DiVecchia:1988jy}
\beq
   d {\bs \mu}_h \, = \, \left[ \frac{\sqrt{ ( \mathbf{v}_1 \dotminus \mathbf{u}_1)
   (\mathbf{u}_1 \dotminus \mathbf{v}_2 ) ( \mathbf{v}_2 \dotminus \mathbf{v}_1 ) } }{d
   \mathbf{v}_1 d \mathbf{u}_1 d \mathbf{v}_2 } \, d \Theta_{\mathbf{v}_1 \mathbf{u}_1
   \mathbf{v}_2} \right] \, \prod_{i = 1}^h \, \frac{d k_i \, \ex{\ii \pi \varsigma_i}}{k_i^{3/2}} \,
   \frac{d \mathbf{u}_i \, d \mathbf{v}_i}{\mathbf{v}_i \dotminus \mathbf{u}_i}  \; ,
\label{eq:muh}
\eeq
where we denote superconformal coordinates in boldface, and the notation $\mathbf{v}_i
\dotminus \mathbf{u}_i$ indicates the supersymmetric difference
\beq
  \mathbf{v}_i \dotminus \mathbf{u}_i \, \equiv \, v_i - u_i + \theta_i  \phi_i  \, .
\label{superdiff}
\eeq
The square parenthesis in \eq{eq:muh} takes into account the super-projective
invariance of the integrand, which allows us to fix three bosonic and two fermionic
variables. $\Theta_{\mathbf{v}_1 \mathbf{u}_1 \mathbf{v}_2}$ is the fermionic
super-projective invariant which can be constructed with three fixed points, defined
in Refs.~\cite{Hornfeck:1987wt,D'Hoker:1988ta}, and given explicitly in \eq{eq:FermInv}.
If we specialize \eq{eq:muh} to $h = 2$ we find
 \beq
  d {\bs \mu}_2 \, = \, \ex{\ii \pi (\varsigma_1 + \varsigma_2)} \frac{\d k_1}{k_1^{3/2}} \,
  \frac{\d k_2}{k_2^{3/2}} \, \frac{d \mathbf{u}_2 \, d \Theta_{\mathbf{v}_1 \mathbf{u}_1
  \mathbf{v}_2}}{\mathbf{v}_2 \dotminus \mathbf{u}_2} \,
  \sqrt{\frac{(\mathbf{u}_1 \dotminus \mathbf{v}_2)(\mathbf{v}_2 \dotminus
  \mathbf{v}_1)}{\mathbf{v}_1 \dotminus \mathbf{u}_1}} \, .
\label{ssc}
\eeq
Let us now examine in turn the various factors in the integrand of \eq{eq:mSetup}.
The ghost contribution is independent of both the magnetic fields and the D-brane
separations, so we can use the result of Ref.~\cite{DiVecchia:1988jy}, which reads
\beq
  \mathbf{F}_{\rm gh} ({\bs \mu}) \, = \, \frac{(1 - k_1)^2 \, (1 - k_2)^2}{\left(1
  + \kso \right)^2 \left(1 + \kst \right)^2} \, \, \, {\prod_\alpha}' \prod_{n = 2}^\infty
  \left( \frac{1 +  k_\alpha^n}{1 + \zn k_\alpha^{n - \frac{1}{2}}} \right)^{\! 2} \, .
\label{eq:ghostsf}
\eeq
In \eq{eq:ghostsf}, the notation $\prod_{\alpha}'$ means that the product is over all
primary classes of the super Schottky group: a primary class is an equivalence class
of primitive super Schottky group elements, \emph{i.e.}~those elements which cannot
be written as powers of another element; two primitive elements are in the same primary
class if one is related to the other by a cyclic permutation of its factors, or by inversion.
The vector $\vec{N}_\alpha$ has $h$ integer-valued components, and is defined
as follows: the $i$-th entry counts how many times the generator ${\bf S}_i$ enters
in the element of the super Schottky group ${\bf T}_\alpha$: more precisely, we
define $N_\alpha^i = 0$ for ${\bf T}_\alpha = \textbf{1}$ and $N_\alpha^i = N_\beta^i
\pm 1$ for ${\bf T}_\alpha = {\bf S}_i^{\pm 1} {\bf T}_\beta$. Finally, also $\vec{\varsigma}$
is a vector with $h$ components, with the $i$-th component denoting the spin
structure along the $b_i$ cycle, as noted above.

In fact, we need to be more precise about the notation in \eq{eq:ghostsf}, because
the half-integer powers of $k_\alpha$ could indicate either of the two branches of the
function. The notation is to be understood in the following way: when the spin structure
is $\vec{\varsigma} = 0$, we define the eigenvalue of the Schottky group element
${\bf T}_\alpha$ with the smallest absolute value to be $- k_\alpha^{{1}/{2}}$, see Eq~\eqref{eq:SuSbraket}. In
particular, we take $k_i^{1/2}$ to be \emph{positive}\footnote{This convention is the opposite to the one used in~\cite{Magnea:2013lna}.} for $i=1,\ldots,h$. This corresponds to the
fact that spinors are anti-periodic around a homology cycle with zero spin structure
(see, for example, Ref.~\cite{AlvarezGaume:1986es}). Furthermore, we expect the
partition function to be symmetric under the exchange of the homology cycles
$b_1$, $b_2$ and $b_1^{-1} \cdot b_2$ (depicted in Fig.~\ref{homology}), and
one can verify that $k^{1/2}({\bf S}_1^{-1} {\bf S}_2)$ is always positive whenever
$k_1^{1/2}$ and $k_2^{1/2}$ have the same sign. Our convention puts all three
multipliers on the same footing. Note that $k_\alpha^{1/2}$ is not in general positive when
${\bf T}_\alpha$ is not a generator: for example, the eigenvalues of ${\bf T}_\alpha
= {\bf S}_1 {\bf S}_2$ are positive when the spin structure is zero, so that $k^{1/2}
({\bf S}_1 {\bf S}_2)$, as computed in \eq{ktwolexp}, is negative.

The scalar contribution to \eq{eq:mSetup} depends on the separation between the
D-branes in the transverse directions, as shown in Fig.~\ref{fig:DBranePositions}.
We can write ${\bf F}_{\text{scal}}^{(\vec{\dd} \,\,)}$ as a product over the super Schottky
group, capturing the non-zero mode contribution, times a new factor ${\cal Y}({\bs \mu},
\vec{\dd} \,)$, as
\beq
  {\bf F}_{\text{scal}}^{(\vec{\dd} \,\,)} ({\bs \mu}) \, = \, {\cal Y} \big( {\bs \mu},
  \vec{\dd} \,\, \big) \, \, {\prod_\alpha}' \prod_{n = 1}^\infty \left(\frac{1 + \zn \,
  k_\alpha^{n - {1}/{2}}}{1 - k_\alpha^n} \right)^{\!\! n_s} \! \! .
\label{eq:scalarf}
\eeq
The explicit form of ${\cal Y}$ can be found by repeating the calculation performed in
Ref.~\cite{Frau:1997mq} for the bosonic theory, and replacing the period matrix $\tau$
with the super-period matrix ${\bs \tau}$ discussed in Appendix \ref{SuperPM}. We find
\beq
  {\cal Y} ({\bs \mu}, \vec{\dd} \,\,) \, \equiv \, \prod_{I = 1}^{n_s} \,
  \exp \left( \frac{\vec{\dd}_I \cdot \bs{\tau} \cdot \vec{\dd}_I}{2 \pi \ii \, \alpha'} \right)  \, .
\label{eq:Ydef}
\eeq
It is instructive, and useful for our later implementation, to consider explicitly the $h = 2$
case. Let the $i = 0, 1, 2$ borders of the world-sheet be on the D-branes labelled by $A$,
$B$ and $C$, respectively. As mentioned above, it is useful in this special $h = 2$ case
to define the $i = 2$ component $\dd_I^2$ with the opposite sign with respect to
\eq{deftewovec}, so we have $\dd_I^1 = Y_I^A - Y_I^B$ and $\dd_I^2 = Y_I^C - Y_I^A$.
By so doing, we can then define an additional (redundant) quantity, describing the
displacement between the D-branes attached to the $i = 1$ and $i = 2$ borders, as
$\dd_I^3 = Y_I^{B} - Y_I^{C}$. Now the three distances $\dd_I^i$ for $i=1, 2, 3$ are
on an equal footing, reflecting the symmetry of the world-sheet topology, and we have
$\dd_I^{1} + \dd_I^{2} + \dd_I^{3} = 0$ (see \Fig{fig:borlab}).
%
%
One may easily verify that the product over the $n_s$ transverse directions in
\eq{eq:Ydef} evaluates to a function of the squared masses $m_{i}^2$, defined as
in \eq{massagain}. One finds
\beq
  {\cal Y}({\bs \mu}, \vec{\dd} \,\, ) \, = \, \exp \bigg[ 2 \pi {\rm i} \alpha' \Big(m_{1}^2 \,
  \bs{\tau}_{11} + m_{2}^2 \, \bs{\tau}_{22} + \big(m_{3}^2 - m_{1}^2 - m_{2}^2
  \big) \, \bs{\tau}_{12} \Big) \bigg] \, .
\label{eq:Yneat}
\eeq
Finally, let us turn to the contribution of the world-sheet fields $X^\mu, \, \psi^\mu$ along
the worldvolume direction of the D-branes. In absence of magnetic fields, the result can
be found in Ref.~\cite{DiVecchia:1988jy} and it reads
 \beq
   \mathbf{F}_{\rm gl}^{(0)} ({\bs \mu}) \, =  \, \Big[ \det \left({\rm Im}\, \bs{\tau} \right)
   \Big]^{- d/2} \, \, {\prod_\alpha}' \prod_{n = 1}^\infty \left(\frac{1 + \zn \,
   k_\alpha^{n - {1}/{2}}}{1 - k_\alpha^n}\right)^{\!\! d} \, .
\label{eq:gluonsf}
\eeq
In the presence of constant background gauge fields, $\mathbf{F}_{\rm gl}^{(0)}$ gets
modified, since string coordinates along the D-branes are sensitive to such backgrounds.
The relevant modification to the bosonic theory was derived in Ref.~\cite{Magnea:2004ai}.
Using the techniques described in Refs.~\cite{Russo:2003tt,Magnea:2004ai,Russo:2007tc},
it is possible to generalize this construction to the Neveu-Schwarz spin structure of the
RNS superstring~\cite{Magnea:2013lna}. The result is that switching on the background
fields amounts to multiplying $\mathbf{F}_{\rm gl}^{(0)}$ by a factor, as
\beq
  {\bf F}_{\rm gl}^{(0)} ({\bs \mu}) \, \longrightarrow \,
  {\bf F}_{\rm gl}^{(\vec{\epsilon} \, )} ({\bs \mu}) \, = \, {\cal R} ({\bs \mu}, \vec{\epsilon} \,) \,
  {\bf F}_{\rm gl}^{(0)} ({\bs \mu}) \, ,
\label{eq:FeR}
\eeq
where, assuming the background fields to be non-zero only in one plane, we have
\beqa
\label{eq:Reps}
  {\cal R} ({\bs \mu}, \vec{\epsilon} \, ) & = & \ex{ - \ii \pi \vec{\epsilon} \cdot \bs{\tau}
  \cdot \vec{\epsilon}} \, \, \frac{ \det \left({\rm Im} \,  \bs{\tau}  \right)}{\det \left( {\rm Im} \,
  \bs{\tau}_{\vec{\epsilon}} \, \right)}  \, \,
  {\prod_\alpha}' \prod_{n = 1}^\infty \Bigg[ \bigg( \frac{1 +  \zn \,
  k_\alpha^{n - {1}/{2}}}{1 - k_\alpha^n} \bigg)^{\!\! -2} \\
  && \hspace{2cm} \times \, \frac{\big({1 +  \ex{ \, \ii \pi
   (2 \vec{\epsilon} \cdot \bs{\tau} + \vec{\varsigma} \, ) \cdot \vec{N}_\alpha} \,
   k_\alpha^{\, n - {1}/{2}}} \big) \big({1 +  \ex{- \ii \pi\,
   (2 \vec{\epsilon} \cdot \bs{\tau} + \vec{\varsigma} \, ) \cdot \vec{N}_\alpha} \,
   k_\alpha^{\, n - {1}/{2}}} \big) }{\big(1 - \ex{\, 2 \pi \ii \vec{\epsilon} \cdot \bs{\tau}
   \cdot \vec{N}_\alpha} \, k_\alpha^n\big) \big(1 - \ex{- 2 \pi \ii \vec{\epsilon} \cdot
   \bs{\tau} \cdot \vec{N}_\alpha} \, k_\alpha^n \big)} \Bigg] \, . \nonumber
\eeqa
The matrix $\boldsymbol{\tau}_{\vec{\epsilon}}$ is the supersymmetric analogue of
the twisted (or Prym) period matrix, the bosonic version of which was computed with
the sewing method in \cite{Russo:2003tt,Russo:2003yk}. Its calculation in outlined in
Appendix \ref{tepsSRS}.

Inspecting \eq{eq:Reps}, we see that ${\bf F}_{\rm gl}^{(\vec{\epsilon} \, )}$ can be
factorized as the product of a term $\mathbf{F}_{\parallel}^{(\vec{\epsilon} \, )}$, capturing
the contribution along the magnetized plane, times an $\epsilon$-independent term
$\mathbf{F}_{\perp}$ arising from the unmagnetized directions. In the field theory limit,
$\mathbf{F}_{\parallel}^{(\veps \, )}$ will generate the contributions of gluons polarized
in the plane of the background field, while $\mathbf{F}_{\perp}$ will give rise to gluons
polarized in the transverse directions. Explicitly, we have
\beqa
\label{eq:Fperp}
  {\bf F}_{\perp} ({\bs \mu}) & = & \Big[ \det \left({\rm Im} \, \bs{\tau} \right)
  \Big]^{- \frac{d - 2}{2}} \, \, {\prod_\alpha}'  \prod_{n = 1}^\infty  \left(
  \frac{1 + \ex{ \ii \pi \vec{\varsigma} \cdot \vec{N}_\alpha} \, k_\alpha^{n - 1/2}}{1 -
  k_\alpha^n} \right)^{\!\! d - 2} \, , \\ \label{eq:Fpar}
  {\bf F}_{\parallel}^{(\veps \, )} ({\bs \mu}) & = & \frac{ {\rm e}^{- {\rm i} \pi \vec{\e}
  \cdot \bs{\tau} \cdot \vec{\e} }}{\det \left({\rm Im} \, \bs{\tau}_{\vec{\epsilon}} \, \right)} \\ \nonumber
  & & \hspace{5mm} \times \,
  {\prod_\alpha}' \prod_{n = 1}^\infty \frac{\big({1  +  \ex{ \, \ii \pi
   (2 \vec{\epsilon} \cdot \bs{\tau} + \vec{\varsigma}) \cdot
   \vec{N}_\alpha} \, k_\alpha^{\, n - {1}/{2}}} \big) \big({1 + \ex{- \ii \pi\,
   (2 \vec{\epsilon} \cdot \bs{\tau} + \vec{\varsigma}) \cdot
   \vec{N}_\alpha} \, k_\alpha^{\, n - {1}/{2}}} \big) }{\big(1 - \ex{\, 2 \pi \ii
   \vec{\epsilon} \cdot \bs{\tau} \cdot \vec{N}_\alpha} \, k_\alpha^n\big) \big(1 -
   \ex{- 2 \pi \ii \vec{\epsilon} \cdot \bs{\tau} \cdot
   \vec{N}_\alpha} \, k_\alpha^n \big)}  \, . \nonumber
\eeqa
Focusing now on the $h = 2$ case, we can use super-projective invariance to fix
three bosonic and two fermionic moduli. A convenient gauge choice in the super
Schottky formalism is to specify the positions of the fixed points, given in terms of
homogeneous coordinates\footnote{The relation between the super-conformal and
homogeneous coordinates is given in \eq{rscdef}.} on $\mathbf{CP}^{1|1}$, as
\beq
  \ket{{\bf u}_1} \, = \, ( 0 , 1| 0)\tran \, , \qquad \ket{{\bf v}_1 }\, = \, (1,0|0)\tran \, , \qquad
  \ket{{\bf u}_2} \, = \, (u,1 | \theta)\tran \, , \qquad \ket{ {\bf v}_2} \, = \,( 1,1 | \phi)\tran \, ,
\label{eq:sfpdef}
\eeq
with $\left( 0 \, < \, u \, < \, 1 \right)$, which leads to
\beq
  \Theta_{\mathbf{v}_1 \mathbf{u}_1 \mathbf{v}_2} \, = \, \phi \, ,  \qquad
  {\bf v}_2 \dotminus {\bf u}_2 \, = \, 1 - u + \theta \phi \, , \qquad
  \sqrt{\frac{(\mathbf{u}_1 \dotminus \mathbf{v}_2) (\mathbf{v}_2 \dotminus
  \mathbf{v}_1)}{\mathbf{v}_1 \dotminus \mathbf{u}_1}} \, = \, 1  \, .
\label{gfix}
\eeq
Implementing this projective gauge fixing in \eq{eq:mSetup}, we can finally express the
$h = 2$ partition function as
\beq
  {\bf Z}_2 \big( \veps, \vec{\dd} \,\, \big) \, = \ex{\ii \pi (\varsigma_1 + \varsigma_2) }
  \int \, \frac{\d k_1}{k_1^{3/2}} \,
  \frac{\d k_2}{k_2^{3/2}} \, \frac{\d u }{y} \, \, \d \theta \, \d \phi \, \,
  {\bf F}_{\rm gh} ({\bs \mu}) \, {\bf F}_{\parallel}^{(\veps \, )} ({\bs \mu}) \,
  {\bf F}_{\perp} ({\bs \mu}) \, {\bf F}_{\rm scal}^{(\vec{\dd} \,\, )} ({\bs \mu}) \, ,
\label{eq:mFull}
\eeq
where we defined
\beq
  y \, \equiv \, \left( {\bf u}_1,  {\bf v}_1,  {\bf u}_2 , {\bf v}_2 \right) \, = \,
  1 - u + \theta \phi \, ,
\label{eq:ydef}
\eeq
in terms of the bosonic super-projective invariant built out of four points, $({\bf z}_1,
{\bf z}_2, {\bf z}_3, {\bf z}_4)$, see Eq.~\eq{eq:crossratioF}.


\section{Taking the field theory limit}
\label{QFTlim}


\subsection{Expanding in powers of the multipliers}
\label{expak}

We are interested in computing the $\alpha' \to 0$ limit of the integrand of the superstring amplitude.
In this limit, we expect massive string states to decouple, so that one is left with
the massless spectrum. Possible contributions from the tachyon ground state
cancel after GSO projection in the superstring case, or should be
discarded by hand in the bosonic case. It is in principle non trivial to take this limit
before integration over (super) moduli, since this requires constructing a map between
the dimensionless moduli of the (super) Riemann surface and the dimensionful
quantities that arise in the computation of field theory Feynman diagrams. This
task is considerably simplified in the Schottky parametrization, where, as discussed
for example in Ref.~\cite{Magnea:2013lna}, the contributions of individual string
states can be identified by performing a Laurent expansion of the integrand of the
string partition function in powers of the multipliers. One finds a correspondence
between the order of expansion and the mass level of the string, and furthermore,
within each mass level, one can track individual states by tracing the origin of each
term to a specific factor in the string integrand.

The main difference between the bosonic string and the RNS superstring is that for
the latter, which we discuss here, the expansion is in powers of $k_i^{{1}/{2}}$ rather
than $k_i$, as is already apparent from our discussion in \secn{dmf}. More precisely,
since the measure of integration contains a factor $k_i^{-3/2}$, a term proportional to
$k_i^{(n - 3)/2}$ corresponds to a contribution from a state belonging to the $n$-th
mass level circulating in the $i$-th string loop (where $n = 0$ corresponds to the
tachyonic ground state). Therefore, all terms with $n > 1$ acquire a positive mass
squared, $m^2 = (n - 1)/(2 \alpha')$, and decouple in the limit $\alpha ' \to 0$. We
conclude that it is necessary to expand the various factors in the integrand of
\eq{eq:mFull} only up to terms of order ${k_i}^{1/2}$, in order to get the complete
massless field theory amplitude.

This task is made possible by the fact that the multipliers of only finitely many
super-Schottky group elements contribute at order $k_1^{{1}/{2}}k_2^{{1}/{2}}$. The
reason is that the leading-order behaviour of the multiplier $k_\alpha = k({\bf
T}_{\alpha})$ is related in a simple way to the index $N_\alpha^i$ introduced in
\secn{dmf}: one may verify that
\beq
  k^{1/2} \left( {\bf S}_i^{\pm 1} {\bf T}_\alpha \right) \, = \,  {\cal O} \left(
  k_i^{1/2} \, k_\alpha^{1/2} \right) \, ,
\label{krecur}
\eeq
unless of course the left-most factor of ${\bf T}_\alpha$ is ${\bf S}_i^{\mp 1}$. Thus,
for every super Schottky group element ${\bf T}_\alpha$ not in the primary class of
an element in the set $\{ {\bf S}_{1}, {\bf S}_{2}, {\bf S}_{1} {\bf S}_2, {\bf S}_1^{-1}
{\bf S}_2 \}$, the multiplier $k^{1/2}_\alpha$ vanishes faster than $k_i^{1/2}$ for
$k_i \to 0$. This enables us to easily compute expressions for all the factors in
\eq{eq:mSetup}, up to the relevant order.

Let us begin with ${\bf F}_{\rm gh}$, defined in \eq{eq:ghostsf}. One immediately
sees that the expansion of the infinite product starts at ${\cal O} (k_i^{3/2})$, and
the numerator of the first factor can similarly be dropped. ${\bf F}_{\rm gh}$ becomes
simply
\beq
  {\bf F}_{\rm gh} ({\bs \mu}) \, =  \, \left(1 - 2 \, \kso \right) \left(1 - 2 \, \kst \right) +
  {\cal O}(k_i) \, .
\label{eq:FghApprox}
\eeq
Next, we compute ${\bf F}_\perp$, defined in \eq{eq:Fperp}. Using the expressions
for the multipliers $k^{1/2} ({\bf S}_1^{-1} {\bf S}_2 )$ and $k^{1/2} ({\bf S}_1 {\bf S}_2 )$,
given in \eq{ktwolexp}, we find
\beqa
\label{eq:FperpK}
  {\bf F}_{\perp} ({\bs \mu}) & =  & \Big[ \det \left({\rm Im} \,  \bs{\tau}  \right)
  \Big]^{- \frac{d - 2}{2}} \, \, \bigg[ 1 + (d - 2) \, \left( \kso + \kst \right)  \\
  & & \hspace{3.3cm} + \, (d - 2) \left ( \frac{y}{u} - y + d - 2 \right) \ksot \bigg]
  + {\cal O}(k_i) \, . \nonumber
\eeqa
The expansion of the determinant of the super period matrix is given in \eq{eq:dettau},
and, substituted here, leads to the factor
\beqa
\label{Fparallelink}
  \big[ \det \left({\rm Im} \, \bs{\tau} \right) \big]^{- \frac{d - 2}{2}} & = &
  \left[ \frac{4 \pi^2}{\log k_1 \log k_2 - \log^2 u} \right]^{\frac{d - 2}{2}} \\
  & & \times \, \, \left[ 1 + (d - 2) \, \frac{y}{u} \, \theta \phi \,
  \frac{ \kso \log k_1 + \kst \log k_2}{\log k_1 \log k_2 - (\log u)^2} \right]
  + {\cal O}(k_i) \, . \nonumber
\eeqa
Notice that logarithmic dependence on (super) moduli must be retained exactly: indeed,
as shown in Ref.~\cite{Magnea:2013lna} and discussed here in \secn{rewrite}, it will turn
into polynomial dependence on Schwinger parameters in the field theory limit.

The expansion of the factor ${\bf F}^{(\veps \,)}_\parallel$, also given in \eq{eq:Fperp},
is more intricate, as well as more interesting, because of the dependence on the
external fields. Writing
\beq
  {\bf F}_\parallel^{(\vec{\epsilon} \, )} ({\bs \mu}) \, = \,
  \frac{ {\rm e}^{- {\rm i} \pi \vec{\e} \cdot \bs{\tau} \cdot
  \vec{\e} }}{\det \left({\rm Im} \, \bs{\tau}_{\vec{\epsilon}} \, \right)} \,\,
  \widehat{\cal R} ({\bs \mu}, \vec{\epsilon} \,) \, ,
\label{threefact}
\eeq
where $\widehat{\cal R}$ is the background-field dependent factor of the infinite product
appearing in \eq{eq:Fpar}, we see that we can separately expand the three factors.
The determinant of the twisted super period matrix is by far the most intricate contribution.
It is discussed in Appendix B, and a complete expression with the exact dependence
on the fields, through $\vec{\epsilon}$, is very lengthy. We will see in \secn{rewrite},
however, that in the field theory limit we must expand in powers of the components
of $\vec{\epsilon}$ as well: at that stage, we will be able to write a completely explicit
expression also for $\det \left({\rm Im} \, \bs{\tau}_{\vec{\epsilon}} \, \right)$. The exponential
factor in the numerator of \eq{threefact} can be computed using the expression for
$\bs{\tau}$ in \eq{eq:stauSch}, and is given by
\beq
  \ex{ - \ii \pi \veps \cdot \bs{\tau} \cdot \veps } \, =  \, k_1^{- \epsilon_1^2/2} \,
  k_2^{- \epsilon_2^2/2} \, u^{- \epsilon_1 \epsilon_2} \bigg[ 1 + \frac{y}{u} \,
  \theta \phi \, \Big(  \kst \epsilon_1^2 + \kso \epsilon_2^2 \Big) \bigg] +
  {\cal O}(k_i) \, .
\label{exponentialterminFglu}
\eeq
Finally, the remaining factor in ${\bf F}^{(\veps \,)}_\parallel$ is given by
\beqa
\label{parallelorbitalk}
    \widehat{\cal R} ({\bs \mu}, \vec{\epsilon} \,) & = &
    1 +  \kso g_{1 2}^+ + \kst g_{2 1}^+ \\
    & & \hspace{2.8mm} + \, \ksot \Bigg[
    g_{1 2}^+ g_{2 1}^+ - 2 \, \theta \phi \, \frac{y}{u} \left( \epsilon_1 g_{1 2}^-
    + \epsilon_2 g_{2 1}^- \right) \nonumber \\
    & & \hspace{2cm} - \, \frac{1}{2} \bigg(
    \left(y - \frac{y}{u} \right) g_{1 2}^+ g_{2 1}^+ +
    \left(y + \frac{y}{u} \right) g_{1 2}^- g_{2 1}^- \bigg)
    \Bigg] \, + \, {\cal O} (k_i) \, , \nonumber
\eeqa
where we defined the factors
\beq
  g_{i j}^\pm \, = \, k_i^{\epsilon_i} u^{\epsilon_j} \pm k_i^{- \epsilon_i} u^{- \epsilon_j} \, .
\label{defgij}
\eeq
The last required ingredient is ${\bf F}_{\rm scal}^{(\vec{\dd} \, )}$, defined in \eq{eq:scalarf}.
Combining \eq{eq:Yneat} with \eq{eq:stauSch}, we get
\beqa
\label{eq:vevfactor}
  {\cal Y} \big( {\bs \mu }, \vec{\dd} \,\, \big) & = & k_1^{\alpha' m_1^2} \, k_2^{\alpha' m_2^2} \,
  u^{\alpha'  \left( m_3^2 - m_1^2 - m_2^2 \right)} \\
  & & \hspace{2cm} \times \, \bigg[ 1 -  2 \, \alpha'  \, \frac{y}{u} \Big( \kso\, m_2^2  +
  \kst \, m_1^2 \Big) \bigg]  + {\cal O}(k_i) \, . \nonumber
\eeqa
The remaining, mass-independent, factor in ${\bf F}_{\rm scal}^{(\vec{\dd} \,\,)}$ in
\eq{eq:scalarf} is easily expanded, getting
\beq
  {\bf F}_{\rm scal}^{(0)}({\bs \mu}) \, = \, 1 + n_s \left( \kso + \kst \right) +
  n_s \, \ksot \left( \frac{y}{u} - y + n_s \right) \, .
\label{eq:FscalorbitK}
\eeq
This completes the list of the factors in \eq{eq:mFull}. It is now straightforward to combine
them, and expand the resulting polynomial in $k_i^{1/2}$ to the relevant order. Before
proceeding, however, we must consider more carefully our choice of variables in view of
the field theory limit.


\subsection{A parametrization for the symmetric degeneration}
\label{symmpar}

In order to go beyond the specification of the mass states circulating in the string loops,
and identify the contribution of individual Feynman diagrams in the field theory limit,
we must refine our parametrization of (super) moduli space. Let us now, in particular,
concentrate on Feynman diagrams with the symmetric topology depicted in the first
two lines of Fig.~\ref{fig:1PIgraphs}. While
individual Feynman diagrams will not be symmetric under the exchange of any two
lines when the propagating states are different, we expect that, when summing
over all states at a given mass level, the result should be fully symmetric, since there
are no features distinguishing the three propagators at the level of the world-sheet
geometry. This symmetry requirement will guide our choice of parametrization
for the region of moduli space close to this degeneration, along the lines already
discussed in Ref.~\cite{Magnea:2013lna}.

\begin{figure}
\begin{center}
\subfloat[]{ 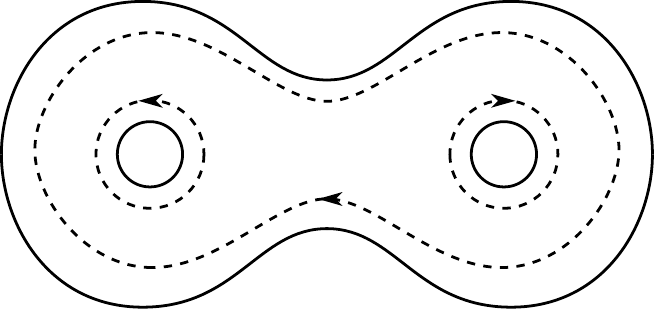 \label{fig:goodcycles} }
\subfloat[]{ 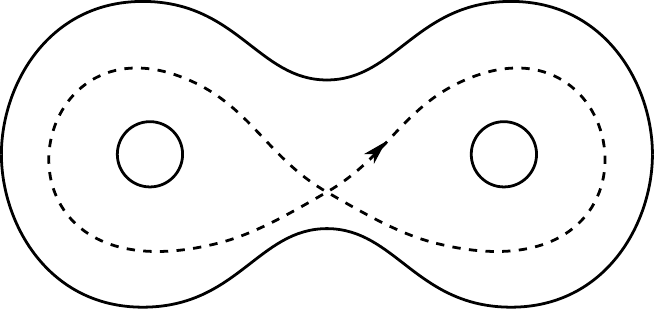 \label{fig:badcycle} }
\caption{Two types of homology cycles on the double annulus.
\label{homology}}
\end{center}
\end{figure}

It is clear that the parametrization in terms of the bosonic moduli $k_1^{1/2}$,
$k_2^{1/2}$ and $u \equiv 1 - y + \theta \phi$ will not be sufficiently symmetric,
since the first two chosen moduli are multipliers of super-Schottky group generators,
while the third one is a cross-ratio of the fixed points. To present the integration
measure in a sufficiently symmetric way, we must parametrize it to be symmetric
under permutations of the super-Schottky group elements ${\bf S}_1$, ${\bf S}_2$
and ${\bf S}_1^{-1} {\bf S}_2$. The reason for this is that the homology cycles
$b_1$, $b_2$ and $(b_1^{-1} \cdot b_2)$ lift to these super-Schottky group
elements on the covering surface ${\bf CP}^{1|1} - \Lambda$, and any two of
$b_1$, $b_2$ and $(b_1^{-1} \cdot b_2)$ (along with the appropriate choice of
$a$-cycles) constitute a good canonical homology basis (see \Fig{fig:goodcycles}).
Our choice of ${\bf S}_1$ and ${\bf S}_2$ as the generators is arbitrary, so, in
order to preserve modular invariance, the measure must be parametrized to
display the symmetry under permutations of ${\bf S}_1$, ${\bf S}_2$ and
${\bf S}_1^{-1} {\bf S}_2$. To reinforce this point, note that any other homology
cycle built out of $b$ cycles will intersect itself, as is the case for example for the
$(b_1 \cdot b_2)$ cycle, depicted in \Fig{fig:badcycle}.

A natural way to symmetrize the measure is to use the multiplier of ${\bf S}_1^{-1}
{\bf S}_2$ as the third bosonic modulus, instead of $u$. Defining $- \, \ex{ \ii \pi
\varsigma_3 } k_3^{{1}/{2}}$ to be the eigenvalue of $\mathbf{S}_1^{-1} \mathbf{S}_2$
with the smallest absolute value, so that $k_3$ is the multiplier of that super Schottky
group element, one can compute $k_3^{1/2}$ using \eq{eq:kFromSTr}. It is related to $y$
implicitly through
\beq
   y \, = \, \frac{(1- k_1)(1- k_2) \, + \, \theta \phi \, \Big[
   \big(1 + \kso \big) \big(1 + \kst \big)  \big(1 + \ksot \big) \Big]}{1 + k_1 k_2 +
   k_1^{1/2} k_2^{1/2} \big( k_3^{1/2} + k_3^{- 1/2} \big)} .
 \label{etapis}
 \eeq
In these definitions, $\varsigma_3$ is the spin structure around the $b_3 \equiv
b_1^{-1} \cdot b_2$ homology cycle, and therefore it is given simply by $\varsigma_3
= \varsigma_1 + \varsigma_2$ (mod 2). $k_3^{1/2}$ is then positive, just as $k_1^{1/2}$
and $k_2^{1/2}$ are.

As discussed in Ref.~\cite{Magnea:2013lna}, the field theory limit becomes particularly
transparent if one factors the three multipliers $k_i$ in order to assign a parameter to
each section of the Riemann surface that will degenerate into an individual field theory
propagator. This is done by defining
\beq
 k_1^{1/2} \, = \, \sqrt{p_1} \sqrt{p_3} \, \qquad
 k_2^{1/2} \, = \, \sqrt{p_2} \sqrt{p_3} \, \qquad
 k_3^{1/2} \, = \, \sqrt{p_1} \sqrt{p_2} \, ,
\label{eq:kToP}
\eeq
where $\sqrt{p_i}$ is defined to be positive. In analogy to the discussion of
Ref.~\cite{Magnea:2013lna}, each $p_i$ will be interpreted, in the field theory
limit, as the logarithm of the Schwinger proper time associated to a propagator.

For bosonic strings, the discussion leading to \eq{eq:kToP} was sufficient to
construct a symmetric measure of integration, prepared for the symmetric degeneration
in the field theory limit. In the present case, instead, one must also worry about fermionic
moduli: our current choice of $\theta$ and $\phi$ as moduli will not yield a symmetric
measure, since they are super-projective invariants built out of the fixed points of
$\mathbf{S}_1$ and ${\bf S}_2$ only. In order to find the proper Grassmann variables
of integration, we take advantage of the fact that we are allowed to rescale $\theta$
and $\phi$ with arbitrary functions of the moduli, since such a rescaling automatically
cancels with the Berezinian of the corresponding change of integration variables.
Such a rescaling of course leaves the integral invariant, but it can be used to move
contributions between the various factors of the integrand, in such a way that individual
factors respect the overall exchange symmetry of the diagram, as we wish to do here.
In order to find a pair of odd moduli invariant under permutations of ${\bf S}_1$,
${\bf S}_2$, ${\bf S}_1^{-1} {\bf S}_2$, we proceed as follows. Define
\beq
  \hat \theta_{i j} \, = \, c_{i j} \, \Theta_{\mathbf{v}_i \mathbf{u}_i
  \mathbf{u}_j} \, , \qquad
  \hat \phi_{i j} \, = \, c_{i j} \, \Theta_{\mathbf{v}_i \mathbf{u}_i
  \mathbf{v}_j} \, ,
\label{eq:newtheta}
\eeq
 for $(ij) = (12),(23),(31)$. For the factors $c_{i j}$ we make the choice
\beq
  c_{1 2} \, = \,  \Big[ \, \ex{\ii \pi \varsigma_3}\left(1 +  q_1 q_2 \right)
  \left(1 - q_1 q_3 \right) \left(1 - q_2 q_3 \right) \Big]^{-1/2} \, ,
\label{c12}
\eeq
with $c_{2 3}$ and $c_{3 1}$ obtained by permuting the indices $(123)$, and where
$\mathbf{u}_3$ and $\mathbf{v}_3$ are the fixed points of the transformation
$\mathbf{S}_1^{-1} \mathbf{S}_2$. In \eq{c12}, we have introduced the symbols
$q_i$, $i = 1, 2, 3$, defined by\footnote{Note that the spin structures of $q_1$ and $q_2$
are swapped compared with what one might expect. This, however, is reasonable,
because the $q_i$ defined in this way factorize the NS sewing parameters $\ex{\ii \pi
\varsigma_i} \, k_i^{1/2}$ as follows: $\ex{\ii \pi \varsigma_1} \, k_1^{1/2} = q_1 q_3$,
\quad $\ex{\ii \pi\varsigma_2} \, k_2^{1/2} = q_2 q_3$, \quad $\ex{\ii \pi\varsigma_3} \,
k_3^{1/2} = q_1 q_2$.}
\beq
q_1 \, = \, \ex{ \ii \pi \varsigma_2} \sqrt{p_1} \, , \qquad
q_2 \, = \, \ex{ \ii \pi \varsigma_1} \sqrt{p_2} \, , \qquad
q_3 \, = \, \ex{ \ii \pi \varsigma_3} \sqrt{p_3} \, .
\label{qdef}
\eeq
With this choice for $c_{ij}$, one can check that
\beq
  \ex{\ii \pi \varsigma_3 } \,
  \d \hat\theta_{12} \, \d \hat\phi_{12} \, =
  \, \ex{\ii \pi \varsigma_1 } \,
  \d \hat\theta_{23} \, \d \hat\phi_{23}
  \, = \, \ex{\ii \pi \varsigma_2 } \,
  \d \hat\theta_{31} \, \d \hat\phi_{31} \, ,
\label{cyclfermeas}
\eeq
so that the Grassmann measure of integration has the required symmetry.

It is not difficult to rewrite the various objects computed in \secn{expak} in terms
of the new variables, and expand the results to the required order in $p_i$.
In order to do so, we use
\beq
  \theta \phi \, = \, q_3 \left( 1 + q_1q_2 \right) \,
  \hat{\theta}_{12} \hat{\phi}_{12} + {\cal O} (p_i) \, ,
\label{eq:oddToHat}
\eeq
as well as
\beq
  u \, = \, p_3 \bigg[ 1 + \hat{\theta}_{12} \hat{\phi}_{12} \Big( q_3  -
  q_1 -q_2  + q_1q_2 q_3 \Big) \bigg] \, + \, {\cal O} \left( p_1, p_2, p_3^2 \right)
  \, . \nonumber
\label{eq:uToP}
\eeq
With these results, it is straightforward to verify the symmetry of the full string integrand.
In particular, we find that the product of the two-loop measure of integration times the
ghost factor is given by
\beqa
\label{eq:FghPser}
  d {\bs \mu}_2 \, {\bf F}_{\rm gh} ({\bs \mu}) & = & \prod_{i = 1}^3
  \left[ \, \frac{\d p_i}{p_i^{3/2}} \, \frac{1 - \ex{\ii \pi \varsigma_i} k_i^{1/2}}{\sqrt{1 + p_i}}
  \right]  \d \hat\theta_{12} \d \hat\phi_{12}
  \frac{1}{\sqrt{1 + p_1 p_2 p_3}} \\
  & = & \prod_{i = 1}^3 \left[ \, \frac{\d p_i}{q_i^3} \, \right] \, \d \hat\theta_{12} \,
  \d \hat\phi_{12} \, \left( 1 - q_1 q_3 -
  q_2q_3 - q_1q_2 \, \right)
  + {\cal O}(p_i ) \, , \nonumber
\eeqa
where the contribution of the spin structure to $\d {\bs \mu}_2$ in \eq{ssc} has been
absorbed in $\d \hat\theta_{12} \, \d \hat\phi_{12}$. Similarly, the contribution of the
orbital modes defined in \eq{parallelorbitalk} becomes
\beqa
\label{eq:FparOrbitP}
  \widehat{\cal R} ({\bs \mu}, \vec{\epsilon} \,) & = & 1 + \bigg\{q_1 q_2 \,
  \left( p_1^{\epsilon_1} p_2^{- \epsilon_2} - p_1^{- \epsilon_1} p_2^{\epsilon_2}
  \right) \, \Big[ \, 1 - \hat{\theta}_{12} \, \hat{\phi}_{12} \, q_3 \left( \epsilon_1 -
  \epsilon_2 \right) \Big] \nonumber \\
  & & \hspace{2cm} + \, \text{cyclic permutations} \bigg\} + {\cal O} (p_i) \, .
\eeqa
Here, and in the rest of this section, we understand `cyclic permutations' to mean
cyclic permutations of the indices $(1,2,3)$ for $p_i$, $q_i$, $\epsilon_i$ and $m_i^2$,
where $\epsilon_3 \equiv - \epsilon_1 - \epsilon_2$.\footnote{Recall that, in the $h = 2$
case, we define $\epsilon_2$ with the opposite sign with respect to \eq{deftewovec},
to exploit the symmetry of the worldsheet.} The indices of $\hat{\theta}_{12} \,
\hat{\phi}_{12}$, on the other hand, are not permuted.

In order to reconstruct the full contribution of fields in the directions parallel to the
magnetized plane, we still need the other factors appearing in \eq{exponentialterminFglu}.
The exponential factor takes the form
\beqa
\label{eq:FparExpP}
  \ex{ - \ii \pi \vec{\epsilon} \cdot \bs{\tau} \cdot \vec{\epsilon}} & = & p_1^{ -
  \epsilon_1^2/2} p_2^{ - \epsilon_2^2/2} p_3^{ - \epsilon_3^2/2} \bigg[ 1 -
  \frac{1}{2} \, \hat{\theta}_{12} \hat{\phi}_{12} \Big( q_1 \, \left( \epsilon_1^2
  - \epsilon_2^2 - \epsilon_3^2 \right)  \nonumber \\
  & & \hspace{2cm} + \, \, q_1 q_2 q_3 \, \epsilon_1^2
  + \text{cycl. perm.} \Big) \bigg] + {\cal O} (p_i) \, .
\eeqa
The last factor in \eq{exponentialterminFglu} is the twisted determinant $\det \left({\rm Im}
\, \bs{\tau}_{\vec{\epsilon}}  \right)$, whose calculation is described in Appendix~\ref{Appb}.
The result for generic values of $u$ is a lengthy combination of hypergeometric functions,
which however simplifies drastically in the limit we are considering here, where $u$,
proportional to $p_3$, is small.

In this limit~\eqref{eq:b.40} reads
\beqa
 \label{eq:twisdetP}
  \det \left( {\rm Im} \, \bs{\tau}_{\vec{\epsilon}} \right) & = & \frac{1}{4 \pi^2} \,
  \Gamma(- \epsilon_1) \Gamma(- \epsilon_2) \Gamma(- \epsilon_3) \, \Bigg\{
  p_1^{\epsilon_1/2} p_2^{\epsilon_2/2} p_3^{\epsilon_3/2} \Big(  \epsilon_1 \,
  p_1^{- \epsilon_1/2} + \epsilon_2 \, p_2^{- \epsilon_2/2} + \epsilon_3 \,
  p_3^{- \epsilon_3/2} \Big) \nonumber \\
  & & \hspace{-1.8cm} + \, \, \hat{\theta}_{12} \hat{\phi}_{12}
  \bigg[ \, q_1 \, \Big( p_1^{\epsilon_1/2} p_2^{ - \epsilon_2/2} p_3^{- \epsilon_3/2} +
  p_1^{- \epsilon_1/2} p_2^{\epsilon_2/2} p_3^{\epsilon_3/2} \Big) \,
  p_1^{\epsilon_1/2} \, \epsilon_2 \epsilon_3 \, + \, \text{cycl. perm.} \bigg] \nonumber \\
  & & \hspace{-1.8cm} + \, \, \hat\theta_{12} \hat{\phi}_{12} \, \, q_1 q_2 q_3 \, \,
  p_1^{- 3 \epsilon_1/2} p_2^{- 3 \epsilon_2/2} p_3^{- 3 \epsilon_3/2}
  \nonumber \\ & & \hspace{-1.5cm} \times \bigg[
  p_1^{2 \epsilon_1} p_2^{2 \epsilon_2} p_3^{\epsilon_3} \big( 2 \epsilon_3^2 -
  \epsilon_1 \epsilon_2 \big) + p_3^{3 \epsilon_3} \Big( 2 \epsilon_3 \big( p_1^{2 \epsilon_1}
  \epsilon_2 + p_2^{2 \epsilon_2} \epsilon_1 \big) - p_1^{\epsilon_1}
  p_2^{\epsilon_2} \epsilon_1 \epsilon_2 \Big) \, + \, \text{cycl. perm.}
  \bigg] \Bigg\} \nonumber \\
  & & \hspace{3cm} + \, \Big( \epsilon_i \leftrightarrow - \epsilon_\mu \Big)
  + {\cal O} (p_i) \, .
\eeqa
Next, we need the contribution of the untwisted gluon sector, given in \eq{eq:FperpK}.
In the current parametrization it reads
\beqa
\label{eq:FperpP}
  {\bf F}_{\perp} ({\bs \mu}) & = & \Big[ \det ({\text{Im }} \bs{\tau}) \Big]^{ - (d - 2)/2} \\
  & & \hspace{5mm} \times \, \Big[ 1 - (d - 2) \Big( q_1 q_3 +
  q_2q_3 + q_1q_2 \Big) \Big] + {\cal O} (p_i) \, , \nonumber
\eeqa
where the determinant of the period matrix, given by \eq{eq:dettau}, becomes
\beqa
\label{eq:dettauP}
  \det(\text{Im }\boldsymbol{\tau}) & = & \frac{1}{4 \pi^2} \, \bigg\{ \log p_1 \log p_2 +
  \log p_2 \log p_3 + \log p_3 \log p_1 \\
  & & \hspace{-4mm} - \, 2  \, \hat{\theta}_{12} \hat{\phi}_{12} \, \Big[ \big(
  q_1 - q_1 q_2 q_3 \big) \log p_1 \, + \, \text{cycl. perm.} \Big] \bigg\} +
  {\cal O}(p_i) \, . \nonumber
\eeqa
Finally, we need the ingredients for the scalar sector, given above in \eq{eq:vevfactor}
and \eq{eq:FscalorbitK}. The mass contribution takes the form
\beqa
\label{eq:vevTermp}
  {\cal Y} \big( {\bs \mu }, \vec{\dd} \,\, \big) & = & p_1^{\alpha ' m_1^2 } \,
  p_2^{\alpha' m_2^2 } \, p_3^{\alpha ' m_3^2} \bigg[ \, 1 +
  \alpha' \, \hat{\theta}_{12} \hat{\phi}_{12} \,
  \Big( q_1 \, \big( m_1^2 - m_2^2 - m_3^2 \big) \\
  & & \hspace{1cm} + \, \, q_1q_2q_3 \, \, m_1^2  \, + \,
  \text{cycl. perm.}  \Big) \, \bigg] + {\cal O}(p_i) \, , \nonumber
\eeqa
while the mass-independent factor is given by
\beq
  {\bf F}_{\rm scal}^{(0)} ({\bs \mu}) \, = \, 1 + n_s \,
  \Big( q_1 q_3 + q_2 q_3 + q_1 q_2 \Big) + {\cal O} (p_i) \, .
\label{eq:scalorbitP}
\eeq
This completes the list of ingredients needed for the analysis of the symmetric
degeneration of the surface. We now turn to the calculation of the $\alpha' \to 0$
limit.


\subsection{Mapping moduli to Schwinger parameters}
\label{rewrite}

The last, crucial step needed to take the field theory limit is the mapping between
the dimensionless moduli and the dimensionful quantities that enter field theory
Feynman diagrams. This $\alpha'$-dependent change of variables sets the
space-time scale of the scattering process and selects those terms in the string
integrand that are not suppressed by powers of the string tension. The basic ideas
underlying the choice of field theory variables have been known for a long time
(see for example Ref.~\cite{DiVecchia:1996uq}), and were recently refined for
the case of multi-loop gluon amplitudes in Ref.~\cite{Magnea:2013lna}. The
change from bosonic strings to superstrings does not significantly affect those
arguments: in the present case we will see that integration over odd moduli
will simply provide a more refined tool to project out unwanted contributions, once
the Berezin integration is properly handled.

Following Ref.~\cite{Magnea:2013lna}, we introduce dimensionful field-theoretic
quantities with the change of variables
\beq
  p_i \, = \, \exp \left[ - \frac{t_i}{\alpha'} \right] \, , \qquad
  \epsilon_i \, = \, 2 \alpha' g  B_i + {\cal O} ({\alpha ' }^3)  \, ,  \qquad i \, = \, 1,2,3 \, .
\label{eq:epsToB}
\eeq
These definitions make it immediately obvious that terms of the form $p_i^{\, c \,
\epsilon_i}$ must be treated exactly, as we have done. On the other hand, terms
proportional to high powers of $\epsilon_i$ are suppressed by powers of $\alpha'$
in the field theory limit, which is the source of further simplifications in our final
expressions.

For completeness, we give here the results for the various factors in \eq{eq:mSetup}
as Taylor expansions powers of $q_i$ (that is, in half-integer powers of $p_i$), but
with the field- and mass-dependent coefficients of the leading terms worked out
exactly. Beginning with the contribution of gluon modes perpendicular to the
magnetic fields, ${\bf F}_{\perp} ({\bs \mu})$, we find
\beqa
\label{eq:FperpField}
  {\bf F}_{\perp} ({\bs \mu}) & = & \left[ \frac{(2 \pi \alpha')^2}{\Delta_0} \right]^{d/2 - 1}
  \bigg\{ 1 + (d - 2) \bigg[ \, q_1 q_2 + q_2 q_3 + q_3 q_1 \\
  & & \hspace{-1.5cm} - \, \alpha'  \, \hat{\theta}_{12} \hat{\phi}_{12} \, \frac{1}{\Delta_0} \,
  \Big( t_1 \, q_1 + t_2 \, q_2 + t_3 \, q_3
  \, + \, (d - 3) \, (t_1 + t_2 + t_3) \, q_1 q_2 q_3 \Big) \bigg]
  + {\cal O} \left( (\alpha')^2, p_i \right) \bigg\} \nonumber \\
  & \equiv & \left[ \frac{(2 \pi \alpha')^2}{\Delta_0} \right]^{d/2 - 1}
  \sum_{m,n,p = 0}^\infty q_1^m \, q_2^n \, q_3^p \, \, \widehat{\bf F}_{\perp}^{(m n p)}
  \left( t_i \right) \, , \nonumber
\eeqa
where we defined
\beq
  \Delta_0 \, = \, t_1 t_2 + t_2 t_3 + t_3 t_1 \, ,
\label{eq:DelZero}
\eeq
which we recognize as the first Symanzik polynomial~\cite{Vanhove:2014wqa} of
graphs with the topology of those in the first two lines of \Fig{fig:1PIgraphs}, expressed in terms of standard
Schwinger parameters.

The result for the contribution of gluon modes parallel to the magnetic field is more
interesting, as one begins to recognize detailed structures that are known to arise
in the corresponding field theory. Multiplying \eq{eq:FparOrbitP} by \eq{eq:FparExpP},
and dividing by \eq{eq:twisdetP}, one finds
\beqa
\label{eq:FparField}
  {\bf F}_{\parallel}^{(\veps \, )} ({\bs \mu}) & = & \frac{(2 \pi \alpha')^2}{\Delta_B}
  \bigg\{ 1 + 2 \, \Big[ \, \cosh \Big( 2 g \, ( B_1 t_1 - B_2 t_2) \Big)  \, q_1 q_2 \nonumber \\
  & & - \, \alpha' \, \hat{\theta}_{12} \hat{\phi}_{12} \,
  \frac{1}{\Delta_B} \, \frac{\sinh(g B_1 t_1)}{g B_1} \,
  \cosh \Big( g \, ( 2 B_1 t_1 - B_2 t_2 - B_3 t_3 ) \Big) \big( q_1 + q_1 q_2 q_3 \big)
  \nonumber \\
  & & + \, \text{cycl. perm.} \Big] + {\cal O} ({\alpha'}^2) + {\cal O}(p_i) \bigg\} \\
  & \equiv & \frac{(2 \pi \alpha')^2}{\Delta_B}
  \sum_{m,n,p = 0}^\infty q_1^m \, q_2^n \, q_3^p \, \,  \widehat{\bf F}_{\parallel}^{(m n p)}
  \left( t_i, B_i \right) \, , \nonumber
\eeqa
where
\beq
  \Delta_B \, = \, \frac{ \cosh \big[ g \left( B_1 t_1 - B_2 t_2 - B_3 t_3 \right)
  \big]}{2 g^2 B_2 B_3} + \text{cycl. perm.} \, ,
\label{eq:DelF}
\eeq
Using the fact that $B_1 + B_2 + B_3 =0$, one can verify that $\Delta_B$ can be
understood as the charged generalization of the first Symanzik polynomial for
this graph topology, and indeed for vanishing fields $\Delta_B$ tends to $\Delta_0$.
It is then easy to see that ${\bf F}_{\parallel}^{(\veps \, )} ({\bs \mu})$, in the same
limit, reproduces ${\bf F}_{\perp} ({\bs \mu})$ with the replacement $d - 2 \to
2$, as expected.

The contribution from the D-brane world-volume scalars can be obtained combining
\eq{eq:vevTermp} and \eq{eq:scalorbitP}. One finds
\beqa
\label{eq:FscalField}
  {\bf F}_{\text{scal}}^{(\vec{\dd} \,\, )} ({\bs \mu}) & = & \ex{- t_1 m_1^2} \,
  \ex{- t_2 m_2^2} \, \ex{- t_3 m_3^2} \bigg[ 1 + n_s \big(q_1 q_2 + q_2 q_3 + q_3 q_1
  \big)  \\ & & \hspace{-1cm}
  + \, \alpha' \, \hat{\theta}_{12} \hat{\phi}_{12} \, \Big( \big( m_1^2 - m_2^2 - m_3^2
  \big) \, q_1 - \big( n_s - 1 \big) \, m_1^2 \, \, q_1 q_2 q_3 \, + \, \text{cycl. perm.}
  \Big) \bigg] + {\cal O}(p_i) \nonumber \\
  & \equiv & \prod_{i = 1}^3 \Big[ \ex{ - t_i m_i^2} \Big] \,
  \sum_{m,n,p = 0}^\infty  q_1^m \, q_2^n \, q_3^p\, \,
  \widehat{\bf F}_{\text{scal}}^{(m n p)} \left( t_i, m_i \right) \, , \nonumber
\eeqa
where one recognizes the exponential dependence on particle masses, each multiplied
by the Schwinger parameter of the corresponding propagator, which is characteristic
of massive field-theory Feynman diagrams. Note that the masses $m_i^2$ appearing
in \eq{eq:FscalField} arise via symmetry breaking from the distance between D-branes,
and therefore represent classical shifts of the string spectrum: below, for brevity, we
will often call `massless' all string states that would be massless in the absence of
symmetry breaking.

The final factor, including the integration measure and the contribution from the ghosts,
can be read off \eq{eq:FghPser}, and can be organized as
\beq
  d {\bs \mu}_2 \, {\bf F}_{\text{gh}} ({\bs \mu}) \, = \,\prod_{i = 1}^3
  \bigg[ \frac{\d p_i}{q_i^3} \bigg] \, d \hat{\theta}_{12} \, d \hat{\phi}_{12}
  \sum_{m,n,p = 0}^\infty q_1^m \, q_2^n \, q_3^p \, \,
  \widehat{\bf F}_{\text{gh}}^{(m n p)} \, .
\label{eq:Fghmeas}
\eeq
The complete integrand of \eq{eq:mSetup} is the product of ${\bf F}_{\perp}$ from
\eq{eq:FperpField}, ${\bf F}_{\parallel}$ from \eq{eq:FparField}, ${\bf F}_{\text{scal}}$
from \eq{eq:FscalField} and $d {\bs \mu}_2 \, {\bf F}_{\text{gh}}$ from \eq{eq:Fghmeas}.
In the proximity of the symmetric degeneration, it can be organized in a power
series in terms of the variables $q_i$, as we have done for individual factors.
We write
\beqa
  \hspace{-2mm}
  d \, {\bf Z}_2^{\rm \, sym} ({\bs \mu}) & = & \prod_{i = 1}^3 \bigg[ \frac{\d p_i}{q_i^3} \,
  \ex{ - t_i m_i^2} \bigg] \, \d \hat{\theta}_{12} \, \d \hat{\phi}_{12} \, \,
  \frac{(2 \pi \alpha')^d}{\Delta_0^{d/2 - 1} \Delta_B} \nonumber \\
  & &  \hspace{1cm} \times \, \sum_{m,n,p = 0}^\infty q_1^m \, q_2^n \, q_3^p  \, \,
  \widehat{\bf F}^{(m n p)} \left( t_i, m_i, B_i \right) .
\label{eq:integrandSeries}
\eeqa
It is now straightforward to extract the contribution of massless states, which is contained
in the coefficient $\widehat{\bf F}^{(1 1 1)} \left( t_i, m_i, B_i \right)$. For bosonic strings,
one had to discard the contribution of the tachyonic ground state by hand: in this case,
one can simply implement the GSO projection and observe the expected decoupling
of the tachyon. We now turn to the analysis of this point.


\subsection{The symmetric degeneration after GSO projection}
\label{GSO}

Starting with the expression in \eq{eq:integrandSeries}, we can now describe more
precisely the connection between the powers of the multipliers and the mass eigenstates
circulating in the loops. For the symmetric degeneration, we now see that the power of
$p_i$ corresponds to the mass level of the state propagating in the $i$-th edge of
the diagram. Indeed one observes that
\beq
  \frac{\d p_i}{p_i^{3/2}} \, \big( p_i^{1/2} \, \big)^n \, =  \,  - \frac{1}{\alpha'} \, d t_i \,
  \ex{ - \frac{n - 1}{2 \alpha'} t_i} \, ,
\label{p_to_t}
\eeq
and one recognizes that $d t_i \, \ex{ - \frac{n-1}{2 \alpha'} t_i}$ is a factor one would
expect to see in a Schwinger-parameter propagator for a field with squared mass
$m^2 = \frac{n - 1}{2 \alpha'}$. In particular, if $n = 0$, then the state propagating in
the $i$-th edge will be a tachyon, and will have to be removed by the GSO projection.

A cursory look at ${\bf F}_{\parallel}$ in \eq{eq:FparField}, ${\bf F}_{\perp}$ in
\eq{eq:FperpField}, ${\bf F}_{\text{scal}}$ in \eq{eq:FscalField} and $\d {\bs \mu}_2
{\bf F}_{\text{gh}}$ in \eq{eq:FghPser}, would suggest that tachyons can propagate
simultaneously in any number of edges of the diagram: indeed, we can find terms
proportional to $\prod_i \d p_i /q_i^3$ times $1$, $q_1$, $q_1q_2$, $q_1q_2q_3, \ldots$,
which correspond respectively to three, two, one or no edges with propagating
tachyons. A closer inspection shows, however, that the nilpotent object $\hat{\theta}_{12}
\hat{\phi}_{12}$ multiplies only terms with an \emph{odd} number of factors of
$q_i$, a property which is preserved when we multiply terms together.
Since the Berezin integral over $d \hat{\theta}_{12} d \hat{\phi}_{12}$ picks out
the coefficient of $\hat{\theta}_{12} \hat{\phi}_{12}\, $, it follows that, after carrying
out the Berezin integration, each term must contain an odd number of factors of
$q_i$.

As a consequence, after Berezin integration and truncation of the integrand to
${\cal O}(p_i^0)$, \eq{eq:integrandSeries} will be written as a sum of four terms,
proportional to $\prod_{i=1}^3 \d p_i / p_i^{3/2}$ multiplied by the factors
\begin{align}
  q_1 & =  \ex{ \ii \pi \varsigma_2} \sqrt{p_1} \, , &
   q_2 & = \ex{ \ii \pi \varsigma_1} \sqrt{p_2} \, , &
   q_3 & = \ex{ \ii \pi \varsigma_3} \sqrt{p_3} \, , &
   q_1 q_2 q_3 & =
  \sqrt{p_1 p_2 p_3} \, .
\label{eq:oddPs}
\end{align}
The first three terms in \eq{eq:oddPs} carry the contributions of tachyons propagating
in loops: since we wish to excise tachyons from the spectrum, we need to implement
a GSO projection in such a way that these three terms vanish. This is achieved by
simply averaging the amplitude over the four spin structures $\left( \varsigma_1,
\varsigma_2 \right) \, \in \, \big\{ (0,0),(1,0),(0,1),(1,1) \big\}$; one clearly sees
that the first three terms in \eq{eq:oddPs} vanish while the fourth term is independent
of $\vec{\varsigma}$ and thus unaffected. Therefore the GSO-projected amplitude
is free of tachyons while the massless sector is intact, as desired.

We are now in a position to take the field theory limit for the symmetric degeneration.
The only missing ingredient is the normalization factor ${\cal N}_h^{\, (\veps \,)}$
introduced in \eq{eq:mSetup}. It is given by
\beq
  {\cal N}_2^{\, (\veps \,)}  \, = \, \frac{C_2}{\cos \left( \pi \epsilon_1 \right)
  \cos \left( \pi \epsilon_2 \right)} \, ,
\label{neps}
\eeq
where $C_h$ is the normalization factor for an $h$-loop string amplitude in terms
of the $d$-dimensional Yang-Mills coupling $g_d$, calculated in Appendix A of
Ref.~\cite{DiVecchia:1996uq}. For $h = 2$ it is given by
\beq
  C_2 \, = \, \frac{g_d^2}{(4 \pi)^d} \, \frac{(\alpha' )^2}{(2 \pi \alpha')^d} \, .
\label{eq:normalization}
\eeq
The denominator in \eq{neps} arises from the Born-Infeld contribution to the
normalization of the boundary state (see for example Ref.~\cite{DiVecchia:1999fx}).
It does not contribute to the field theory limit, since $\cos( \pi \epsilon_1)  \cos( \pi
\epsilon_2) = 1 + {\cal O} ( {\alpha'}^2)$, after expressing the twists $\epsilon_i$
in terms of the background field strengths via \eq{eq:epsToB}.

Applying the GSO projection to \eq{eq:integrandSeries}, and using $d p_i / p_i = -
\d t_i / \alpha'$, we finally find that the QFT limit of the partition function can be
represented succinctly by
\beqa
\label{eq:AQFTa}
  Z_{2, QFT}^{\rm \, sym} \left( m_i, B_i \right) & = & \frac{g_d^2}{(4 \pi)^d} \int \,
  \prod_{i = 1}^3  d t_i \, \ex{ - t_i m_i^2} \, \frac{1}{\Delta_0^{(d - 2)/2} \Delta_B}
  \nonumber \\ & & \hspace{1cm} \times \,  \lim_{\alpha' \to 0} \,
  \bigg[ - \frac{1}{\alpha'} \int \d \hat{\theta}_{12} \, \d \hat{\phi}_{12}
  \, \, \widehat{\bf F}^{(111)} \left( t_i, m_i, B_i \right) \bigg] \, ,
\eeqa
where the limit on the second line is finite after Berezin integration. In order to
see that, and in order to give our results as explicitly as possible, we define (for simplicity we omitt the arguments of the functions ${\bf f}$)
\begin{align}
  {\bf f}_{\rm gh}^{(m n p)}  & =
  \begin{cases}
    - \frac{1}{\alpha ' } \, \, \partial_{\hat{\theta}_{12}} \partial_{\hat{\phi}_{12}}
    \widehat{\bf F}_{\rm gh}^{(m n p)}
    & \text{ if } \, m + n + p \, \text{ is odd} \, ,\\
    \widehat{\bf F}_{\rm gh}^{(m n p)}
    & \text{ if } \, m + n + p \, \text{ is even} \, ,\\
  \end{cases}
\label{bercases}
\end{align}
and similarly for ${\bf f}_{\parallel}^{(m n p)} (t_i, B_i)$, ${\bf f}_{\perp}^{(m n p)} (t_i)$
and ${\bf f}_{\rm scal}^{(m n p)} (t_i, m_i)$. With our definitions, one easily sees that
\beq
  {\bf f}_{\text{gh}}^{(000)} \, = \, {\bf f}_{\parallel}^{(000)} \, = \,
  {\bf f}_{\perp}^{(000)} \, = \, {\bf f}_{\text{scal}}^{(000)} \, = \, 1 \, .
\label{zerocomp}
\eeq
Performing the Berezin integration is then a simple matter of combinatorics, and one
finds
\beqa
\label{eq:partialA}
  Z_{2, QFT}^{\rm \, sym} \left( m_i, B_i \right)  & =  & \frac{g_d^2}{(4 \pi)^d}
  \int  \, \prod_{i=1}^3  d t_i \, \ex{ - t_i m_i^2} \, \frac{1}{\Delta_0^{(d - 2)/2} \Delta_B} \\
  & & \hspace{1cm} \times \, \bigg[ {\bf f}^{(111)}_\parallel (t_i, B_i) + \, {\bf f}^{(110)}_\parallel
  (t_i, B_i) \, {\bf f}^{(001)}_\perp (t_i) \, + \,
  \big( \text{25 more terms} \big ) \bigg] \, . \nonumber
\eeqa
We can read off the various terms in the integrand by picking the coefficients of the
appropriate factors of $q_i$ from ${\bf F}_{\parallel}$ in \eq{eq:FparField}, ${\bf
F}_{\perp}$ in \eq{eq:FperpField}, ${\bf F}_{\text{scal}}$ in \eq{eq:FscalField} and $d
{\bs \mu}_2 \, {\bf F}_{\text{gh}}$ in \eq{eq:FghPser}, and then selecting the coefficient
of $\hat{\theta}_{12} \,\hat{\phi}_{12}$, divided by $\alpha ' $. We find
\beqa
  {\bf f}^{(111)}_\parallel & = & \frac{2}{\Delta_B} \, \frac{\sinh \big( g B_1 t_1
  \big)}{g B_1} \, \cosh \Big[ g \big( 2 B_1 t_1 - B_2 t_2 - B_3 t_3 \big) \Big] + \,
  \text{cycl. perm.} \, , \nonumber \\
  {\bf f}^{(110)}_\parallel \, {\bf f}^{(001)}_\perp & =  & \frac{2}{\Delta_0} \, (d - 2) \,
  t_3 \, \cosh \Big[ 2 g \big( B_1 t_1 - B_2 t_2 \big) \Big]  \, , \nonumber  \\
  {\bf f}^{(001)}_\parallel \, {\bf f}^{(110)}_\perp & = & \frac{2}{\Delta_B} \, (d-2) \,
  \frac{\sinh \big( g B_3 t_3 \big)}{g B_3} \, \cosh \Big[ g \big( 2 B_3 t_3 - B_1 t_1 - B_2 t_2
  \big) \Big] \, , \nonumber \\
  {\bf f}^{(111)}_\perp & = & \frac{1}{\Delta_0} \, (d - 2) \, (d - 3) \, \big( t_1 + t_2 + t_3 \big) \, ,
  \label{eq:Fnnn}  \\
  {\bf f}_{\text{gh}}^{(110)} \, {\bf f}^{(001)}_\parallel & = & - \, \frac{2}{\Delta_B} \,
  \frac{\sinh \big( g B_3 t_3 \big)}{g B_3} \, \cosh \Big[ g \big( 2 B_3 t_3 - B_1 t_1 - B_2 t_2
  \big) \Big] \, , \nonumber \\
  {\bf f}_{\text{gh}}^{(110)} \, {\bf f}^{(001)}_\perp & = & - \, \frac{1}{\Delta_0} \, (d - 2) \, t_3
  \, , \nonumber \\
  {\bf f}_{\text{gh}}^{(110)} \, {\bf f}^{(001)}_{\text{scal}} & = & m_3^2 - m_1^2 - m_2^2
  \, , \nonumber \\
  {\bf f}_{\text{scal}}^{(110)} {\bf f}^{(001)}_\parallel & = & \frac{2}{\Delta_B} \, n_s \,
  \frac{\sinh \big( g B_3 t_3 \big)}{g B_3} \, \cosh \Big[ g \big( 2 B_3 t_3  - B_1 t_1 - B_2 t_2
  \big) \Big] \, , \nonumber \\
  {\bf f}_{\text{scal}}^{(110)} \, {\bf f}_\perp^{(001)} & = & \frac{1}{\Delta_0} \, (d - 2) \,
  n_s \, t_3 \, , \nonumber \\
  {\bf f}^{(110)}_\parallel {\bf f}_{\text{scal}}^{(001)} & = & 2 \, \big( m_1^2 + m_2^2 - m_3^2
  \big) \cosh \Big[ 2 g \big( B_1 t_1 - B_2 t_2 \big) \Big] \, , \nonumber \\
  {\bf f}^{(110)}_\perp \, {\bf f}_{\text{scal}}^{(001)} & = & (d - 2) \big( m_1^2 + m_2^2 -
  m_3^2 \big) \, , \nonumber \\
  {\bf f}_{\text{scal}}^{(111)} & = & ( n_s - 1 ) \big( m_1^2 + m_2^2 + m_3^2 \big)  \, .
  \nonumber
\eeqa
The other terms in the integrand can be obtained from the above by cyclic symmetry.

We conclude by noting that \eq{eq:partialA} does not give the complete 2-loop
contribution to the vacuum amplitude with this topology, since the string theory
calculation distinguishes the three D-branes where the world-sheet boundaries
are attached. This is reflected in the integration region over the Schwinger
parameters $t_i$, already discussed in Ref.~\cite{Magnea:2004ai}: they are
not integrated directly in the interval $0 < t_i < \infty$, as would be the case in
field theory, but they are ordered, as $0 < t_3 < t_2 < t_1 < \infty$. In order to
recover the full amplitude, with the correct color factors and integration region,
one must sum over all possible attachments of the string world-sheet to the
D-branes, effectively summing over the different values of the background fields
$B_{i}$ and masses $m_i^2$. In the absence of external fields, this sum amounts
just to the introduction of a symmetry and color factor; for non-vanishing $B_i$,
it reconstructs the correct symmetry properties of the amplitude under permutations.


\subsection{The incomplete degeneration}
\label{NSeighthandle}

\begin{figure}
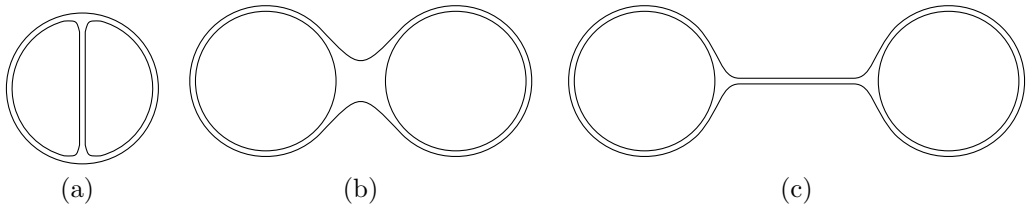

\begin{center}
\subfloat[]{ \stringapplediagram \label{fig:stringapple}}
\subfloat[]{\stringeightdiagram \label{fig:stringeight}}
\subfloat[]{\stringhandlediagram \label{fig:stringhandle}}
\caption{The symmetric (\Fig{fig:stringapple}),
incomplete (\Fig{fig:stringeight}), and separating (\Fig{fig:stringhandle}) degenerations of the two-loop vacuum amplitude.}
\label{fig:thinworldsheets}
\end{center}
\end{figure}

In the last three sections, \ref{symmpar}, \ref{rewrite} and \ref{GSO}, we have
given the tools to compute the field theory limit of the partition function in the
vicinity of the symmetric degeneration, see~\Fig{fig:stringapple}: our final result is summarized in \eq{eq:partialA}.
The field theory two-loop effective action, however, includes also the Feynman diagrams with a quartic vertex depicted in the last
two lines of Fig.~\ref{fig:1PIgraphs}.


%

The main feature of vacuum graphs with a four point vertex, which drives
the corresponding choice of parametrization for the neighborhood of moduli
space depicted in \Fig{fig:stringeight}, is the fact that such graphs have only two
propagators, each one encompassing a complete loop, and furthermore they
are symmetric under the exchange of the two loops. It is natural therefore
to associate to each propagator a Schwinger parameter linked to the Schottky
multiplier of the corresponding string loop. The fact that there are no further
Schwinger parameters implies also that the third bosonic modulus must be
integrated over its domain except for a small region around each boundary. We therefore call the configuration depicted in \Fig{fig:stringeight}
the {\it incomplete} degeneration.

To compute the field theory limit for the incomplete degeneration, we must
retrace our steps back to \secn{expak}, where the various factors in the partition
function were expressed in terms of $k_i$ and $u$ (or $y$). We then relate the
multipliers to Schwinger parameters as
\beq
  k_i \, = \, \ex{- t_i/\alpha '} \, , \qquad \qquad \left( i \, = \, 1 , \, 2 \right) \, ,
\label{eq:kTot}
\eeq
and replace $\epsilon_i$ according to \eq{eq:epsToB}.

As may be expected from the simplicity of the target graph,
the string partition function simplifies drastically when the $\alpha' \to 0$ limit is taken
in this way. One finds for example that the determinant of the (twisted) period
matrix reduces to
\beqa
\label{dettauinc}
  \Big[ \det \left({\rm Im} \, \bs{\tau} \right) \Big]^{-1} & = & \, \frac{(2 \pi \alpha')^2}{t_1 t_2} \, + \,
  {\cal O} \Big( ({\alpha'})^3 \Big) \\
  \Big[ \det \left({\rm Im} \, \bs{\tau}_{\veps} \right) \Big]^{-1} & = & \, (2 \pi \alpha')^2 \,
  \frac{g B_1}{\sinh \big( g B_1 t_1 \big)} \, \frac{g B_2}{\sinh \big( g B_2 t_2 \big)}
  \, + \, {\cal O} \Big( ({\alpha'})^3 \Big) \, , \nonumber
\eeqa
while the hyperbolic functions appearing in the field theory limit arise in a direct way from
combinations like
\beqa
\label{samphyp}
  k_1^{\epsilon_1} u^{\epsilon_2} + k_1^{- \epsilon_1} u^{- \epsilon_2} & = &
  2 \cosh \big( 2 g B_1 t_1 \big) \, + \, {\cal O} (\alpha ') \, , \\
  k_1^{\epsilon_1} k_2^{\epsilon_2} u^{\epsilon_1 + \epsilon_2} +
  k_1^{- \epsilon_1} k_2^{- \epsilon_2} u^{- \epsilon_1 - \epsilon_2}  & = &
  2 \cosh \Big( 2 g \, ( B_1 t_1 + B_2 t_2 ) \Big) \, + \, {\cal O}(\alpha' ) \, . \nonumber
\eeqa
The resulting expressions are very simple because, with no Schwinger parameter
associated to $u$, factors of the form $u^{\pm \, \epsilon_i}$ do not contribute to the field
theory limit. This is what makes it possible to perform the integration over the third
bosonic modulus, and over the two fermionic moduli: indeed, in the parametrization considered here,
the entire partition function can be written explicitly, in the $\alpha' \to 0$ limit, in terms
of just three simple integrals over the non-degenerating coordinates of super-moduli
space. After GSO projection, one finds
\beqa
\label{eq:dmSepField}
   d \, {\bf Z}_2^{\rm \, inc + sep} ({\bs \mu}) & = & \,
   \frac{(2 \pi \alpha')^d}{(\alpha')^2} \, \prod_{i=1}^2  \, \Bigg[ \frac{d t_i \,
   \ex{ - t_i m_i^2} \,  g B_i }{t_i ^{d/2 - 1} \sinh \big( g B_i t_i \big)}  \Bigg] \,
   d u \, d \theta \, d \phi \, \\
   & & \hspace{-1cm} \times \,
   \Bigg\{ \frac{1}{y} \Big( d - 2 + 2 \cosh \big( 2 g B_1 t_1 \big) + n_s - 2 \Big)
   \Big( d - 2 + 2 \cosh \big( 2 g B_2 t_2 \big) + n_s - 2 \Big) \nonumber \\
   &  &\hspace{5mm} - \, \bigg[ d - 2 + 2 \cosh \Big( 2 g (B_1 t_1 + B_2 t_2) \Big) + n_s \bigg]
   \nonumber \\ &  & \hspace{5mm} + \, \frac{1}{u} \, \bigg[ d - 2 + 2 \cosh \Big( 2 g
   (B_1 t_1 - B_2 t_2) \Big) + n_s \bigg] \Bigg\} \, + \,
   {\cal O} \left( \ex{ - {t_i}/{\alpha'}}, \alpha' \right) \, . \nonumber
\eeqa
As the notation suggests, \eq{eq:dmSepField} in principle contains contributions
from both the incomplete and the separating degenerations, and we now turn to
the problem of disentangling them. We also see that in order to complete the
calculation one just needs to determine three numerical constants, given by the
following integrals over the non-degenerating super-moduli,
\beq
  I_1 \, = \, \int_{\widehat{\frak M}_{1|2}} \! d u \, d \theta\, d \phi \, \, \frac{1}{y} \, , \qquad
  I_2 \, = \, - \, \int_{\widehat{\frak M}_{1|2}}  \! d u \, d \theta \, d \phi \, , \qquad
  I_3 \, = \, \int_{\widehat{\frak M}_{1|2}}  \! d u \, d \theta\, d \phi \, \,\frac{1}{u} \, .
\label{eq:I1I2I3}
\eeq
To determine the domain of integration $\widehat{\frak M}_{1|2}$ in \eq{eq:I1I2I3},
and to identify the different degeneration limits, note that the separating, symmetric and
incomplete degenerations all come from the region of super-moduli space in which the
two Schottky multipliers $k_1$ and $k_2$ are small. In this limit, we can think of
super moduli space as a $1|2$-dimensional space parametrized by $(u|\theta,\phi)$.
The separating degeneration corresponds to the limit $y \to 0$, while the symmetric
degeneration corresponds to the limit $u \to 0$, and the incomplete degeneration
comes from the region of super moduli space interpolating between the two limits.
As pointed out in Refs.~\cite{Witten:2012bg,Witten:2012ga}, however, this simple
characterization must be made more precise, in particular with regards to the choice
of parameters near the two degenerations.

First of all, let us briefly consider the first term in braces on the right-hand side of
\eq{eq:dmSepField}. This dominates in the limit $y \to 0$, and we expect it to represent
the contributions of the one-particle reducible (1PR) Feynman diagrams,
which we neglect.
We can now concentrate on the evaluation of the integrals relevant for
our purposes, which are $I_2$ and $I_3$ in \eq{eq:I1I2I3}. They can be calculated using
Stokes' theorem for a super-manifold with a boundary (see section 3.4 of \cite{Witten:2012bg}),
since the integrands are easily expressed as total derivatives. We write
\beqa
\label{eq:nu3}
  - d u \, d \theta \, d \phi & \equiv & d \nu_2  \, , \qquad  \nu_2 \, = \, - \, u \,
  d \theta  \, d \phi \, , \\
  \frac{d u}{u} \, d \theta \, d \phi & \equiv & d \nu_3 \, , \qquad
  \nu_3 \, = \, \log(u) \, d \theta \, d \phi \, . \nonumber
\eeqa
These expressions mean that the corresponding integrals are localized on the boundary
of $\widehat{\frak M}_{1|2}$, which consists of two loci associated with the two distinct
ways in which the double annulus world-sheet can completely degenerate: the symmetric
degeneration of \fig{fig:stringapple}, and the separating degeneration of \fig{fig:stringhandle}.

To use Stokes' theorem, it is important to characterize precisely the $0|2$-dimensional
boundary of the super-moduli-space region over which we are integrating. More precisely,
we need to find bosonic functions of the worldsheet moduli $\xi_i (u, \theta, \phi)$, defined
near the boundaries of $\widehat{\frak M}_{1|2}$, such that the vanishing of $\xi_i$ defines a
compactification divisor $\EuScript{D}_i$. Such functions are called {\it canonical parameters}
in section 6.3 of \cite{Witten:2012ga}.
%
%
%
%
%
%
It is important to note that, for singular integrands such as those of $I_1$ and $I_3$, it is
not sufficient to define the canonical parameter $\xi$ up to an overall factor, which may
include nilpotent terms. For example, if we attempt to rescale $\xi \, = \, (1 + \theta \phi)
\xi'$, then $\log \xi = \log \xi' + \theta \phi$, so that the Berezin integral $\int d \theta \, d
\phi \, \log \xi$ does not coincide with $\int d \theta \, d \phi \, \log \xi'$.

In the small-$u$ region, the proper choice of the canonical parameter $\xi_{\rm sym}$ is
dictated by our parametrization of the symmetric degeneration: we must take $\xi_{\rm
sym} = p_3$, as defined in \eq{eq:kToP}, in order to properly glue together the two regions.
Although $p_3$ and the cross-ratio $u$ vanish at the same point, they are related by a
non-trivial rescaling at leading order in the multipliers. Indeed
\beq
  u \, = \, \frac{p_3}{1 + p_3} \big( 1 +  \theta \phi \big)  + {\cal O} \left( k_i^{1/2} \right) \, ,
\label{uresc}
\eeq
which affects the Berezin integral of \eq{eq:nu3}, as discussed above. Note that, not
having introduced a parametrization for the separating degeneration, we would not
have a similar guideline in the small-$y$ region. Furthermore, the fact that the
corresponding field theory diagram needs to be regulated\footnote{Notice that in the gauge we use this Feynman diagram would not automatically vanish in a $U(N)$ theory as the 3-point vertices contain also terms proportional to the symmetric color tensor $d^{abc}$.} in order to make sense
of the vanishing momentum flowing in the intermediate propagator introduces an
ambiguity also in the field theory result.

With this choice of parametrization, we can now use Stokes' theorem to determine the values
of $I_2$ and $I_3$. Taking $\xi_{\text{sep}} = y$ as a canonical parameter for the separating degeneration, we find
\beqa
\label{eq:I2ans}
  I_2 & = & \lim_{\epsilon \to 0} \Bigg[ \int_{y = \epsilon} \nu_2 - \int_{p_3 = \epsilon} \nu_2
  \Bigg]  \, , \\
  & = & \lim_{\epsilon \to 0} \Bigg[ - \int  d \theta \, d \phi \, \big( 1 - \epsilon + \theta \phi \big)
  \, + \, \int d \theta \, d \phi \,\frac{\epsilon}{1 + \epsilon} \big(1 + \theta \phi \big)
  \Bigg] \, = \, 1 \, . \nonumber
\eeqa
where we used $\int d \theta \, d \phi \, \theta \phi = - 1$. Similarly
\beqa
\label{eq:I3ans}
  I_3 & = & \lim_{\epsilon \to 0} \Bigg[ \int_{y = \epsilon} \nu_3 - \int_{p_3 = \epsilon} \nu_3
  \Bigg]  \, , \\
  & = & \lim_{\epsilon \to 0} \Bigg[ \int  d \theta\, d \phi \, \log( 1 - \epsilon + \theta \phi )
   \, - \, \int d \theta \, d \phi \, \log \left[  \frac{\epsilon}{1 + \epsilon}
  (1+ \theta \phi)\right] \Bigg] \, = \, 0 \, . \nonumber
\eeqa
Inserting these results into \eq{eq:dmSepField}, discarding the separating degeneration,
and introducing the overall normalization given in \eq{eq:normalization}, we obtain our
final expression for the contribution of diagrams with a four-point vertex to the field-theory
effective action. It is given by
\beqa
\label{eq:dmSepfinal1}
  Z_{2, QFT}^{\rm \, inc} \left( m_i, B_i \right)  & = & \frac{g^2}{(4 \pi)^d}
  \int_0^\infty \, \prod_{i = 1}^2 \Bigg[ \frac{d t_i \, \ex{ - t_i m_i^2} \, g B_i }{t_i ^{d/2 - 1}
  \sinh \big( g B_i t_i \big)} \Bigg] \\
  & & \hspace{2cm}
  \Big( d - 2 + 2 \cosh \big( 2 g (B_1 t_1 + B_2 t_2) \big) + n_s \Big)
  \, . \nonumber
\eeqa
In order to identify the contributions of individual Feynman diagrams to \eq{eq:dmSepfinal1},
we can retrace the steps of the calculation and assign each term in our result to the appropriate
world-sheet conformal field theory, as we did for the symmetric degeneration in \eq{eq:partialA}.
We find that we can rewrite \eq{eq:dmSepfinal1} as
\beq
  Z_{2, QFT}^{\rm \, inc} \left( m_i, B_i \right)  \, = \, - \, \frac{g^2}{(4 \pi)^d}
  \int_0^\infty \, \prod_{i = 1}^2 \Bigg[ \frac{d t_i \, \ex{ - t_i m_i^2} \, g B_i }{t_i ^{d/2 - 1}
  \sinh \big( g B_i t_i \big)} \Bigg] \, \Big( {\bf f}_{\parallel}^{11} + {\bf f}_{\perp}^{11} +
  {\bf f}_{\text{scal}}^{11} + {\bf f}_{\text{gh}}^{11} \Big) \, ,
\label{eq:dmSepfinal}
\eeq
where here the superscripts denote the powers of $k_i^{1/2}$ $(i = 1,2)$ from which the
coefficients were extracted, and we have omitted the arguments of the functions ${\bf f}$
for simplicity. The precise identification is
\beq
  {\bf f}_{\parallel}^{11} \, = \, - \, 2 \cosh \big( 2 g (B_1 t_1 + B_2 t_2) \big) \, \quad
  {\bf f}_{\perp}^{11} \, = \, - \, (d - 2) \, , \quad {\bf f}_{\text{scal}}^{11} \, = \,- \, n_s \, ,
  \quad {\bf f}_{\text{gh}}^{11} \, = \, 0 \, .
\label{eq:F01}
\eeq
A few remarks are in order. First of all we note that ${\bf f}_{\text{gh}}^{11}$ vanishes; this
corresponds to the fact that, in the infinite product in ${\bf F}_{\text{gh}} (k_i, \eta)$ in
\eq{eq:gluonsf}, $n$ ranges from 2 to $\infty$, not from $1$ to $\infty$ as in the case
of ${\bf F}_{\text{gl}}$ and ${\bf F}_{\text{scal}}$. As a consequence, there is no term
proportional to $k^{1/2} ( {\bf S}_1 {\bf S}_2 )$ in the partition function for the ghost
systems. We will see that this corresponds to the fact that there is no quartic ghost
vertex in the associated Yang-Mills theory. Next, we note that all terms associated
with the four-point vertex diagram are not factorizable into the product of two contributions,
proportional to $k_1^{1/2}$ and $k_2^{1/2}$ respectively. If, on the other hand, we
had traced the origin of the terms associated with the separating degeneration, and
proportional to the integral $I_1$, we would have found that the factor multiplying $1/y$
in \eq{eq:dmSepField} can be written as
\beq
  \Big( {\bf f}_{\parallel}^{10} + {\bf f}_{\perp}^{10} + {\bf f}_{\text{scal}}^{10}
  + {\bf f}_{\text{gh}}^{10} \Big) \Big( {\bf f}_{\parallel}^{01} + {\bf f}_{\perp}^{01} +
  {\bf f}_{\text{scal}}^{01} + {\bf f}_{\text{gh}}^{01} \Big) \, .
\label{septerms}
\eeq
This means that no contributions arise from the Schottky group elements  ${\bf S}_1
{\bf S}_2$ and  ${\bf S}_1^{-1} {\bf S}_2$, which would imply a genuine correlation
between the two loops. Rather, as expected, these terms are simply the product
of factors rising from individual disconnected loops. Finally, we note that the result
$I_3  = 0$ is crucial in order to recover the correct field theory limit: indeed, as will
be verified in the next section and shown in Appendix~\ref{Appc}, no field theory
diagram yields hyperbolic functions with the parameter dependence displayed on
the last line of \eq{eq:dmSepField}. We see once again that the field theory limit,
once the contributions of individual diagrams have been identified, provides
non-trivial checks of the procedures used to perform the integration over super-moduli.


\section{Yang-Mills theory in the Background Field Gervais-Neveu gauge}
\label{YMQFT}

In order to make a precise comparison between string theory and field theory at
the level of individual Feynman diagrams, as was done in a simple case in
Ref.~\cite{Magnea:2013lna}, we need a precise characterization of the field-theory
Lagrangian we are working with, including gauge fixing and ghost contributions.
In principle, this presents no difficulties, since our target is a $U(N)$ Yang-Mills
theory, albeit with a rather special gauge choice. There are however a number
of subtleties, ranging from the special features of the background field framework,
to issues of dimensional reduction, and to the need to break spontaneously the
gauge symmetry in order to work with well-defined Feynman diagrams in the
infrared limit, which altogether lead to a somewhat complicated and unconventional
field theory setup. We will therefore devote this section to a detailed discussion
of the field theory Lagrangian which arises from the field theory limit of our chosen
string configuration.

The first layer of complexity is due to the fact that the string theory setup naturally
corresponds to a field theory configuration with a non-trivial background field. In
general, such a background field breaks the gauge symmetry: in our case, since
we are working with mutually commuting gauge fields with constant field strengths,
and we have a string configuration with separated $D$-brane sets, one will generically
break the $U(N)$ gauge symmetry down to $U(1)^N$.
We will
have to adjust our notation to take this into account. Notice also that our background
fields break Lorentz invariance as well, since only certain polarizations are non-vanishing.
As a consequence, the polarizations of the quantum field will also be distinguished as
parallel or perpendicular to the given background.

Furthermore, it is interesting to work in a generic space-time dimension $d$,
and we will find it useful to work with massive scalar fields giving infrared-finite
Feynman diagrams. We will therefore work with a $d$-dimensional gauge theory
obtained by dimensional reduction from the dimension ${\cal D} > d$ appropriate to the
string configuration. This yields $n_s = {\cal D} - d$ adjoint scalar fields minimally coupled
to the $d$-dimensional gauge theory, and we will choose our background fields such
that these fields acquire a non-vanishing expectation value, giving mass to some of
the gauge fields.

Finally, as suggested originally in Ref.~\cite{Bern:1991an}, and recently confirmed by
the analysis of Ref.~\cite{Magnea:2013lna}, covariantly quantized string theory picks
a very special gauge in the field theory limit: a background field version of the non-linear
gauge first introduced by Gervais and Neveu in Ref.~\cite{Gervais:1972tr}. This gauge
has certain simplifying features: for example at tree level and at one loop it gives
simplified color-ordered Feynman rules which considerably reduce the combinatoric
complexity of gauge-theory amplitudes~\cite{Bern:1991an}. Only at the two-loop level,
however, the full complexity of the non-linear gauge fixing becomes apparent. One
effect, for example, is that the diagonal $U(1)$ `photon', which ordinarily is manifestly
decoupled and never appears in `gluon' diagrams, in this case has non-trivial,
gauge-parameter dependent couplings to $SU(N)$ states, and the decoupling only
happens when all relevant Feynman diagrams are summed.

In what follows, we adopt the following notations: we use calligraphic letters for
matrix-valued $\mathfrak{u}(N)$ gauge fields, and ordinary capital letters for their
component fields; we use $M,N, \ldots = 1, \ldots, {\cal D}$ for Lorentz indices in
${\cal D}$-dimensional Minkowski space before dimensional reduction, and $\mu, \nu,
\ldots = 1, \ldots, d$ for Lorentz indices in the $d$-dimensional reduced space-time;
finally, $I,J, \dots = 1, \ldots, n_s$ indices enumerate adjoint scalars, and $A,B, \ldots
 = 1, \ldots, N$ indices enumerate the components of $\mathfrak{u}(N)$ vectors
 and matrices. In this language, ${\cal A}_M$ will denote the ${\cal D}$-dimensional
 classical background field, ${\cal Q}_M$ the corresponding quantum field, while
 ${\cal C}$ and $\overline{\cal C}$ are ghost and anti-ghost fields.

We will now proceed to write out the quantum lagrangian (including gauge-fixing
and ghost terms) in terms of matrix-valued fields. We will then comment on
the form taken by various terms in component notation, which  is more directly
related to the vertices appearing in diagrammatic calculations.


\subsection{The $\mathfrak{u}(N)$ Lagrangian}
\label{lagra}

We begin by constructing the ${\cal D}$-dimensional Yang-Mills Lagrangian, which, in the
presence of a background gauge field, depends on the combination ${\cal  A}_M +
{\cal Q}_M$. The field-strength tensor ${\cal F}_{M N}$ can be expressed in terms
of the covariant derivative of the quantum field with respect to the background field,
${\frak D}_M = \partial_M +  \ii \, g \left[ {\cal A}_M , \cdot \right]$, as
\beqa
  {\cal F}_{M N} \left( {\cal  A} + {\cal Q} \right) & = & - \,
  \frac{\ii}{g} \left[ {\frak D}_M^{ \, {\cal A} + {\cal Q}} , {\frak D}_N^{ \, {\cal A} + {\cal Q}}
  \right]  \nonumber \\
  & =  & {\cal F}_{M N} \left( {\cal  A} \right) + {\frak D}_M {\cal Q}_N - {\frak D}_N {\cal Q}_M
  + \ii \, g \left[ {\cal Q}_M , {\cal Q}_N \right] \, ,
\label{fmn}
\eeqa
where ${\cal F}_{M N} \left( {\cal  A} \right)$ is the field strength tensor for the background
field only, while ${\frak D}_M^{ \, {\cal A} + {\cal Q}}$ is the covariant derivative with respect
to the complete gauge field. The classical Lagrangian for the quantum gauge field ${\cal Q}$
can then be written as
\beqa
  {\cal L}_{\rm cl} & = & {\rm Tr} \bigg[ {\frak D}^M {\cal Q}^N {\frak D}_N {\cal Q}_M -
  {\frak D}^{M} {\cal Q}^N {\frak D}_M {\cal Q}_N
  + 2 \, \ii \, g \, {\cal F}_{M N} {\cal Q}^M {\cal Q}^N
  \nonumber \\ & & \hspace{40pt}
  + 2 \, \ii \, g \, {\frak D}^M {\cal Q}^N
  \left[ {\cal Q}_M , {\cal Q}_N \right] + \frac{1}{2} \, g^2  \left[ {\cal Q}_M , {\cal Q}_N \right]
  \left[ {\cal Q}^M , {\cal Q}^N \right] \bigg] \, ,
\label{classicallagrangiantrace}
\eeqa
where ${\rm Tr}$ here denotes the trace over the ${\frak u}(N)$ Lie algebra, and we have
removed terms independent of ${\cal Q}$, as well as terms linear in ${\cal Q}$, because
they are not relevant for our effective action calculation.

In anticipation of the string theory results, we now wish to fix the gauge using a
background field version of the non-linear gauge condition introduced by Gervais
and Neveu in Ref.~\cite{Gervais:1972tr}, setting
\beq
  {\cal G} \left( {\cal A}, {\cal Q} \right) \, = \,
  {\frak D}_M {\cal Q}^M + \ii \, \gamma \, g \, {\cal Q}_M {\cal Q}^M \, = \, 0 \, ,
\label{YMgaugecon}
\eeq
where $\gamma$ is a gauge parameter. The gauge-fixing Lagrangian ${\cal L}_{\rm gf}$
is then given by
\beqa
  {\cal L}_{\rm gf} & = & - \, {\rm Tr} \bigg[ \Big( {\cal G} \left( {\cal A}, {\cal Q} \right)
  \Big)^2 \bigg]  \nonumber \\
  & = & - \, {\rm Tr} \bigg[ {\frak D}_M {\cal Q}^M {\frak D}_N {\cal Q}^N + \,
  2 \, \ii \, \gamma \, g \, {\frak D}_M {\cal Q}^M {\cal Q}_N {\cal Q}^N -
  \gamma^2 g^2 {\cal Q}_M {\cal Q}^M {\cal Q}_N {\cal Q}^N \bigg] \, .
\label{gaugelagrangiantrace}
\eeqa
Notice that the overall covariant gauge-fixing parameter which would appear in front of
\eq{gaugelagrangiantrace} has been set equal to one. Note also that this gauge fixing
modifies not only the gluon propagator, as expected, but also the three- and four-gluon
vertices. In particular, the symmetric nature of the quadratic term in the gauge-fixing
function ${\cal G}$ will generate Feynman rules involving the symmetric ${\frak u}(N)$
tensors $d_{a b c}$, which in turn will induce spurious couplings  between gluons and
${\frak u}(1)$ photons.

Finally, we need the Lagrangian for the Faddeev-Popov ghost and anti-ghost fields,
$\mathcal{C}$ and  $\overline{\mathcal{C}}$. It is defined as usual in terms of the gauge
transformation of the gauge-fixing function, as
\beq
  {\cal L}_{\rm gh} \, = \, {\rm Tr} \Big[ \overline{\cal C} \, \delta_{{\cal C}}
  {\cal G} \left( {\cal A}, {\cal Q} \right) \Big] \, ,
\label{vargf}
\eeq
using ${\cal C}$ as parameter of the gauge transformation. The result is
\beqa
  {\cal L}_{\rm gh} & = & 2 \, {\rm Tr} \bigg[ - \overline{\mathcal{C}} \, {\frak D}_M
  {\frak D}^{M} \mathcal{C} + \ii \, g \, {\frak D}_M \overline{\mathcal{C}} \,
  \left[ {\cal Q}^M , \mathcal{C} \right] \nonumber \\
  && - \, \ii \, \gamma \, g \, \overline{\mathcal{C}} \,
  \big\{ Q_M , {\frak D}^M \mathcal{C} \big\} + \gamma \, g^2 \, \overline{\mathcal{C}}
  \big\{ {\cal Q}_M , \left[ {\cal Q}^M , \mathcal{C} \right] \big\} \bigg] \, .
\label{lgh}
\eeqa
This completes the construction of the pure Yang-Mills Lagrangian in ${\cal D}$ dimensions;
next, we want to dimensionally reduce it to $d$ dimensions. The reduction splits the
${\cal D}$-dimensional gauge fields (both classical and quantum) into a $d$-dimensional
field and $n_s \equiv {\cal D} - d$ adjoint scalars, according to
\beq
  \big\{ {\cal A}_M \big\} \, \to \,  \big\{ {\cal A}_\mu \, , \frac{1}{g} {\cal M}_I \big\} \, , \qquad
  \big\{ {\cal Q}_M \big\} \, \to \,  \big\{ {\cal Q}_\mu \, , \Phi_I \big\} \, ,
\label{splitQ}
\eeq
with $\mu = 0, \ldots, d - 1$ and $I = 1, \ldots, n_s$, and we have assumed that the classical
background scalars take on constant values ${\cal M}_I$, which we will use to spontaneously
break the gauge symmetry and give masses to selected components of the gauge field.
Similarly, since we are neglecting the dependence of the fields on the reduced coordinates,
the covariant derivative splits into a $d$-dimensional covariant derivative and a pure
commutator with the background scalar fields, as
\beq
   \big\{ {\frak D}_M \big\} \, \to \, \Big\{ {\frak D}_\mu \equiv \partial_\mu + \ii \, g
   \left[ {\cal A}_\mu , \cdot \right] , \, \ii \, \left[ {\cal M}_I \, , \cdot \right] \Big\} \, .
\label{splitD}
\eeq
Indeed, the ${\cal D}$-dimensional d'Alembertian differs from the $d$-dimensional one by
a mass term: for any field $X$,
\beq
  {\frak D}_M {\frak D}^M X \, = \, {\frak D}_\mu {\frak D}^\mu X +
  \Big[ {\cal M}_I , \big[ {\cal M}^I , X \big] \Big] \, .
\label{dalemb}
\eeq
Notice that in this section we work with the metric $\eta = {\rm diag}(+, -, \ldots, -)$.
However, when summing over reduced dimensions, our summation convention does
not include the negative signature of the metric, and must be understood simply as a
summation over flavor indices $I$. With these conventions, the gauge condition in
\eq{YMgaugecon} becomes
\beq
  {\frak D}_\mu {\cal Q}^\mu +  \ii \, \gamma \, g \, {\cal Q}_\mu {\cal Q}^\mu
  - \ii \, \big[ {\cal M}_I, \Phi^I \big] - \ii \, \gamma \, g \, \Phi_I \Phi^I \, = \, 0 \, .
\label{scalargaugecon}
\eeq
When these further changes are implemented in the Lagrangian, a number of
non-trivial interaction vertices are generated. It is then useful to organize the
dimensionally-reduced Lagrangian as a sum of terms with different operator
content. One can write
\beqa
  {\cal L}_{{\cal Q}^2} & = &{\rm Tr} \bigg[ { \cal Q}^{\mu} \Big( {\frak D}_\nu {\frak D}^\nu
  {\cal Q}_\mu + 4 \ii \,g F_{\mu \rho } {\cal Q}^\rho + \big[ {\cal M}_I, [ {\cal M}^I ,
  {\cal Q}_\mu ] \big]  \Big) \bigg] \, ,
  \nonumber \\
  {\cal L}_{\Phi^2} & = & {\rm Tr} \bigg[ - \Phi_I {\frak D}_\nu {\frak D}^\nu \Phi^I -
  \Phi^I \Big[ {\cal M}_J, \big[ {\cal M}^J , \Phi_I \big]  \Big]  \bigg] \, ,
  \nonumber  \\
  {\cal L}_{\overline{\mathcal{C}} \mathcal{C}} & = & {\rm Tr} \bigg[ - 2 \,
  \overline{\mathcal{C}} {\frak D}_\mu {\frak D}^\mu \mathcal{C} - 2 \,
  \overline{\mathcal{C}} \Big[ {\cal M}_J, \big[ {\cal M}^J , \mathcal{C}
  \big] \Big] \bigg] \, ,
  \nonumber \\
  {\cal L}_{{\cal Q}^3} & = & - 2 \, \ii \, g \gamma {\rm Tr} \Big[ {\frak D}_\mu {\cal Q}^\mu
  {\cal Q}_\nu {\cal Q}^\nu \Big] - 2 \, \ii \, g {\rm Tr} \Big[ {\frak D}_\mu {\cal Q}_\nu
  \big[ {\cal Q}^\mu , {\cal Q}^\nu \big] \Big] \, ,
  \nonumber  \\
  {\cal L}_{{\cal Q} \Phi^2} & = & 2 \, \ii \, g \gamma {\rm Tr} \Big[ {\frak D}_\mu
  {\cal Q}^\mu \, \Phi_I  \Phi^I \Big] + 2 \, \ii \, g {\rm Tr} \Big[ {\frak D}_\mu \Phi_I
  \big[ {\cal Q}^\mu , \Phi^I \big] \Big] \, ,
  \nonumber  \\
  {\cal L}_{\overline{\mathcal{C}} \mathcal{C} {\cal Q} } & = & 2 \, \ii \, g
  {\rm Tr} \Big[ {\frak D}_\mu {\overline{\mathcal{C}} } \big[ {\cal Q}^\mu,
  \mathcal{C} \big] \Big] - 2 \, \ii \, \gamma \, g {\rm Tr} \Big[ \overline{\mathcal{C}} \,
  \big\{ {\cal Q}_\mu, {\frak D}^\mu \mathcal{C} \big\} \Big] \, ,
  \label{matrilagra} \\
  {\cal L}_{ \Phi {\cal Q}^2} & = & - 2 \, \gamma \, g {\rm Tr} \Big[
  \big[ {\cal M}_I , \Phi^I \big] {\cal Q}_\mu {\cal Q}^\mu \Big] - 2 \, g {\rm Tr} \Big[
  \big[ {\cal M}_I, {\cal Q}_\mu \big] \big[ \Phi^I, {\cal Q}^\mu \big] \Big] \, ,
  \nonumber  \\
  {\cal L}_{\Phi^3} & = & 2 \, \gamma \, g \Big[
  \big[ {\cal M}_I, \Phi^I \big] \Phi_J \Phi^J \Big] + 2 \, g {\rm Tr} \Big[
  \big[ {\cal M}_I, \Phi_J \big] \big[ \Phi^I, \Phi^J \big] \Big] \, ,
  \nonumber  \\
  {\cal L}_{\Phi \overline{\mathcal{C}} \mathcal{C}} & = & 2 \, g {\rm Tr} \Big[
  \big[{\cal M}_I , \overline{\mathcal{C}} \big] \big[ \Phi^I , \mathcal{C} \big] \Big] -
  2 \, \gamma \, g {\rm Tr} \Big[ \overline{\mathcal{C}} \, \big\{ \Phi^I , \big[ {\cal M}_I ,
  \mathcal{C} \big] \big\} \Big] \, ,
  \nonumber  \\
  {\cal L}_{{\cal Q}^4} & = & g^2 \Big( \eta_{\rho \mu} \eta_{\nu \sigma} -
  \left(1 - \gamma^2 \right) \eta_{\rho \nu} \eta_{\sigma \mu} \Big)
  {\rm Tr} \big[ {\cal Q}^\mu {\cal Q}^\nu {\cal Q}^\rho {\cal Q}^\sigma
  \big] \, ,
  \nonumber \\
  {\cal L}_{{\cal Q}^2 \Phi^2} & = & - 2 \, g^2 {\rm Tr} \big[ \Phi_I {\cal Q}^\mu
  \Phi^I {\cal Q}_\mu \big] + 2 \left(1 - \gamma^2 \right) g^2
  {\rm Tr} \big[ \Phi_I \Phi^I {\cal Q}^\mu {\cal Q}_\mu \big] \, ,
  \nonumber  \\
  {\cal L}_{\Phi^4} & = & g^2 {\rm Tr} \left[ \Phi_I \Phi_J \Phi^I \Phi^J \right] -
  \left(1 - \gamma^2 \right) g^2 {\rm Tr} \big[ \Phi_I \Phi^I \Phi_J \Phi^J
  \big] \, ,
  \nonumber  \\
  {\cal L}_{\overline{\cal C} {\cal C} {\cal Q}^2 } & = & 2 \, \gamma \, g^2
  {\rm Tr} \Big[ \overline{\cal C} \, \big\{ {\cal Q}_\mu ,
  [ {\cal Q}^\mu , {\cal C} ] \big\} \Big] \, ,
  \nonumber  \\
  {\cal L}_{\overline{\cal C} {\cal C} \Phi^2 } & =  & - 2 \, \gamma \, g^2 {\rm Tr}
  \Big[ \overline{\cal C} \, \big\{ \Phi_I , [ \Phi^I , {\cal C} ] \big\} \Big] \, .
  \nonumber
\eeqa
As is typical in cases of broken symmetry, the Lagrangian in \eq{matrilagra} displays a
variety of interactions, and is considerably more intricate than the combination of
Eqns.~(\ref{classicallagrangiantrace}), (\ref{gaugelagrangiantrace}) and (\ref{lgh}).
In order to compute Feynman diagrams, and to compare with the string theory calculation,
it is useful to write down an expression for the Lagrangian in terms of component fields
as well. In order to do so, we now assume that the matrices ${\cal A}_\mu$ and ${{\cal
M}}_{I}$ are all mutually commuting: we can then pick a basis of $\mathfrak{u}(N)$ in
which they are diagonal. In this basis, we write
\beq
  {\cal M}_I \, = \, {\rm diag} \left\{ m_I^A \right\} \, , \qquad
  {\cal A}_\mu \, = \, {\rm diag} \left\{ A_\mu^A \right\} \, , \qquad A = 1, \ldots, N  \, .
\label{diagmatr}
\eeq
Similarly, we write the quantum matrix fields as
\beqa
 \big[ {\cal Q}_\mu \big]^{A B}  & = & \frac{1}{\sqrt{2}} \, Q_\mu^{A B}  \, ,
 \qquad
 \big[ \Phi_I \big]^{A B} \, = \, \frac{1}{\sqrt{2}} \, \phi_I^{A B} \nonumber \\
 \Big[ \overline{\cal C} \Big]^{AB} & = & \frac{1}{\sqrt{2}} \, \overline{c}^{\, A B} \, ,
 \qquad \hspace{5mm}
 \Big[ {\cal C} \Big]^{A B} \, = \, \frac{1}{\sqrt{2}} \, c^{A B}  \, ,
\label{CCdef}
\eeqa
all satisfying $X^{AB} = (X^{BA})^*$, since $\mathfrak{u}(N)$ matrices are Hermitian; the
factors of $1/\sqrt{2}$ ensure that the matrix element fields are canonically normalized.
Notice that, thanks to diagonal form of the classical field ${\cal A}_\mu$, the covariant
derivative ${\frak D}_\mu$ does not mix matrix entries. Indeed, defining
\beq
  A_{\mu}^{AB} \, \equiv \, A_\mu^A - A_\mu^B  \, ,
\label{diffA}
\eeq
one can write
\beq
  \Big[ {\frak D}_{\mu} X \Big]^{A B} \, = \, \left( \partial_\mu + \ii \, g \, A_\mu^{A B}
  \right) X^{A B} \, ,
\label{covdermatr}
\eeq
where indices on the right-hand side are not summed. In particular, the covariant
derivative of diagonal entries reduces to the ordinary derivative. Motivated by this,
we can define a covariant derivative $D_\mu$ acting directly on matrix entries, as
opposed to $\mathfrak{u}(N)$ elements. Suppressing the $A,B$ indices on the derivative
symbol, we write
\beq
  D_\mu X^{A B} \, = \, \left( \partial_\mu + \ii \, g \, A_\mu^{AB} \right) X^{AB}
  \quad \longrightarrow \quad \Big[ {\frak D}_\mu X \Big]^{A B} \, = \,
  D_\mu  X ^{A B} \, .
\label{covderel}
\eeq
Note that $D_\mu$ is a derivation, obeying the Leibnitz rule
\beq
  D_\mu \left( X Y \right)^{A B}  \, = \, \left( D_\mu X^A_{\phantom{A} C} \right) \, Y^{C B} +
  X^A_{\phantom{A} B} \left( D_\mu Y^{C B} \right) \, ,
\label{leibni}
\eeq
so it can be partially integrated in any integrand with contracted color indices.
Treating the mass matrices in a similar way, we define
\beq
  m^{A B}_I \, \equiv \, m_I^A - m_I^B \, , \qquad
  m_{A B}^2 \, \equiv \, \sum_{I = 1}^{n_s} \left( m^{AB}_I \right)^2 \, .
\label{eq:masssum}
\eeq
This implies
\beq
  \Big[ {\cal M}_I , X \Big]^{A B} \, = \, m_I^{A B} \, X^{A B} \, ,  \quad
  \Big[ {\cal M}_I , \big[ {\cal M}^I , X \big] \Big]_{A B} \, = \, m_{A B}^2 \, X_{A B} \, ,
\label{masscomm}
\eeq
where again on the right-hand side the indices $A$ and $B$ are fixed and not summed.
As an example, the term quadratic in $\Phi$ in \eq{matrilagra} can be written in component
notation as
\beqa
\label{scalcomp}
  {\cal L}_{\Phi^2} & = & - \, \frac{1}{2} \phi_I^{A B} D_\mu D^{\mu} \phi^{B A, I} -
  \frac{1}{2} \phi_I^{AB} m_{ij}^2 \, \phi^{BA, I} \\
  & = & - \, \frac{1}{2} \, \sum_{A = 1}^N \phi_I^{A A} \, \partial_\mu \partial^\mu \phi^{A A, I} -
  \sum_{1 \leq A < B}^N \Big[ \left( \phi_I^{A B} \right)^* D_\mu D^{\mu} \phi^{A B, I}
  + m_{A B}^2 \left| \phi_I^{A B} \right|^2 \Big] \, , \nonumber
\eeqa
which is the correctly normalized quadratic part of the Lagrangian for $N$ massless real
scalar fields $\phi^{AA}$, and $\frac{1}{2} N (N - 1)$ complex scalars $\phi^{A B}$, $A < B$,
with mass $|m_{AB}|$. To give a second example, the gauge-fixing condition in component
notation reads
\beq
  D^\mu Q_\mu^{A B} + \ii \, \gamma \, g \,  Q_\mu^{A C} Q_C^{\mu, B} -
  \ii \, m_I^{A B} \phi_I^{A B} - \ii \, \gamma g \, \phi_{I, C}^A \phi_I^{C B} \, = \, 0 \,  ,
\label{componentgaugecon}
\eeq
where $C$ is summed over but there is no summation over $A$ or $B$. Note that after
dimensional reduction and spontaneous symmetry breaking the gauge fixing has
become more unconventional from the $d$-dimensional point of view, involving
scalar fields as well as gauge fields, and mass parameters.

We conclude this section by giving the explicit expression for the background field that we
will be working with. We choose it so that, for each $A$, the abelian field strength $F_{\mu
\nu}^A = \partial_\mu A^A_\nu - \partial_\nu A^A_\mu$ is a $U(1)$ magnetic field in the $\{x_1,
x_2\}$ plane. A possible choice, already employed in Ref.\cite{Magnea:2013lna}, is
\beq
  A_\mu^A (x) \, = \, x_1 \eta_{\mu 2} B^A  \quad \longrightarrow \quad
  F_{\mu \nu}^A \, = \, f_{\mu \nu} B^A \, ,
\label{backfield}
\eeq
where we defined the antisymmetric tensor
\beq
  f_{\mu \nu} \, = \, \eta_{\mu 1} \eta_{\nu 2} - \eta_{\nu 1} \eta_{\mu 2} \, .
\label{Amndef}
\eeq
We now turn to the evaluation of selected two-loop vacuum diagrams, contributing to the
effective action, which we can then compare with the corresponding expressions derived
from string theory. Preliminarily, we collect useful expressions for the relevant coordinate-space
propagators in the presence of the background field.


\subsection{Propagators in a constant background field}
\label{propagators}

The quantum field theory objects that we wish to compute, in order to compare with
string theory results, are two-loop vacuum diagrams contributing to the effective action,
and computed with our chosen background field, \eq{backfield}. At two loops, these
diagrams can be computed in a straightforward manner in coordinate space, directly
from the path integral definition of the generating functional,
\beqa
\label{eq:ZoPert}
  Z \Big[ J_\mu^{A B}, \eta^{A B}, \overline{\eta}^{A B}, J_I^{A B} \big] & = &
  \int \, \big[ D Q_\mu^{A B} \, D \overline{c}^{A B} \, D c^{A B} \, D \phi^{A B}_I \big] \\
  & & \hspace{1cm} \times
  \exp \bigg[ \ii \int \d^d x \, \Big( {\cal L} \big[ Q_\mu^{A B}, \overline{c}^{A B} , c^{A B} ,
  \phi^{A B}_I \big] \nonumber \\
  & & \hspace{2cm} + J_\mu^{A B} Q^\mu_{B A } + \overline{c}^{B A} \eta^{A B} +
  \overline{\eta}^{A B} c^{B A} + J_I^{A B} \phi_I^{B A} \Big) \bigg] \, , \nonumber
\eeqa
where $J_\mu$, $J_I$, $\eta$ and $\overline{\eta}$ are matrix sources for the fields
in the complete Lagrangian ${\cal L}$. The only non-trivial step is the computation
of the quantum field propagators in the presence of the background field: diagrams
are then simply constructed by differentiating the free generating functional with respect
to the external sources. For a background field of the form of \eq{backfield}, the solution
is well-known for the scalar propagator (see, for example, Ref.~\cite{Magnea:2004ai}):
we briefly describe it here, and discuss the generalization to vector fields.

For scalar fields, the propagator in the presence of the background in \eq{backfield}
can be expressed in terms of a heat kernel as
\beq
  G^{A B} (x, y) \, = \, \int_0^\infty \! \d t \, {\cal K}^{A B} \left(x, y; t \right) \, ,
\label{scalarGdef}
\eeq
where, defining $B_{A B} \equiv B_A - B_B$, one can write
\beqa
  {\cal K}^{A B} \left(x, y; t \right) & = & \frac{1}{(4 \pi t )^{d/2}} \,
  \ex{ - \frac{\ii}{2} g B^{A B} (x_1 + y_1)(x_2 - y_2) - t m_{A B}^2} \,
  \frac{g B^{A B} t}{\sinh (g B^{A B} t)} \nonumber \\
  & & \hspace{2cm} \times \, \exp \left[ \frac{1}{4 t} \, \left( x_\mu - y_\mu \right) \,
  \Sigma^{\mu \nu} \left(g B^{A B} t \right) \left( x_\nu - y_\nu \right) \right] \,  .
\label{heatkernel}
\eeqa
In \eq{heatkernel}, we have introduced the tensor
\beq
  \Sigma_{\mu \nu} \left( g B^{A B} t \right) \, =
  \frac{g B^{A B} t}{\tanh \left( g B^{AB} t \right)}  \, \eta_{\mu \nu}^\parallel + \,
  \eta_{\mu \nu}^\perp \, ,
\label{eq:betadef}
\eeq
where the projectors $\eta^{\mu \nu}_\parallel$ and $\eta^{\mu \nu}_\perp$ identify
components parallel and perpendicular to the background field, and are given by
\beq
  \eta^{\mu \nu}_\parallel \, = \, f^{\mu \rho} f_\rho{}^\nu \, = \,
  \delta_1{}^\mu \delta_1{}^\nu + \delta_2{}^\mu \delta_2{}^\nu \, , \qquad
  \eta^{\mu \nu}_\perp \, = \, \eta^{\mu \nu} - \eta^{\mu \nu}_\parallel \, .
\label{eq:etaperpar}
\eeq
The propagator $G^{A B} (x,y)$ in \eq{scalarGdef} satisfies
\beq
  \Big( D_\mu^{(x)} D^\mu_{(x)} + m_{A B}^2 \Big) G^{A B} (x,y) \, = \, - \, \ii \,
  \delta^d (x - y) \, ,
\label{eq:scalpropeq}
\eeq
where we noted explicitly the variable on which the derivatives act. In fact, covariant
derivatives act on a propagator with color indices $(AB)$ as
\beqa
  D_\mu^{(x)} G^{A B} (x, y) & \equiv & \left[ \frac{\partial}{\partial x^\mu} + \ii \, g
  A_\mu^{A B} (x) \right] G^{A B} (x, y) \, , \nonumber \\
  D_\mu^{(y)} G^{A B} (x, y) & \equiv & \left[ \frac{\partial}{\partial y^\mu}  - \ii \, g
  A_\mu^{A B} (y) \right] G^{A B} (x, y) \, .
\label{covderG}
\eeqa
For real scalar fields, or vanishing backgrounds, one recovers the well-known
expression for the scalar propagator as a Schwinger parameter integral,
\beq
  G_0^{\, A B} (x,y) \, \equiv \, \lim_{B^{A B} \to 0} G^{A B} (x, y) \, = \,
  \int_0^\infty d t \, \, \frac{\ex{ - t \, m_{AB}^2}}{(4 \pi t)^{d/2}} \,
  \exp \left[ \frac{(x - y)^2}{4 t} \right] \, .
\label{G0}
\eeq
Ghosts fields are scalars, and they share the same propagator. For gluons, on the
other hand, the background field strength $F_{\mu \rho}$ enters the kinetic term, given
in the first line of \eq{matrilagra}. The propagator must then satisfy
\beq
  \bigg[ \eta_{\mu \rho} \Big( D_\sigma^{(x)} D_{(x)}^\sigma + m_{A B}^2 \Big)
  + 2 \, \ii \, g \, F_{\mu \rho}^{A B} \bigg] G^{A B, \rho \nu} (x, y) \, = \,
  \ii \, \delta_\mu^\nu \delta^d (x - y) \, .
\label{glupropeq}
\eeq
To diagonalize this equation, one can introduce the projection operators
\beq
  (P_\pm)_{\rho \sigma} \, = \, \frac{ \eta^\parallel_{\rho \sigma}
  \pm f_{\rho \sigma} }{2} \, , \qquad (P_\perp)_{\rho \sigma}
  \, = \, \eta^\perp_{\rho \sigma} \, ,
\label{projvec}
\eeq
satisfying
\beq
  \left( P_+ + P_- + P_\perp \right)_{\mu \nu} \, = \, \eta_{\mu \nu} \, \qquad
  \left(P_+ - P_- \right)_{\mu \nu} \, = \, f_{\mu \nu} \, .
\label{projprop}
\eeq
It is then easy to show that the function
\beq
  G^{A B, \sigma \alpha} (x,y) \, = \, - \, \eta_\perp^{\sigma \alpha} \, G^{A B}(x,y)
  - P_+^{\sigma \alpha} \, G^{A B}_+(x,y) - P_-^{\sigma \alpha} \, G^{AB}_-(x,y)
\label{solvec}
\eeq
satisfies \eq{glupropeq}, provided the functions $G^{A B}_\pm (x,y)$ satisfy
\beq
  \Big[ D^{(x)}_\mu D_{(x)}^\mu + m_{A B}^2 \pm 2 \, \ii \, g B^{A B} \Big]
  G^{A B}_\pm (x,y) \, = \, - \ \ii \, \delta^d (x - y) \, .
\label{chargeq}
\eeq
\eq{chargeq} simply gives a scalar propagator with a mass shifted by the
appropriate background field. It's easy therefore to write the solution for the complete
gluon propagator explicitly as
\beq
\hspace{-2mm}
  G^{AB}_{\mu \nu} (x,y) = - \! \int_0^\infty d t \, \left[ \eta^\perp_{\mu \nu} +
  \eta^\parallel_{\mu \nu} \cosh \left( 2 g B^{A B} t \right) +
  f_{\mu \nu} \sinh \left( 2 g B^{A B} t \right) \right] {\cal K}^{A B} (x, y; t) \, .
\label{solvec2}
\eeq
Note that this can be written also in the more compact and elegant form
\beq
  G^{A B}_{\mu \nu} (x,y) \, = \, - \, \int_0^\infty \d t \,
  \left[ {\rm e}^{\, - \, g \, t \, F^{AB}_{\alpha \beta} \, S_{\bf 1}^{\alpha \beta}}
  \right]_{\mu \nu} \, {\cal K}^{A B}(x,y; t) \, ,
\label{gluonpropexp}
\eeq
where $S_{\bf 1}^{\alpha \beta}$ are the Lorentz generators in the spin one representation
appropriate for gauge bosons,
\beq
  \left[ S_{\bf 1}^{\, \alpha \beta} \right]_{\mu \nu} \, = \, - \, \ii \, \left(
  \delta^\alpha_{\phantom{\alpha} \mu} \delta^\beta_{\phantom{\beta} \nu} -
  \delta^\alpha_{\phantom{\alpha} \nu} \delta^\beta_{\phantom{\beta} \mu}
  \right) \, .
\label{onelorentz}
\eeq
The propagator in \eq{gluonpropexp} naturally generalizes to other representations of
the Lorentz group, simply changing the form of the generators. The spin one-half case,
where $S_{\bf 1/2}^{\alpha \beta} = \ii [\gamma^\alpha, \gamma^\beta]/4$, will be useful
for example when studying the gluino contribution to the effective action in the
supersymmetric case.


\subsection{Selected two-loop vacuum diagrams}
\label{seldiag}

We will now illustrate the structure of the field theory calculation of the effective
action by outlining the calculation of a selection of the relevant two-loop diagrams.
A complete list of the result for all 1PI diagrams depicted in Fig.~\ref{fig:1PIgraphs}
is given in Appendix C.

We begin by considering the ghost-gluon diagram given by the sum of \eq{fgng} and
\eq{fgmg}, which we denote by $H_b (B_{A B}, m_{A B})$. The relevant interaction vertex,
involving ghost, antighost and gluon fields, arises from the sixth line in \eq{matrilagra},
and may be written explicitly in component language using \eq{CCdef}. Upon integrating
by parts, it can be rewritten as
\beqa
  {\cal L}_{\overline{c} c Q} & = & \frac{\ii \, g}{\sqrt{2}} \bigg[
  \Big( \delta_{B C} \delta_{DE} \delta_{F A} - \delta_{B E} \delta_{F C}
  \delta_{DA} \Big) \, D_\mu \overline{c}^{A B} \, Q^{\mu, CD} c^{E F} \nonumber \\
  & & \hspace{50pt} - \, \gamma \Big( \delta_{B C} \delta_{DE} \delta_{F A} +
  \delta_{B E} \delta_{F C} \delta_{DA} \Big)  \, \overline{c}^{A B} \, Q^{\mu, C D   } \,
  D_\mu c^{E F} \bigg] \, .
\label{cbarcQ}
\eeqa
Sewing two copies of this vertex together to obtain the desired diagram, one
first of all observes that terms linear in the gauge parameter $\gamma$, which
involve double derivatives of scalar propagators, cancel out upon contracting color
indices. Next, one notices that some of the color contractions lead to a non-planar
configuration, which would correspond to an open string diagram with only one
boundary. We are not interested in these contributions, since the corresponding
diagram is built of propagators which are neutral with respect to the background
field, and does not contribute to the effective action. Furthermore, we do not expect
to obtain this diagram from our string configuration, since we start with a planar
worldsheet. Discarding non-planar contributions, one finds that the remaining
planar terms can be written as
\beqa
  H_b (B_{A B}, m_{A B}) & = & - \, g^2 \, \frac{1 + \gamma^2}{4}
  \int \d^d x \, \d^d y \,  \Big[ D_\mu^{(x)} G^{A B} (x, y) \, D_\nu^{(y)} G^{B C} (x,y) \,
  G^{\mu \nu, C A} (x,y) \, \nonumber \\
 & &\hspace{150pt} + \, (A B C) \leftrightarrow (C B A) \Big] \, .
\label{Hbin}
\eeqa
Inserting the expressions for the scalar and gluon propagators given by \eq{scalarGdef}
and \eq{gluonpropexp}, one can immediately rewrite \eq{Hbin} in terms of covariant
derivatives of the heat kernels ${\cal K}$. These, in turn, can be written as
\beqa
  D_\mu^{(x)} {\cal K}_{AB} (x, y; t) & = & \frac{\Sigma_{\mu \rho} (g B^{A B}  t) + \ii \, t \,
  F^{AB}_{\mu \rho}}{2 \, t} \, \left( x^\rho - y^\rho \right) \, {\cal K}_{AB}(x, y; t) \, , \\
  D_\nu^{(y)} {\cal K}_{BC} (x, y; t) & = & - \frac{\Sigma_{\nu \sigma} (g B^{BC}   t) - \ii \, t \,
  F^{BC}_{\nu \sigma}}{2\, t} \, \left( x^\sigma - y^\sigma \right) \, {\cal K}_{BC}(x, y; t) \, ,
  \nonumber
\eeqa
where $\Sigma(g B t)$ is defined in \eq{eq:betadef}. The integrand in \eq{Hbin} is then
proportional to the product of three heat kernels, which we write as
\beq
  \prod_{i = 1}^3  {\cal K}_i (x, y ; t_i) \, = \, \exp \left[ {\frac{1}{4} \left( x^\mu - y^\mu \right)
  \overline{\Sigma}_{\mu \nu} \left( x^\nu - y^\nu \right)} \right] \prod_{i = 1}^3 \,
  \frac{\ex{ - t_i m_i^2}}{(4 \pi t_i)^{\frac{d}{2}}} \, \frac{g B_i t_i}{\sinh(g B_i t_i)} \, .
\label{Kproduct}
\eeq
In \eq{Kproduct} we have simplified the notation by using a single index $i = 1,2,3$ in
place of the pairs of color indices $(AB),(BC),(CA)$, respectively. Furthermore, we have
defined $\overline{\Sigma}_{\mu \nu} = \sum_{i = 1}^3 \Sigma_{\mu \nu} ( g B^i t_i)/t_i$,
and we have taken advantage of the fact that the complex phases in each ${\cal K}_{i}(x,y;t_i)$
cancel due to the fact that $\sum_{i = 1}^3 B_i = 0$. At this point one sees that the integrand
in \eq{Hbin} is translationally invariant, depending only on the combination $z = x - y$.
We can then, for example, replace the integral over $x$ with an integral over $z$ while
the integral over $y$ gives a factor of the volume of spacetime, which we will not write
explicitly. One needs finally to evaluate the gaussian integral
\beq
  \int d^d z \, z^\rho z^\sigma \exp \left[ \frac{1}{4} \, z^\mu \, \overline{\Sigma}_{\mu \nu}
  \, z^\nu \right] \, = \, - 2 \, \left( \overline{\Sigma}^{-1} \right)^{\rho \sigma}
  \int d^d z \, \exp \left[ \frac{1}{4} \, z^\mu \, \overline{\Sigma}_{\mu \nu} \, z^\nu \right] \, .
\label{gaussint}
\eeq
Note that taking the inverse of $\overline{\Sigma}_{\mu \nu}$ is trivial because it is a diagonal
matrix, which can be written as
\beq
  \overline{\Sigma}_{\mu \nu} \, =  \, \Delta_0 \, \eta_{\mu \nu}^\perp \prod_{i = 1}^3
  \frac{1}{ t_i}  + \Delta_B \, \eta_{\mu \nu}^\parallel \prod_{i = 1}^3 \,
  \frac{g B_i }{\sinh(g B_i t_i)} \, ,
\label{BtensorDeltas}
\eeq
where $\Delta_0$ and $\Delta_B$ were defined in \eq{eq:DelZero} and \eq{eq:DelF}
respectively, while $\eta^\perp_{\mu \nu}$ and $\eta^{\parallel}_{\mu \nu}$ are given
in \eq{eq:etaperpar}. One finds then
\beq
  \int \d^d z \, \exp \left[ \frac{1}{4} \, z^\mu \, \overline{\Sigma}_{\mu \nu} \, z^\nu \right]
  \, = \, - \, \ii \, (4 \pi )^{d/2} \, \Delta_0^{1 - d/2} \, \Delta_B^{-1} \,
  \prod_{i = 1}^3  \, \frac{\sinh(g B_i t_i)}{g B_i t_i} \, \, t_i^{d/2} \, .
\label{gaussianintegral}
\eeq
Putting together all these ingredients, one may evaluate \eq{Hbin}. Using the symmetry
of the Schwinger parameter integrand under the exchange $t_1 \leftrightarrow t_2$ one
can write
\beqa
  H_b (B_i, m_i) & = & - \, \ii \, \frac{g^2}{(4 \pi )^d} \, \frac{1 + \gamma^2}{2} \int_0^\infty
  \prod_{i = 1}^3 d t_i  \,  \frac{ \ex{ - t_1 m_1^2 - t_2 m_2^2 - t_3 m_3^2}}{\Delta_0^{d/2 - 1}
  \Delta_B} \, \bigg[ \frac{d - 2}{\Delta_0} t_3
  \label{eq:glghfinal}  \\
  & & \hspace{2cm} + \, \frac{2}{\Delta_B} \frac{\sinh(g B_3 t_3)}{g B_3} \,
  \cosh(2 g B_3 t_3 - g B_1 t_1 - g B_2 t_2) \bigg] \, ,
  \nonumber
\eeqa
which can be directly matched to the string theory result.

We conclude this section by briefly describing the calculation of two further Feynman
diagrams, which arise in our theory because of the pattern of symmetry breaking and
dimensional reduction. First of all, there are diagrams, like \eq{fmsm}, with the
same topology as \eq{fgmg}, but with an odd number of scalar propagators, and
vertices proportional to the scalar vacuum expectation values $m_{ij}$, characteristic
of the broken symmetry phase. The relevant vertex can be found by expanding
${\cal L}_{\Phi {\cal Q}^2}$ from \eq{matrilagra} in terms of the component fields
defined in the \eq{CCdef}. Upon relabeling the indices and using $m_I^{CB} + m_I^{BA}
+ m_I^{AC}  = 0$, it reads
\beq
  {\cal L}_{Q^2 \phi} \, = \, \frac{g}{\sqrt{2}} \big[ (1 + \gamma) m_I^{CB} -
  (1 - \gamma) m_I^{AC} \big] \, \phi^{AB} \, Q_\mu^{BC} \, Q^{\mu , CA} \, .
\label{QQphitext}
\eeq
Labeling this diagram as $H_d (B_{A B}, m_{A B})$, we find for it the coordinate space
expression
\beqa
  H_d (B_{A B}, m_{A B}) & = & \frac{g^2}{4} \big[ (1 - \gamma) m_I^{CB} - (1 + \gamma)
  m_I^{AC} \big] \big[ (1 - \gamma) m_I^{AC} - (1 + \gamma) m_I^{CB} \big]  \nonumber \\
  & & \hspace{1cm} \times \int d^d x \, d^d y \,  G^{AB} (x,y) \, G_{\mu \nu}^{BC} (x,y) \,
  G^{\mu \nu, CA} (x,y) \, ,
\eeqa
where we neglected non-planar contributions, and we used $m_I^{BC}  = - m_I^{CB}$.
Manipulations similar to those leading to \eq{eq:glghfinal}, simplified by the absence of
derivative interactions, yield the result
\beqa
  H_d (B_i, m_i) & = & - \frac{\ii}{(4 \pi)^d} \, \frac{g^2}{2} \, \big[
  (1 + \gamma^2) \, m_3^2 - 2 \, (m_1^2 + m_2^2 ) \big]
  \label{finmassapp} \\
  & & \hspace{2mm} \times \int_0^\infty \prod_{i = 1}^3 \d t_i \, \,
  \frac{ \ex{- t_1 m_1^2 - t_2 m_2^2 - t_3 m_3^2} }{\Delta_0^{{d/2} - 1} \Delta_B} \,
  \Big[  d - 2 + 2 \cosh(2 g B_1 t_1 - 2 g B_2 t_2) \Big]
  \nonumber \, ,
\eeqa
where we relabeled double indices as was done for \eq{eq:glghfinal}.

Finally, we briefly consider a diagram with a quartic vertex: the figure-of-eight scalar
self-interaction shown in \eq{fss}, which we label $E_i (B_k, m_k)$. The relevant
interaction term in the Lagrangian comes from ${\cal L}_{\Phi^4}$ in \eq{matrilagra}
and can be written as
\beq
  {\cal L}_{\phi^4} \, = \, \frac{g^2}{4} \, \left[ \delta_{KI} \delta_{LJ} - (1 - \gamma^2)
  \delta_{IL} \delta_{JK} \right]  \phi^{AB}_K \phi^{BC}_L \phi^{CD   }_I \phi^{D    A}_J \, ,
\label{phifour}
\eeq
which immediately gives
\beqa
\label{hadforgotten}
  E_i (B_k, m_k) & = & \ii \, \frac{g^2}{4} \, \left[ \delta_{KI} \delta_{LJ} - (1 - \gamma^2)
  \delta_{IL}\delta_{JK} \right] \int d^d x \,  \big[ G^{DA } (x,x) G^{BC} (x,x) \, \delta_{AC}
  \delta^{IJ} \delta^{LK} \nonumber \\
  & & \hspace{2cm} + \, G^{C D   }(x,x) G^{D    A} (x,x) \delta_{D    B}
  \delta^{LI} \delta^{JK} \big] \, .
\eeqa
Contracting flavor indices we get, as expected, the product of two one-loop integrals,
\beq
  E_i (B_k, m_k) \, =  \, \ii \, \frac{g^2}{\, (4 \pi )^d} \left[ 1 - \frac{1 -
  \gamma^2}{2} (1 + n_s) \right] n_s
  \int_0^\infty \prod_{i = 1}^2 \left[ \frac{d t_i}{t_i^{d/2 - 1}} \,
  \frac{g \, B_i \, \ex{- t_i m_{i}^2}}{\sinh(g B_i t_i)} \right] \, .
\eeq
The diagrams in Eqs.~\ref{fmn} -- \ref{fns} can be calculated similarly. One easily sees that all
these results, and those for the remaining diagrams, given in Appendix C, are directly
comparable with the ones obtained from the field theory limit of the string effective action.


\section{Discussion of results}
\label{Disc}


\subsection{Comparison between QFT and string theory}
\label{cfQFT}

We have now assembled all the results that we need to establish and verify a precise
mapping between the degeneration limits of the string world-sheet and the 1PI Feynman
diagram topologies in the field theory limit. Furthermore, as announced, we can trace
the contributions of individual string states propagating in each degenerate surface,
and these can be unambiguously mapped to space-time states propagating in the
field theory diagrams. This diagram-by-diagram mapping allows us in particular to
confirm that covariantly quantized superstring theory\footnote{On the other hand,
it has been shown in Refs.~\cite{Goddard:1973qh,Thorn:2002fj} that light-cone
quantization of string theory results in quantum field theory amplitudes computed in
a light-cone gauge.} naturally selects a specific gauge in the field theory limit, and  the
gauge condition is given here in \eq{YMgaugecon}.

More precisely, our string theory results for the symmetric degeneration are given
in \eq{eq:partialA} and \eq{eq:Fnnn}. A careful inspection shows that these results
reproduce all the Feynman diagrams in the first two lines of Fig.~\ref{fig:1PIgraphs},
which are listed in Eqs.~(\ref{fnnn}) through~(\ref{fsss}) in Appendix C, with the choice
$\gamma^2 = 1$. Similarly, our string theory results for the incomplete degeneration
are given in \eq{eq:dmSepfinal} and \eq{eq:F01}, and one may verify that one recovers
all Feynman diagrams in the last two lines of Fig.~\ref{fig:1PIgraphs}, given in Eqs.~(\ref{fmm})
through~(\ref{fss}). It is easy to identify each term in \eq{eq:partialA} with a particular
Feynman diagram: for example, the term ${\bf f}_{\perp}^{111}$ in \eq{eq:Fnnn} matches
the diagram resulting in \eq{fnnn}, in which all three lines correspond to gluons that
are polarized in directions perpendicular to the external magnetic field; on the other
hand, the term ${\bf f}_{\parallel}^{001} {\bf f}_{\perp}^{110}$, plus its cyclic permutations,
matches the result of \eq{fnmn}, in which one line carries a gluon polarized in the
plane parallel to the magnetic field, while the other two gluons are polarized in the
directions perpendicular to the magnetic fields. One may easily continue through
the list, identifying the other available combinations of gluons, ghosts and scalars.
In a similar vein, one can associate individual Feynman graphs to each term in
\eq{eq:dmSepfinal}: for example, ${\bf f}_{\parallel}^{11}$ corresponds to \eq{fmm},
${\bf f}_{\perp}^{11}$ to \eq{fnn}, while ${\bf f}_{\text{scal}}^{11}$  gives the diagram
with two scalar propagators given in \eq{fss}. Note also that diagrams with a quartic
vertex where the two propagators come from different sectors, such as for example
Eqs.~(\ref{fmn}), (\ref{fms}) and (\ref{fns}), all vanish for $\gamma^2 = 1$, so it comes
as no surprise that no contribution of this kind arises on the string theory side.

We finally note that we can also characterize the contributions to all 1PI Feynman diagrams
according to the Schottky multiplier they originate from. As an example, consider the infinite
product over the Schottky group which arises from the determinant of the non-zero modes
of the Laplacian, appearing in \eq{eq:gluonsf}. Tracing the gluon contributions to different
Feynman diagrams back to that product, one may verify that all terms appearing in 1PI
diagrams originate from at most one value of the index $\alpha$ in the product. More
precisely, gluon contributions in \eq{eq:partialA} come from ${\bf T}_\alpha = \{ {\bf S}_1,
{\bf S}_2, {\bf S}_1^{-1}{\bf S}_2 \}$, and if, say, a factor of $k_1^{1/2} = \sqrt{p_1} \sqrt{p_3}$
comes from the infinite product, then the necessary factor of $\sqrt{p_2}$ must come from
elsewhere in the amplitude. On the other hand, all terms in \eq{eq:dmSepfinal} come from
${\bf T}_\alpha = {\bf S}_1 {\bf S}_2$. This is not surprising from a world-sheet point of
view: in fact, one may recall from \Fig{homology} that ${\bf S}_1 {\bf S}_2$ corresponds
to a homology cycle which passes around both handles with a self-intersection between
them, and furthermore this is the only Schottky group element with this property which
survives in the field theory limit. We see that this homological property is directly related
to the graphical structure of the resulting Feynman diagram.


\subsection{Comparison with bosonic string theory}
\label{cfBos}

It is interesting to compare our results with the field theory limit of the bosonic string
effective action which was studied in Refs.~\cite{Russo:2003yk,Magnea:2004ai,
Russo:2007tc}. This comparison was discussed also in  \cite{Magnea:2013lna}, but
we are now in a position to make a more detailed analysis.

Bosonic strings are clearly a simpler framework, since the world-sheet is an ordinary
two-dimensional manifold, and not a super-manifold: one can then use the techniques
applying to ordinary Riemann surfaces, and specifically the (purely bosonic) Schottky
parametrization discussed in detail for example in Ref.~\cite{Magnea:2013lna}. At two
loops, one can use the ${\rm SL}(2,\mathbf{R})$ invariance of the amplitude to choose
the fixed points of the two Schottky group generators as
\beq
  \eta_1 \, = \, 0 \, , \qquad  \xi_1 \, = \, \infty \, , \qquad
  \eta_2 \, \equiv \, u \, , \qquad  \xi_2 \, = \, 1 \, .
\label{bosfix}
\eeq
The two-loop partition function can then be written as
\beq
  Z_2 \big( \veps, \vec{d} \,\, \big) \, = \, \int  \frac{d k_1 \, d k_2 \, d
     u}{k_1^2 \, k_2^2 \, (1 - u)^2} \, F_{\rm gh} (\mu) \, F_{\parallel}^{(\veps \, )}
     (\mu) \, { F}_{\perp} (\mu) \, F_{\rm scal}^{(\vec{d} \,\, )} (\mu) \, ,
\label{aritwomeas}
\eeq
where $\mu$ denotes the set of bosonic moduli, $\mu = \{k_1, k_2, u\}$, and
one may compare with the corresponding two-loop superstring expression, given
in \eq{eq:mFull}. Note that the integration variable $u$ is equal to the gauge-fixed
value of a projective-invariant cross ratio of fixed points, $u = \frac{\eta_1 - \eta_2}{\eta_1
- \xi_2}\frac{\xi_1 - \xi_2}{\xi_1 - \eta_2}$.  The various factors in \eq{aritwomeas}, already
discussed in \cite{Magnea:2004ai,Magnea:2013lna}, are given by
\beqa
\label{bosfactors}
  F_{\rm gh} (k_i, u)  & = & (1 - k_1)^2 \, (1 - k_2)^2 \,\,
  {\prod_\alpha}' \prod_{n = 2}^\infty \big( 1 - k_\alpha^n \big)^2 \, , \nonumber  \\
  F_{\parallel}^{(\veps \, )} (k_i, u) & = &  {\rm e}^{ - {\rm i} \pi \vec{\e} \cdot \tau
  \cdot \vec{\e} } \, \Big[ \det \left({\rm Im} \, \tau_{\vec{\epsilon}} \right) \Big]^{- 1} \,
  {\prod_\alpha}' \prod_{n = 1}^\infty \big( 1 - \ex{\, 2 \pi \ii
  \vec{\epsilon} \cdot \tau \cdot \vec{N}_\alpha} \, k_\alpha^n \big)^{-1} \,
  \big(1 - \ex{- 2 \pi \ii \vec{\epsilon} \cdot \tau \cdot \vec{N}_\alpha} \, k_\alpha^n \big)^{-1} \! ,
  \nonumber \\
  F_{\perp} (k_i, u) & = &  \Big[ \det \left({\rm Im} \, \tau
  \right) \Big]^{- (d - 2)/2} \, {\prod_\alpha}' \prod_{n = 1}^\infty
  \big( 1 - k_\alpha^n \big)^{- d + 2} \, , \\
  F_{\rm scal}^{(\vec{d} \,\, )} (k_i, u) & = & \prod_{I = 1}^{n_s} \, \ex{ \, \vec{d}_I
  \cdot \tau \cdot \vec{d}_I/( 2 \pi {\rm i} \alpha' ) } \, \, {\prod_\alpha}'
  \prod_{n = 1}^\infty \big(1 - k_\alpha^n \big)^{- n_s} \, . \nonumber
 \eeqa
Here $\tau$ is the period matrix of the Riemann surface, whose expression in the Schottky parametrization can be found, for instance, in Eq.~(A.14) of~\cite{DiVecchia:1988cy}.
Similarly, ${\tau}_{\vec{\epsilon}}$ is the twisted period matrix, the bosonic equivalent of
$\bs{\tau}_{\vec{\epsilon}}\,$, computed here in Appendix \ref{Appb}.

The most obvious difference between the measures in \eq{eq:mFull} and \eq{aritwomeas}
is the occurrence of half-integer powers of the multipliers in the former. In the bosonic string,
the mass level of states propagating in the $i$-th loop increases with the power of $k_i$,
whereas in the superstring it increases with the power of $k_i^{1/2}$. Necessarily, the
propagation of a massless state must correspond to terms of the form $d k_i/k_i = d \log
k_i$ in the integrand, so tachyons propagating in loops correspond to terms of the form
$\d k_i / k_i^2$ in the bosonic theory and $d k_i/ k_i^{3/2}$ in the superstring, as seen
explicitly in \eq{aritwomeas} and in \eq{eq:mFull}, respectively. These tachyonic states
must be removed by hand in the bosonic theory, whereas they are automatically
eliminated from the spectrum of the superstring upon integrating over the odd moduli
and carrying out the GSO projection.

The identification of the symmetric degeneration proceeds in the same way for the two
theories: in particular, the symmetry of \Fig{fig:stringapple} leads to the choice of the
parameters $p_i$, defined by \eq{eq:kToP}. The cross-ratio $u$ can then be written
as
\beq
  u \, = \, \frac{(1 + p_1)(1 + p_2) \, p_3}{(1 + p_3) (1 + p_1 p_ 2 p_ 3)} \, ,
\label{etapi}
\eeq
and the integration measure takes the symmetric form
\beq
  \frac{d k_1}{k_1^2} \, \frac{d k_2}{k_2^2} \,
  \frac{d u}{(1 - u)^2} (1 - k_1)^2 (1 - k_2)^2 \, = \,
  \frac{d p_1}{p_1^2} \, \frac{d p_2}{p_2^2} \, \frac{d p_3}{p_3^2} \,
  (1 - p_2 p_3)(1 - p_1 p_3)(1 - p_1 p_2) \, .
\label{eq:ZkZp}
\eeq
It is interesting to note that in the field theory limit a number of contributions arise in
slightly different ways in the two approaches. As an example, let us consider the
twisted determinant of the period matrix for the bosonic string. To lowest order in
$k_i$, it is given by a combination of hypergeometric functions with argument $u$,
in a manner similar to what happens for its supersymmetric counterpart. In the neighborhood
of the symmetric degeneration, the hypergeometric functions can be expanded
in powers of $p_3$, and the bosonic string determinant reduces to
\beqa
\label{eq:bostep3}
  \det \left({\rm Im} \, \tau_{\vec{\epsilon}} \right) & = & \frac{1}{4 \pi^2 (\alpha')^2}
  \Bigg[ \Delta_B - 2 \alpha' p_3 \cosh \Big( g \, \big( 2 B_3 t_3 -  B_1 t_1 - B_2 t_2 \big)
  \Big) \, \frac{\sinh \big( g B_3 t_3 \big)}{g B_3} \nonumber \\
  & & \hspace{2cm} + \, {\cal O} \Big( p_1, p_2, p_3^2; (\alpha')^2 \Big) \Bigg]  \, ,
\eeqa
where $\Delta_B$ is defined in \eq{eq:DelF}.  We note that the term proportional to
$p_3$ in \eq{eq:bostep3} receives a contribution from the series expansion of the
hypergeometric functions, and contributes to Feynman diagrams with a gluon polarized
parallel to the magnetic field propagating in the leg parametrized by $t_3$.

For the superstring, the situation changes: one needs to keep terms only up to order
$q_i$, which implies that all the hypergeometric functions appearing in the
expression for the supersymmetric twisted determinant can be replaced by unity.
Since the first-order term in the expansion of the hypergeometric functions is crucial
in order to get the correct coefficient of $p_3$ in \eq{eq:bostep3}, and in turn to match
the field theory diagrams, it is necessary that terms proportional to $q_3$ arise
from the nilpotent contributions to $\det \left({\rm Im} \, \bs{\tau}_{\vec{\epsilon}} \right)$.
This is indeed what happens: expanding the supersymmetric twisted determinant in
powers of $q_i$ one finds
\beqa
\label{eq:fermtep}
  \det \left({\rm Im} \, \bs{\tau}_{\vec{\epsilon}} \right) & = & \frac{1}{4 \pi^2 (\alpha')^2}
  \Bigg[ \Delta_B - 2 \alpha' q_3 \, \hat{\theta}_{12} \hat{\phi}_{12} \,
  \cosh \Big( g \big( 2 B_3 t_3 -  B_1 t_1 - B_2 t_2 \big) \Big) \frac{\sinh \big( g B_3 t_3
  \big)}{g B_3} \nonumber \\
  & & \hspace{2cm} + \, {\cal O} \Big( q_1,q_2, q_3^2; (\alpha')^2 \Big) \Bigg] \, .
\eeqa
To be precise, we note that terms proportional to $p_3$ and $q_3 \, \hat{\theta}_{12}
\hat{\phi}_{12}$ in $\det \left({\rm Im} \, {\tau}_{\vec{\epsilon}} \right)$ and $\det \left({\rm Im}
\, \bs{\tau}_{\vec{\epsilon}} \right)$, respectively, also receive contributions from sources
other than the ones we have discussed, specifically from factors of the form $u^{{n_i
\epsilon_i/2}}$, with $n_i$ integers. It is easy to see, however, that these contribute in
the same way in the two cases, since in the bosonic case we have
\beq
  u^{{n_i \epsilon_i/2}} \, =  \, p_3^{\, {n_i \epsilon_i/2}} \Big(1 + \frac{n_i \epsilon_i}{2}
  \, p_3 \Big) + {\cal O} \left( p_1, p_2,p_3^2 \right) \, ,
\label{expeta1}
\eeq
while in the superstring case we get
\beq
  u^{{n_i \epsilon_i/2}} \, = \, p_3^{\, {n_i \epsilon_i/2}} \Big(1 + \frac{n_i \epsilon_i}{2} \,
  q_3 \, \hat{\theta}_{12} \hat{\phi}_{12} \Big) + {\cal O} \left( q_1 , q_2,q_3^2
  \right) \, .
\label{expeta2}
\eeq
As a consequence, and as required, when all of the other factors are inserted, the
coefficient of $p_1 p_2 p_3$ in the bosonic string measure is the same as the coefficient
of $q_1 q_2 q_3 \, \hat{\theta}_{12} \hat{\phi}_{12}$ in the superstring measure, and the
same field theory amplitude is obtained for the massless sectors of the bosonic and
supersymmetric theories.

The terms computed in section \ref{NSeighthandle}, which correspond to field theory
diagrams with the topology of the diagrams in the bottom two rows of \Fig{fig:1PIgraphs} as well as 1PR graphs, also appear
in the bosonic theory. In fact, one gets once more an expression of the form of
\eq{eq:dmSepField} in the field theory limit, but the integrals $I_1$, $I_2$ and $I_3$
get replaced by
\beq
  \tilde{I}_1 \, = \, \int_0^1 \frac{\d u}{(1 - u)^2} \, , \qquad
  \tilde{I}_2 \, = \, \int_0^1 \d u \, , \qquad \tilde{I}_3 \, = \, \int_0^1 \frac{\d u}{u^2} \, .
\label{eq:I1I2bos}
\eeq
When using the bosonic string, the integral $\tilde{I}_3$ has to be discarded by hand,
either by arguing that it corresponds to tachyon propagation, or by explicitly matching to
the field theory result. In the case of the superstring, on the other hand, the correct result
emerges automatically, provided a consistent integration procedure in super-moduli space
is followed. The complete answer for the four-point vertex diagrams emerges in both cases
from the terms proportional to $I_2 = \tilde{I}_2 = 1$. As in the superstring case, the contribution $\tilde{I}_1$ is related to the sepating degeneration.


\section*{Acknowledgements}

We would like to thank Paolo Di Vecchia for discussions. The work of LM was partially funded by MIUR (Italy),
under contract 2010YJ2NYW$\_$006, and by the University of Torino and the
Compagnia di San Paolo under contract ORTO11TPXK. The work of SP was partially supported by an STFC studentship and partially supported by
the Compagnia di San Paolo contract “MAST: Modern Applications of String Theory”
TO-Call3-2012-0088.
The work of RR was partially supported by the Science and
Technology Facilities Council (STFC) Consolidated Grant ST/L000415/1 {\em
``String theory, gauge theory \& duality''}.
We made use of the Mathematica package \verb=grassmann.m= \cite{grassmann} in our computations.

\appendix



\section{Appendix A}
\label{Appa}

In this appendix we discuss the super-Schottky parametrization of super moduli space
in the Neveu-Schwarz sector, and we compute some relevant geometric quantities
in this parametrization. Bosonic Schottky groups are discussed, for example, in Section 2
of Ref.~\cite{Magnea:2013lna}: here we focus only on the supersymmetric case.


\subsection{Super-projective transformations}
\label{superproj}

Super-projective transformations are automorphisms of the super-Riemann sphere
${\bf CP}^{1|1}$, which is defined in terms of homogeneous coordinates in ${\bf C}^{2|1}$
by the equivalence relation $(z_1,z_2 | \theta) \sim (\lambda z_1, \lambda z_2| \lambda
\theta)$ for non-zero complex $\lambda$, where the bosonic coordinates $z_1$ and
$z_2$ are not allowed to vanish simultaneously. To fix the superconformal structure,
we introduce a skew-symmetric quadratic form, using a bra-ket notation $\langle {\bf u}
| {\bf v} \rangle$ defined by\footnote{Notice that $\langle {\bf u} | {\bf v} \rangle = -
\langle {\bf u } , {\bf v} \rangle$ where $ \langle {\bf u } , {\bf v} \rangle$ is the
skew-symmetric quadratic form introduced in Eq.~(5.54) of Ref.~\cite{Witten:2012ga}.}
\beq
  \langle {\bf u} | \, = \, (u_2, - u_1, \theta) \, , \qquad \qquad
  | {\bf u} \rangle \, = \, (u_1 , u_2 | \theta )\tran \, ,
\label{braketdef}
\eeq
satisfying $\langle {\bf u} | {\bf v} \rangle = - \langle {\bf v} | {\bf u} \rangle$. This bracket
is related to the super-difference between two points, ${\bf z} - {\bf w}$, defined in
\eq{superdiff}, as follows: if $\ket{\bf z} = (\lambda_1 z, \lambda_1, \lambda_1 \psi)\tran$,
and $\ket{\bf w} = (\lambda_2 w, \lambda_2, \lambda_2 \theta)\tran$, for $\lambda_1,\,
\lambda_2 \neq 0$, then
\beq
  \langle {\bf w} | {\bf z} \rangle \, = \, \lambda_1 \lambda_2 ({\bf z} - {\bf w}) \, = \,
  \lambda_1 \lambda_2 ( z - w - \psi \theta) \, .
\label{braktodiff}
\eeq
The group of transformations which preserves the skew-symmetric quadratic form
is ${\rm OSp}(1|2)$, which can be realised by $\text{GL}(2|1)$ matrices of the form
\beq
  {\bf S } \, = \, \left( \begin{array}{cc|c} a & b & \alpha \\ c & d & \beta  \\
  \hline \gamma & \delta & e \end{array} \right) \, ,
\label{GLmatrix}
\eeq
where the five even and four odd variables are subject to the two odd and two even
constraints,
\beq
  \left( \begin{array}{c} \alpha \\ \beta \end{array} \right) \, = \,
  \left( \begin{array}{cc} a & b \\ c & d \end{array} \right)
  \left( \begin{array}{c} - \delta \\ \gamma \end{array} \right) \, ,  \qquad
  a d - b c - \alpha \beta \, = \, 1 \, , \qquad  e \, = 1 - \alpha \beta \, ,
\label{OSpConstraints}
\eeq
so that the group has dimension $3|2$.

We can define a map from homogeneous coordinates to superconformal coordinates by
\beq
  \mathfrak{f}  : \{ (z_1 , z_2 | \theta)\tran \, | \, z_2 \neq 0 \} \, \to \, \mathbf{C}^{1|1} \, ,
  \qquad (z_1 , z_2 | \theta)\tran \, \mapsto \, \mathfrak{f} \big( (z_1 , z_2 |
  \theta)\tran \big) \, \equiv \, \left( \frac{z_1}{z_2} \bigg| \frac{\theta}{z_2} \right) \, .
\label{rscdef}
\eeq
Then, any other map of the form $\mathfrak{f} \circ {\bf S}$, with ${\bf S}$ an ${\rm
OSp}(1|2)$ matrix, also defines superconformal coordinates. Recall~\cite{Friedan:1986rx}
that two ${\bf C}^{1|1}$ charts $(z|\theta)$ and $(\hat{z} | \hat{\theta})$ belong to the
same superconformal class whenever $D_\theta \hat{z} = \hat{\theta} \, D_\theta
\hat{\theta}$, where
\beq
  D_\theta \, \equiv \, \partial_\theta + \theta \partial_z \, ,
\label{superD}
\eeq
is the super derivative which satisfies $D_\theta^2 = \partial_z$. In particular, we can
cover ${\bf CP}^{1|1}$ with two superconformal charts ${\bf z}_1 = \mathfrak{f} \big(
(z_1, z_2 | \theta)\tran \big)$ and ${\bf z}_2 = (\mathfrak{f} \circ {\bf I}) \big( (z_1, z_2 |
\theta)\tran \big)$, where ${\bf I}$ is the ${\rm OSp}(1|2)$ matrix
\beq
  {\bf I} \, = \, \left( \begin{array}{cc|c} 0 &  -1 & 0 \\ 1 & 0 & 0 \\ \hline 0 & 0 & 1 \end{array}
  \right) \, .
\label{Idef}
\eeq
In general, one can find an ${\rm OSp}(1|2)$ matrix taking two given points $|{\bf u} \rangle
= (u_1, u_2 | \theta)\tran$ and $|{\bf v}\rangle = (v_1, v_2 | \phi)\tran$ to $|\bs{0}
\rangle \equiv (0, 1 | 0)\tran$ and $|\bs{\infty} \rangle \equiv (1, 0 | 0)\tran \sim
{\bf I} | \bs{0}\rangle$ respectively; one such matrix is
\beq
  {\bf \Gamma}_{\bf uv} \, = \, \frac{1}{\sqrt{\langle {\bf v} | {\bf u} \rangle }}
  \left( \begin{array}{cc|c}
  u_2  & - u_1 & \theta \\ \phantom{\Big|} v_2 & - v_1 & \phi \\ \hline
  \frac{\phantom{\big|} u_2 \phi - v_2 \theta}{\sqrt{\langle {\bf v} | {\bf u} \rangle }} &
  \frac{\phantom{\big|} v_1 \theta - u_1 \phi}{\sqrt{\langle {\bf v} | {\bf u} \rangle }} &
  \sqrt{\langle {\bf v} | {\bf u} \rangle } - \frac{\phantom{\big|} \theta \phi}{\sqrt{\langle
  {\bf v} | {\bf u} \rangle }}
  \end{array} \right) \, .
\label{Gammauvdef}
\eeq
We can further stipulate that a point $|{\bf w}\rangle = (w_1, w_2 | \omega)\tran$ be mapped
to a point equivalent to $(1,1 | \Theta_{\bf u w v } )\tran$, where now there is no freedom
in choosing the fermionic co-ordinate, which is therefore a super-projective invariant
built out of the triple $\left\{| {\bf u} \rangle, | {\bf v} \rangle, | {\bf w} \rangle \right\}$.
The image of $| {\bf w} \rangle$ under $\bf \Gamma_{uv}$ is then
\beq
  {\bf \Gamma_{uv}} \, | {\bf w} \rangle \, = \, \frac{1}{\sqrt{\langle {\bf v} | {\bf u} \rangle}}
  \bigg( \langle {\bf u} | {\bf w} \rangle, \langle {\bf v} | {\bf w} \rangle \, \bigg| \,
  \frac{ \theta \langle {\bf w} | {\bf v} \rangle + \phi \langle {\bf u} | { \bf w} \rangle +
  \omega \langle {\bf v} | {\bf u} \rangle + \omega \theta \phi}{\sqrt{\langle {\bf v} | {\bf u}
  \rangle }} \bigg) \tran \, .
\label{Gammaw}
\eeq
A general dilatation of the superconformal coordinates corresponds to the ${\rm OSp}
(1|2)$ matrix\footnote{Note that the same symbol ${\bf P}$ was used in Eq.~(4.10) of
Ref.~\cite{Magnea:2013lna} for a square root of this definition.}
\beq
  {\bf P} (\varepsilon) \, = \, \left( \begin{array}{cc|c}
  \varepsilon & 0 & 0 \\ 0 & \varepsilon^{-1} & 0 \\ \hline 0 & 0 & 1
  \end{array} \right) \, ,
\label{eq:Pdef}
\eeq
which has $|\bs{0} \rangle $ and $|\bs{\infty} \rangle$ as fixed points. Note that for
$|\varepsilon| < 1$, $| \bs{0} \rangle$ is an attractive fixed point and $| \bs{\infty} \rangle$
is a repulsive fixed point.

We may use such a dilatation to scale the bosonic coordinates of ${\bf \Gamma_{uv} } |
{\bf w}\rangle$ as desired, obtaining for example
\beq
  {\bf P} \left( \frac{\sqrt{\langle {\bf v} | {\bf w} \rangle}}{\sqrt{\langle {\bf u} | {\bf w}}
  \rangle} \right) {\bf \Gamma_{uv}} | {\bf w} \rangle \sim \bigg( 1 , 1 \, \bigg| \,
  \frac{ \theta \langle {\bf v} | {\bf w} \rangle + \omega \langle {\bf u} | {\bf v} \rangle +
  \phi \langle {\bf w} | { \bf u} \rangle  + \theta \omega \phi}{\sqrt{\langle {\bf u} | {\bf v}
  \rangle \langle {\bf w} | {\bf u} \rangle \langle {\bf v} | {\bf w} \rangle }} \bigg)\tran \, ,
\label{dlatgamw}
\eeq
which gives us an explicit expression for the odd super-projective invariant $\Theta_{\bf
z_1 z_2 z_3 }$, as
\beq
  \Theta_{\bf z_1 z_2 z_3 } \, = \, \frac{ \zeta_1 \langle {\bf z}_3 | {\bf z}_2 \rangle +
  \zeta_2 \langle {\bf z}_1 | {\bf z}_3 \rangle + \zeta_3 \langle {\bf z}_2 | { \bf z}_1 \rangle +
  \zeta_1 \, \zeta_2 \, \zeta_3 }{\sqrt{\langle {\bf z}_2 | {\bf z}_1 \rangle \langle {\bf z}_3 |
  {\bf z}_2 \rangle \langle {\bf z}_1 | {\bf z}_3 \rangle }} \, ,
\label{eq:FermInv}
\eeq
where ${\bf z}_i = (z_i | \zeta_i)$, as in Eq.~(3.4) of \cite{Hornfeck:1987wt} and Eq.~(3.222)
of \cite{D'Hoker:1988ta}.

As with projective transformations, super-projective transformations preserve cross-ratios
of the form
\beq
  \left( {\bf z}_1, {\bf z}_2, {\bf z}_3, {\bf z}_4 \right) \, \equiv \,
  \frac{ \langle {\bf z}_1 | { \bf z}_2 \rangle \, \langle {\bf z}_3 | { \bf z}_4
  \rangle } {\langle {\bf z}_1 | { \bf z}_4 \rangle \, \langle {\bf z}_3 | { \bf z}_2 \rangle }
  \, = \, \frac{ \left( {\bf z}_1 - {\bf z}_2 \right) \left( {\bf z}_3 - {\bf z}_4 \right) }{
  \left( {\bf z}_1 - {\bf z}_4 \right) \left( {\bf z}_3 - {\bf z}_2 \right) } \, \, ;
\label{eq:crossratioF}
\eeq
one must  keep in mind, however, that the simple relations between the three possible
cross ratios that can be constructed with four points are modified by nilpotent terms.
For example, one finds that
\beq
  \left( {\bf z}_1, {\bf z}_2, {\bf z}_3, {\bf z}_4 \right) + \left( {\bf z}_1, {\bf z}_3, {\bf z}_2,
  {\bf z}_4 \right) - \left( {\bf z}_1, {\bf z}_3, {\bf z}_2, {\bf z}_4 \right)^{1/2} \,
  \Theta_{ {\bf z}_1 {\bf z}_3 {\bf z}_2 } \, \Theta_{ {\bf z}_1 {\bf z}_4 {\bf z}_2 } \, = \, 1 \, ,
\label{eq:suproj}
\eeq
which can be checked quickly by noting that the left-hand side is ${\rm OSp}(1|2)$-invariant,
so that one can fix $3|2$ parameters, for example by choosing $| {\bf z_1} \rangle = | \bs{0}
\rangle$, $| {\bf z}_2 \rangle = | \bs{\infty} \rangle$, and $| {\bf z}_4 \rangle = (1, 1 | \phi)\tran$.

With these ingredients, it is now easy to construct a super-projective transformation
with chosen fixed points and multiplier: using ${\bf \Gamma_{u v}}$ to map a pair of
points $| {\bf u} \rangle$ and $| {\bf v \rangle}$ to $| \bs{0} \rangle$ and $| \bs{\infty}
\rangle$ respectively, one easily verifies that the transformation
\beq
  {\bf S} \, = \, {\bf \Gamma}_{\bf uv}^{-1} \, {\bf P} \left(- \ex{ \ii \pi \varsigma} k^{1/2} \right) \,
  {\bf \Gamma}_{\bf uv}
\label{SchottkySdef}
\eeq
has $| {\bf u} \rangle$ as an attractive fixed point and $| {\bf v} \rangle$ as a repulsive
fixed point. Here $k$, for which we take $|k| < 1$, is called the multiplier\footnote{The
sign $\ex{ \ii \pi \varsigma}$ is related to the spin structure: see the discussion between
\eq{eq:ghostsf} and \eq{eq:scalarf} for the conventions.} of the super-projective
transformation ${\bf S}$.

The bracket notation has the benefit of allowing us to write ${\bf S}$ as
\beq
  {\bf S} \, = \, \one  \, + \, \frac{1}{ \langle \mathbf{v} | \mathbf{u} \rangle }
   \bigg[ \left( 1 + \ex{\ii \pi \varsigma} k^{\frac{1}{2}} \right) \ket{\mathbf{v}} \bra{\mathbf{u}}
   \, - \, \left( 1 + \ex{- \ii \pi \varsigma} k^{- \frac{1}{2}} \right) \ket{\mathbf{u}}
   \bra{\mathbf{v}} \, \bigg] \, ,
 \label{eq:SuSbraket}
\eeq
which satisfies
\beq
  \frac{ \langle {\bf S}({\bf z}) | {\bf u} \rangle }{ \langle {\bf S}({\bf z}) | {\bf v}
  \rangle} \, = \, k \, \frac{ \langle {\bf z} | {\bf u} \rangle }{ \langle {\bf z} | {\bf v} \rangle} \, .
\label{Srat}
\eeq


\subsection{The super Schottky group}
\label{SuScho}

Taking the quotient of ${\bf CP}^{1|1}$ by the action of ${\bf S}$, defined in \eq{eq:SuSbraket},
is equivalent to the insertion of a pair of NS punctures at $| {\bf u} \rangle$ and $| {\bf v}
\rangle$, which are then sewed with a sewing parameter related to $k$. To see this,
recall that sewing of NS punctures at $P_1$ and $P_2$ is defined by taking two sets
of superconformal coordinates, say $(x | \theta)$ and $(y | \psi)$, which vanish
respectively at the two points, $(x | \theta)(P_1) = (0|0) = (y|\psi)(P_2)$, and then
imposing the conditions~\cite{Witten:2012bh}
\beq
  x y \, = \, - \, \varepsilon^2 \, , \qquad y \theta \, = \, \varepsilon \psi \, , \qquad
  x \psi \, = \, - \, \varepsilon \theta \, , \qquad \theta \psi \, = \, 0 \, .
\label{nssew}
\eeq
Now, let $| {\bf x} \rangle$ and $| {\bf y} \rangle$ be homogeneous coordinates satisfying
$\mathfrak{f} | {\bf x} \rangle = (x | \theta)$ and $\mathfrak{f} | {\bf y} \rangle = (y | \psi)$,
where $\mathfrak{f}$ is the map defined in \eq{rscdef}. If we make the identification
\beq
  | {\bf x} \rangle \, \sim \, \left( {\bf P} (\varepsilon) \circ {\bf I} \right) | {\bf y} \rangle \, ,
\label{homogequiv}
\eeq
where ${\bf P}$ and ${\bf I}$ are defined in \eq{eq:Pdef} and \eq{Idef}, respectively.
Then, by acting on both sides with $\mathfrak{f}$, we get $(x | \theta) \sim ( -
\varepsilon^2/y \, \big| \, \varepsilon \psi/y )$, which can easily be found to satisfy
\eq{nssew}.

Let us take $(z | \zeta)$ to be a superconformal coordinate on ${\bf CP}^{1|1}$, with
$(z|\zeta)(P_1) = \mathfrak{f} | {\bf u} \rangle$ and $(z | \zeta)(P_2) = \mathfrak{f} |
{\bf v} \rangle$. Recall that the super-projective transformation ${\bf \Gamma}_{\bf u v}$
defined in \eq{Gammauvdef} simultaneously maps $| {\bf u} \rangle$ and $| {\bf v} \rangle$
to $| \bs{0} \rangle$ and $| \bs{\infty} \rangle$, respectively. Then if $|{ \bf x} \rangle
= {\bf \Gamma}_{\bf u v} \circ \mathfrak{f}^{-1} \circ (z | \zeta)$ and $| {\bf y} \rangle
= {\bf I}^{-1} \circ {\bf \Gamma}_{\bf u v} \circ \mathfrak{f}^{-1} \circ (z | \zeta)$, we
have that $(x | \theta) = \mathfrak{f} |\bf x \rangle$ and $(y | \psi) = \mathfrak{f} |
{\bf y} \rangle$ are local superconformal coordinates which vanish at $P_1$ and
$P_2$ respectively, since ${\bf I}^{-1} | \bs \infty \rangle = | \bs 0 \rangle$ and
$\mathfrak{f}|\bs 0\rangle = (0|0)$. As a consequence, we can perform a NS
sewing by making the identification in \eq{homogequiv} using these expressions
for $| {\bf x} \rangle$ and $| {\bf y} \rangle$, and we find that we need to impose
an equivalence relation on $(z | \zeta)$: we have ${\bf \Gamma}_{\bf u v} \circ
\mathfrak{f}^{-1} \circ (z | \zeta) \sim {\bf P}( \varepsilon) \circ {\bf I} \circ {\bf I}^{-1}
\circ {\bf \Gamma}_{\bf u v} \circ \mathfrak{f}^{-1} \circ (z | \zeta)$, or to put it differently,
\beq
  (z | \zeta) \, \sim \, \mathfrak{f} \, {\bf S} \, \mathfrak{f}^{-1} \, (z |\zeta) \, , \qquad
  {\bf S} \, \equiv  \, {\bf \Gamma}_{\bf u v}^{-1} \circ {\bf P} (\varepsilon) \circ
  {\bf \Gamma}_{\bf u v} \, .
\label{equivrel}
\eeq
This is what we wanted to show, with ${\bf S}$ matching the definition in \eq{eq:SuSbraket},
as long as we identify $\varepsilon = - \ex{\ii \pi \varsigma} k^{1/2}$, so the NS sewing
parameter is directly related to the Schottky group multiplier. Topologically, this sewing
has the same effect (at least on the reduced space ${\bf CP}^1$) as cutting out discs
around ${\bf u}$ and ${\bf v}$ and identifying their boundaries, so this quotient adds a
handle to the surface, increasing the genus by one.

To build a genus-$h$ SRS, we may repeat this sewing procedure $h$ times, choosing
$h$ pairs of attractive and repulsive fixed points ${\bf u}_i = (u_i | \theta_i)$, ${\bf v}_i =
(v_i | \phi_i)$, and $h$ multipliers $k_i$, for $i = 1, \ldots, h$. The super-Schottky
group $\overline{\cal S}_h$ is the group freely generated by
\beq
  {\bf S}_i \, = \, {\bf \Gamma}_{{\bf u}_i {\bf v}_i}^{- 1} \, {\bf P} \left(- \, \ex{ \ii \pi \varsigma_i}
  k_i^{1/2} \right) \, {\bf \Gamma}_{{\bf u}_i {\bf v}_i} \, , \qquad  i \, = \, 1 , \ldots , h \, .
\label{suSchogen}
\eeq
We then subtract the limit set $\Lambda$ (the set of accumulation points of the orbits
of $\overline{\cal S}_h$) from ${\bf CP}^{1|1}$, and we quotient by the action of the
super Schottky group. this leads to the definition
\beq
  \bs{\Sigma}_h \, = \, \left( {\bf CP}^{1|1} - \Lambda \right) \! \Big/ \, \overline{\cal S}_h \, .
\label{Sigmah}
\eeq
Note that the fixed points must be sufficiently far from each other, and the multipliers
sufficiently small, to allow for the existence of a fundamental domain with the topology
of ${\bf CP}^{1|1}$ with $2 h$ discs cut out. The fixed points ${\bf u}_i$, ${\bf v}_i$
and the multipliers $k_i$ are moduli for the surface, but for $h \geq 2$ we can use
the ${\rm OSp}(1|2)$ symmetry to fix $3|2$ of these: in our conventions, we take
$| {\bf u}_1 \rangle = | \bs{0} \rangle$, $| {\bf v}_1 \rangle = | \bs{\infty} \rangle$,
$| {\bf v}_2 \rangle =  | 1, 1 | \Theta_{{\bf u}_1 {\bf v}_2 {\bf v}_1} \rangle$, so the
super-moduli space $\widehat{\frak M}_h$ has complex dimension $3 h - 3 | 2 h - 2$.

To build multi-loop open superstring world-sheets in a similar way, we should start with
the super-disc ${\bf D}^{1|1}$ which can be obtained by quotienting ${\bf CP}^{1|1}$ by
the involution $(z | \theta) \mapsto (z^* | \theta)$, so that ${\bf RP}^{1|1}$ becomes
the boundary of the disk. A super-projective map will be an automorphism of
${\bf D}^{1|1}$ if it preserves ${\bf RP}^{1|1}$, so we should build the super Schottky
group from super-projective transformations whose fixed points ${\bf u}_i$, ${\bf v}_i$
are in ${\bf R}^{1|1}$ and whose multipliers $k_i$ are real. If we quotient ${\bf D}^{1|1} -
\Lambda$ by $h$ of these, then we will get a SRS with $(h + 1)$ borders and no
handles. The moduli space $\widehat{\frak M}_h^{\text{open}}$ of such SRSs
has \emph{real} dimension  $3 h - 3 | 2 h - 2$. In the case of $h = 2$ surfaces,
we use the ${\rm OSp}(1|2)$ symmetry to write the fixed points as in \eq{eq:sfpdef}.


\subsubsection{Multipliers}

Every element ${\bf S}_\alpha$ of a super Schottky group is similar to a matrix of the
form
\beq
{\bf P} \left( - \, \ex{\ii \pi \vec{\varsigma} \cdot \vec{N}_\alpha} \, k_\alpha^{1/2} \right) \, ,
\label{Pgen}
\eeq
as in \eq{eq:Pdef}, for some $k_\alpha^{1/2}$. We can find $k_\alpha^{1/2}$ by
setting the spin structure around the $b$-cycles to zero, $\vec{\varsigma} = \vec{0}$,
then using the cyclic property of the supertrace\footnote{Recall that the supertrace
of a $\text{GL}(2|1)$ matrix $M = (M_i{}^j)$ is given by $\text{sTr}(M) = M_1{}^1 +
M_2{}^2 - M_3{}^3$.}. This leads to a quadratic equation, with roots
\beq
  k_\alpha^{1/2} \, = \, - \, \frac{ 1 + {\rm sTr} \left( {\bf S}_\alpha \right) \pm
  \sqrt{ \big( {\rm sTr} \left( {\bf S}_\alpha \big) +1 \right)^2 - 4}}{2} \, ,
\label{eq:kFromSTr}
\eeq
one root being the inverse of the other. We then pick $k_\alpha^{1/2}$ to be the root
whose absolute value satisfies $| k_\alpha^{1/2} | < 1$. With this choice, we can
expand the $k_\alpha^{1/2}$ in powers of $k_i^{1/2}$: for $h = 2$, using the fixed
points in \eq{eq:sfpdef}, we find
\beqa
\label{ktwolexp}
  k^{1/2} \left( {\bf S}_1 {\bf S}_2 \right) & = & - \, y \, k_1^{1/2} \, k_2^{1/2}
  + {\cal O} (k_i) \, = \, - \, \left( {\bf u}_1, {\bf v}_1, {\bf u}_2, {\bf v}_2 \right) \,
  k_1^{1/2} \, k_2^{1/2} \, + {\cal O}(k_i) \nonumber \\
  k^{1/2} \left( {\bf S}_1^{- 1} {\bf S}_2 \right) & = &  \frac{y}{u} \, k_1^{1/2} \, k_2^{1/2} +
  {\cal O} (k_i) \, = \, - \, \left( {\bf v}_1, {\bf u}_1,
  {\bf u}_2, {\bf v}_2 \right) \, k_1^{1/2} \, k_2^{1/2} + {\cal O}(k_i) \, .
\eeqa
where $y$ was defined in \eq{eq:ydef}. Note that $k^{1/2} ({\bf S}_1^{-1} {\bf S}_2)$
can be obtained from $k^{1/2} ({\bf S}_1 {\bf S}_2)$ by swapping the attractive and
repulsive fixed points of ${\bf S}_1$ in the cross-ratio, as might be expected.


\subsubsection{The super period matrix}
\label{SuperPM}

The super abelian differentials are an $h$-dimensional space of holomorphic volume
forms, \emph{i.e.}~sections of the Berezinian bundle, defined on a genus-$h$ SRS.
They are spanned by ${\bf \Omega}_i$, $i = 1 , \ldots, h$, which can be normalized
by their integrals around the $a$-cycles, according to
\beq
  \frac{1}{2 \pi \ii } \, \oint_{a_i} {\bf \Omega}_j \, = \, \delta_{i j} \, ,
\label{normab}
\eeq
while their integrals around the $b$-cycles define the \emph{super period matrix}
\beq
  \frac{1}{2 \pi \ii } \, \oint_{b_i} {\bf \Omega}_j \, \equiv \, \btau_{i j} \, .
\label{defsupeper}
\eeq
Here $a_i$ and $b_i$ are closed cycles on the SRS which are projected to the usual
homology cycles on the reduced space. The ${\bf \Omega}_i$'s can be expressed in
terms of the super Schottky parametrization as in Eq.~(21) of Ref.~\cite{DiVecchia:1988jy}.
In our current notation
\beqa
\label{eq:superabelian}
  {\bf \Omega}_i (z | \psi) & = & \d {\bf z} \, {\sum_\alpha}^{(i)} \, D_\psi \, \log
  \frac{ \langle {\bf z} | {\bf T}_\alpha | {\bf u}_i \rangle}{ \langle {\bf z} | {\bf T}_\alpha |
  {\bf v}_i \rangle} \nonumber \\
  & = & \d {\bf z} \, {\sum_\alpha}^{(i)} \Bigg[ \frac{ \langle {\bf z} | \, \Phi \, {\bf T}_\alpha \, |
  {\bf u}_i \rangle}{ \langle {\bf z} | {\bf T}_\alpha | {\bf u}_i \rangle} -
  \frac{ \langle {\bf z} | \, \Phi \, {\bf T}_\alpha \, | {\bf v}_i \rangle}{ \langle {\bf z} |
  {\bf T}_\alpha | {\bf v}_i \rangle} \Bigg] \, ,
\eeqa
where $\d {\bf z} = (\d z \, | \, \d \psi)$, the sum $\sum_\alpha^{(i)}$ is over all elements
of the super-Schottky group which do not have ${\bf S}_i^{\pm 1}$ as their right-most
factor, $D_\psi$ is the superconformal derivative $D_\psi = \partial_\psi + \psi \partial_z$,
and finally $\Phi$ is the matrix
\beq
  \Phi \, = \, \left( \begin{array}{cc|c} 0 & 0 & 1 \\ 0 & 0 & 0 \\ \hline 0 & -1 & 0 \end{array}
  \right) \, .
\label{eq:Phidef}
\eeq
The matrix $\Phi$ has the property that, if $\mathfrak{f} \, | {\bf z} \rangle = (z|\psi)$, then
\beq
  D_\psi \langle {\bf w} | {\bf z} \rangle \, = \, \langle {\bf w} | \Phi | {\bf z} \rangle \, ,
\label{prop phi}
\eeq
for any $\langle {\bf w} |$, and furthermore for $\ket{\bf w} = (w,1|\omega)\tran$ and $\ket{\bf z} = (z,1|\psi)\tran$ the map $(z | \psi) \mapsto \langle {\bf w} | {\bf z} \rangle | \langle {\bf w} |
\Phi | {\bf z} \rangle$ is superconformal. The super period matrix can be computed as
\beq
  \bs{\tau}_{i j} \, = \, \frac{1}{2 \pi \ii }  \left[ \, \delta_{i j} \log k_i - \phantom{}^{(j)}
  \sum_{\alpha} \phantom{}^{(i)} \, \log \frac{ \bra{ {\bf u}_j} {\bf T}_\alpha \ket{{\bf v}_i}
  \bra{ {\bf v}_j} {\bf T}_\alpha \ket{{\bf u}_i} }{\bra{ {\bf u}_j} {\bf T}_\alpha \ket{{\bf u}_i}
  \bra{ {\bf v}_j} {\bf T}_\alpha \ket{{\bf v}_i}} \right] \, .
\label{eq:supertau}
\eeq
The sum is over all elements of the super Schottky group which do not have
${\bf S}_j^{\pm 1}$ as their left-most element or ${\bf S}_i^{\pm 1}$ as their right-most
element. It is not difficult to compute the leading terms of the super period matrix in the
small-$k_i$ expansion. For $h = 2$, using the fixed points in \eq{eq:sfpdef}, we find
\beq
  2 \pi \ii \, \btau  \, = \, \left( \begin{array}{cc} \log k_1 - 2 \, \theta \phi \, \frac{y}{u} \,
  \ex{\ii \pi \varsigma_2} \, k_2^{1/2} & \log u \\
  \log u & \log k_2 - 2 \, \theta \phi \, \frac{y}{u}  \, \ex{\ii \pi \varsigma_1} \, k_1^{1/{2}}
  \end{array} \right) + {\cal O}(k_i)
\label{eq:stauSch}
\eeq
so that
\beqa
\label{eq:dettau}
  4 \pi^2 \det \left( {\rm Im} \; \btau \right) & = & \log(k_1) \log(k_2) - \log(u)^2  \\
  & & \hspace{1pt} - \, 2 \, \theta \phi \, \frac{y}{u} \, \Big( \ex{\ii \pi \varsigma_2} \,
  k_2^{1/2} \log k_1\,  + \ex{\ii \pi \varsigma_1} \, k_1^{1/2} \log k_2 \Big)
  + {\cal O}(k_i) \, . \nonumber
\eeqa
This completes our review of the super Schottky parametrization. Our next task is to
introduce twisted boundary conditions corresponding to external background gauge fields.

\section{Appendix B}
\label{Appb}


\subsection{The twisted determinant on a Riemann surface}
\label{tepsbos}

\begin{figure}
\centering
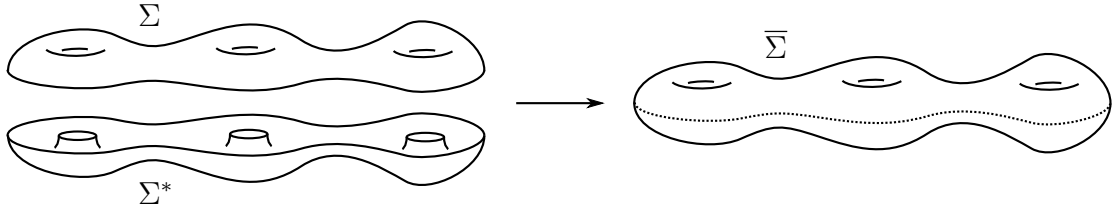
\caption{Recall that the double of a Riemann surface $\Sigma$ is defined by taking two copies of $\Sigma$, replacing the charts on one copy with their complex conjugates, and identifying corresponding points on the boundaries of the two copies \cite{Alessandrini:1971cz}.}
\label{fig:DoubleSurface}
\end{figure}

The worldsheet theory of strings becomes `twisted' in a number of contexts: for example, on orbifolds \cite{Dixon:1986qv}, in electromagnetic fields \cite{Tseytlin:1998kw,Abouelsaood:1986gd,Bachas:1992bh} or when an open string is stretched between a pair of D-branes which have a velocity \cite{Bachas:1995kx} or are at an angle \cite{Berkooz:1996km} with respect to each other. If we appropriately pair up the string spacetime coordinate fields $X^\mu$ as complex coordinates (\emph{e.g.}~in our case, by setting $Z^\pm = (X^1 \pm \ii X^2)/ \sqrt{2}$), then in these backgrounds the worldsheet fields $\partial Z^\pm$ are described by non-integer mode expansions on the upper-half-plane, as in \eq{eq:XmeUa}. This means that on the double of the worldsheet (the double of the upper-half-plane is the complex plane; see \Fig{fig:DoubleSurface}), $\partial Z^\pm(z,\overline{z})$ is no longer a single-valued field but rather it has a monodromy, changing by a factor of $\ex{\pm 2 \pi \ii \epsilon}$ as it is transported anti-clockwise around $z=0$.

Computing multi-loop amplitudes in these backgrounds is complicated because it is not easy to use the sewing procedure when states propagating along plumbing fixture belong to a twisted sector. We must use, instead, the approach of \cite{Russo:2007tc}. This takes advantage of the fact that although the $\partial Z^\pm$ fields have non-trivial monodromies along the $a_i$-cycles of the double worldsheet, the monodromies along the $b_j$-cycles are trivial. Therefore, the idea is to build the double worldsheet by sewing along the $b_i$ cycles, and then to perform the modular transformation swapping the $a_j$ and $b_i$ cycles with each other, in order to obtain the partition function expressed in terms of the Schottky moduli which are the appropriate ones for the worldsheet degeneration we are interested in.

From a more physical point of view, we are using the fact that in a different region of moduli space, the string diagram can be described as a tree-level interaction between three closed strings being emitted or absorbed by the D-branes.
\begin{figure}
\centering
\def\svgwidth{10cm}
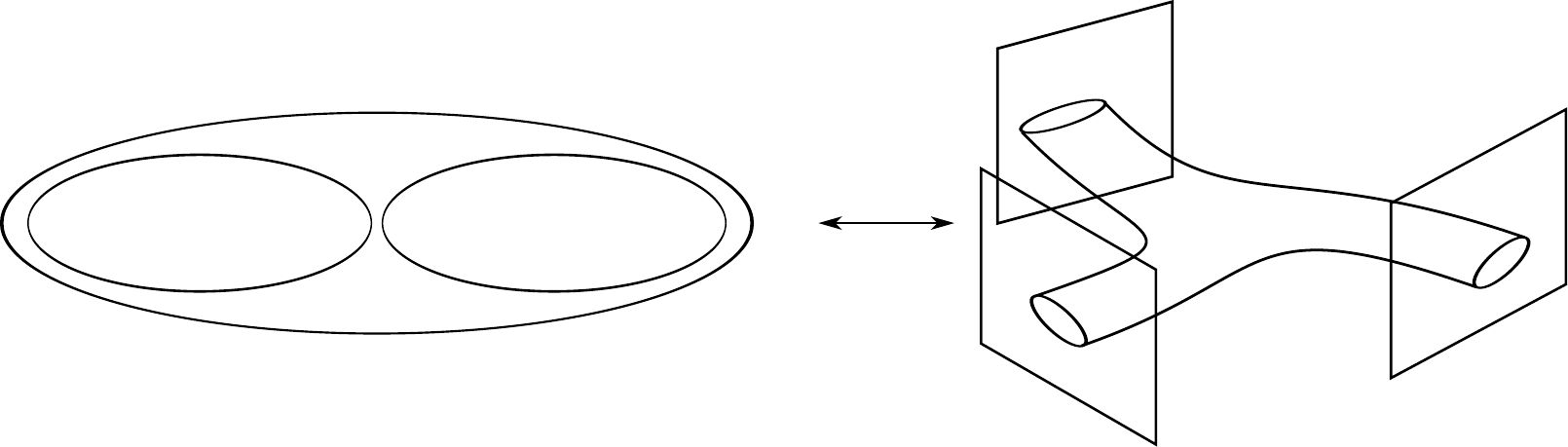
\caption{The same string diagram can be computed as an open string vacuum multi-loop diagram or as the T-dual setup of a tree-level interaction between closed strings emitted from D-branes at angles.}
\label{fig:openclosed}
\end{figure}
In terms of the closed string moduli, the string partition function is given by \cite{Russo:2007tc}
\begin{align}
Z(F_{i}) & = \Big( \prod_{i=0}^h\sqrt{\det(1 - G^{-1} {\cal F}_i )} \Big) \int [ \d Z]_h^{\rm cl}
\;
{\cal R}_h (q_i , \vec{\epsilon}\, )  \, .
\end{align}
The overall factor is just the Born-Infeld lagrangian for the background fields on the D-branes, divided by $\sqrt{G}$ because all of the background-field independent factors are included in the measure $[\d Z]_h^{\rm cl}$.

The factor ${\cal R}_h (q_i , \vec{\epsilon}\, ) $, which is dependent on both the worldsheet moduli and on the background field strengths, has a simple form so long as it is expressed in terms of the closed string Schottky group moduli, in other words, in terms of the multipliers of a Schottky group whose $2h$ Schottky circles are homotopic to the $b_i$ cycles of the worldsheet instead of the $a_j$ cycles which we have been using. Let us denote the multipliers of the elements $T_\alpha$ of this Schottky group as $q_\alpha$, then we have
\begin{align}
{\cal R}_h (q_i , \vec{\epsilon} \, )  & =
\frac{
\prod_{\alpha}' \prod_{n=1}^\infty (1  - q _\alpha^n)^2 }
{\prod_\alpha' \prod_{n=1}^\infty (1 - \ex{- 2 \pi \ii \vec{N}_{\alpha } \cdot \vec{\epsilon}}q_\alpha^n)(1 - \ex{2 \pi \ii \vec{N}_{\alpha } \cdot \vec{\epsilon}}q_\alpha^n)
}
\, ,
\label{eq:Rdef}
\end{align}
where the notation ${\prod}_\alpha ' $ has the same meaning as for the super Schottky group case, defined after \eq{eq:ghostsf}.

The modular transformation that swaps the $a_i$- and $b_j$-cycles, necessary to switch between the open string and closed string channels, acts non-analytically on the Schottky group multipliers.\footnote{This is easiest to see in the $h=1$ case when the single open string multiplier $k$ is related to the torus period $\tau$ via $k = \ex{2 \pi \ii \tau}$, and similarly the multiplier in the closed string channel $q$ is related to $\tau^{\rm cl}$ via $q = \ex{ 2 \pi \ii \tau^{\rm cl}}$, where the closed string torus period $\tau^{\rm cl}$ can be obtained from the open string one $\tau$ via $\tau^{\rm cl} = - 1 / \tau$, so $(\log q )(\log k) = 4 \pi^2 $.} We need to rewrite \eq{eq:Rdef} in terms of the open string moduli, so the following strategy is used: ${\cal R}_h (q_i , \vec{\epsilon} \, )$ is re-expressed in terms of functions which transform in simple ways under modular transformations, the modular transformations are then carried out, and finally the results are re-expressed in terms of the open string Schottky moduli, allowing us to investigate the field theory limit. This analysis was performed in \cite{Russo:2003tt,Russo:2003yk} and the results are summarized in section 2 of \cite{Magnea:2004ai}. Assuming without loss of generality that the $h$'th twist is nonzero, $\epsilon_h \neq 0$, the result is that
\begin{align}
{\cal R}_h( q_\alpha, \vec{\epsilon} \,) & =
{\cal R}_h ( k_\alpha , \vec{\epsilon} \cdot \tau )
\,
\ex{ - \ii \pi \vec{\epsilon} \cdot \tau \cdot \vec{\epsilon}}
\;
\frac{\det( \text{Im}\,\tau)}{\det (\text{Im}\,\tau_{\vec{\epsilon}})}  \, .
\label{eq:qkswap}
\end{align}
where
${\cal R}_h ( k_\alpha , \vec{\epsilon} \cdot \tau )$ is the same as in \eq{eq:Rdef} but with the closed string channel multipliers $q_\alpha$ replaced with open string channel multipliers $k_\alpha$, and the twists $\vec{\epsilon}$ replaced with $\vec{\epsilon} \cdot \tau$, $(\tau_{i j})$ being the period matrix computed in the open string channel. $\tau_{\vec{\epsilon}}$ is the \emph{twisted period matrix}, defined by (Eq.~(3.24) of \cite{Russo:2007tc})
\begin{align}
(\tau_{\vec{\epsilon}}\, )_{j i} & =
\begin{cases}
\frac{1}{2 \pi \ii} \int_w^{S_j(w)} \,
\vphantom{\bigg[}
\Omega_{i }^{\vec{\epsilon} \cdot \tau} (z) \,
\ex{ \frac{2 \pi \ii}{h-1} \vec{\epsilon} \cdot \vec{\Delta}(z) }
\vphantom{\bigg]}
& j \neq h \neq i \\
\frac{{\cal S}_h^{\veps \cdot \tau} - 1}{{\cal S}_h^{\veps} - 1}
& j = i =  h\, , \\
0 &  \text{otherwise.}
\end{cases}
\label{eq:tepsdef}
\end{align}
where
\begin{align}
  {\cal S}^{\veps}_i & \equiv \ex{2 \pi \ii \epsilon_i }  \, . \label{eq:periodicity1}
  \end{align}
The Prym differentials $\Omega_i^{\,\veps}$ appearing in the integrand in the first line of \eq{eq:tepsdef} are $(h-1)$ 1-forms with trivial monodromies along the $a_i$ cycles and twists along the $b_i$-cycles, \emph{i.e.}~they obey
\begin{align}
\Omega_i^{\veps}(S_j(z)) & = {\cal S}_j^{\veps} \, \Omega_i^{\veps}(z) \, ;\label{eq:periodicity2}
\end{align} and they are regular everywhere.\footnote{
 N.B.~that \eq{eq:tepsdef} uses $\Omega_i^{\vec{\epsilon}\cdot \tau}$ which is obtained from $\Omega_i^{\vec{\epsilon}}$ as computed in Eqs.~(\ref{eq:Omegaepsdef}) and (\ref{eq:zetaepsdef}) by making the substitution $\epsilon_i \mapsto (\veps \cdot \tau)_i =  \epsilon_1 \tau_{1 i} + \ldots +  \epsilon_h \tau_{h i}$.}
 Assuming without loss of generality that $\epsilon_h \neq 0$, they can be expressed as (Eq. (3.11) of \cite{Russo:2007tc}):
\begin{align}
\Omega^{\veps}_j(z) & = \zeta^{\veps}_j(z) - \frac{1 - {\cal S}^{\veps}_j }{1 - {\cal S}^{\veps}_h}\,\, \zeta^{\veps}_h(z) \, &  j & = 1 , \ldots , (h-1) \, . \label{eq:Omegaepsdef}
\end{align}
In \eq{eq:Omegaepsdef} the $\zeta_i^{\veps}$ are a basis of $h$ 1-forms which are holomorphic everywhere except some arbitrary base point $z_0$, which can be computed in terms of the Schottky group as (Eq. (3.15) of \cite{Russo:2007tc})
\begin{align}
\zeta_i^{\veps}(z) & = \Big( {\cal S}_i^{\veps} \, {\sum_\alpha }^{(i)} \ex{ 2 \pi i  \veps \cdot {\vec{N}}_\alpha  } \Big[ \frac{1}{z - T_\alpha(\eta_i) } - \frac{1}{z - T_\alpha(\xi_i)} \Big] \label{eq:zetaepsdef} \\
& \phantom{=====}
+(1 - {\cal S}_i^{\veps}\, ) \sum_\alpha \ex{2 \pi i \veps \cdot \vec{N}_\alpha } \Big[ \frac{1}{z - T_\alpha(z_0)} - \frac{1}{z - T_\alpha(a_i^\alpha)} \Big] \Big) \, \d z
\nonumber
\end{align}
where the first sum is over all Schottky group elements which don't have $S_i^{\pm \ell}$ as their right-most factor and the second sum is over all Schottky group elements. $\eta_i$ and $\xi_i$ are the attractive and repulsive fixed points of the generator $S_i$, respectively.
Note that the dependence on $z_0$ cancels out when the $\zeta_i^{\veps}$ are combined as in \eq{eq:Omegaepsdef}. Also in \eq{eq:zetaepsdef},
\begin{align}
a_i^\alpha  & =
\begin{cases}
\eta_i & \text{ if } T_\alpha = T_\beta S_i^{\ell} \text{ with } \ell \geq 1 \\
\xi_i & \text{ otherwise.}
\end{cases}
\end{align}
 The other object appearing in the integrand in the first line of \eq{eq:tepsdef}, $\Delta_i(z)$, is the vector of Riemann constants or Riemann class; it can be expressed in the Schottky parametrization as (Eq. (A.21) of \cite{DiVecchia:1988cy})
\begin{align}
\Delta_i(z) & = \frac{1}{2 \pi \ii } \left\{ - \frac{1}{2} \log k_i + \pi \ii + \sum_{j=1}^h \phantom{}^{(j)} \sum_{\alpha} \phantom{}^{(i)} \log \frac{ \xi_j - T_\alpha(\eta_i)}{\xi_j - T_\alpha(\xi_i) } \frac{ z - T_\alpha(\xi_i)}{z - T_\alpha(\eta_i) } \right\} \label{eq:Deltazdef}
\end{align}
where the second sum $\phantom{}^{(j)} \sum_{\alpha} \phantom{}^{(i)}$ is over all elements of the Schottky group which have neither $S_j^{\pm 1}$ as their left-most element nor $S_i^{\pm 1}$ as their right-most element. Owing to the transformation properties of $\Delta_i(z)$,
\begin{align}
\Delta_i(z) & = \Delta_i(z_0) - \frac{h-1}{2 \pi \ii} \int_{z_0}^z \omega_i \, , \label{eq:Dtran}
\end{align}
where $\omega_i$ are the abelian differentials (Eq.~(A.10) of \cite{DiVecchia:1988cy}), it is easy to check that the integrand of the first line of \eq{eq:tepsdef} has twists along the $a_i$ cycles and trivial monodromies along the $b_i$-cycles (therefore the integrand does not depend on the starting point $w$).

For simplicity, from now on we focus only on the case $h=2$, which yields
\begin{align}
\det(\text{Im}\,\tau_{\veps}) & = \frac{1}{2 \pi \ii } \frac{{\cal S}_2^{\veps \cdot \tau} - 1 }{{\cal S}_2^{\veps} - 1} \int_w^{S_1(w)}\Omega^{\veps \cdot \tau}(z) \ex{2 \pi \ii  \veps \cdot \vec{\Delta}(z)} \, , \label{eq:twdethtwo}
\end{align}
where $\Omega^{\veps} \equiv \Omega_1^{\veps}$ is the sole component of the Prym form.

\begin{figure}
\centering
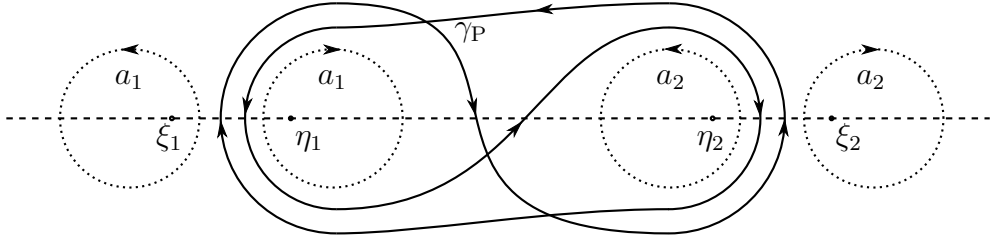
\caption{The Pochhammer contour $\gamma_{\rm P} = a_2 a_1 {a_2}^{-1} {a_1}^{-1}$.}
\label{fig:pochhammer}
\end{figure}

Instead of explicitly evaluating the integral over $z$ in \eq{eq:twdethtwo}, it is possible to find an alternative expression for $\det(\text{Im}\,\tau_{\veps})$ in the following way. First of all, we recall the object ${\bf D}(\veps\,)_{ij}$ defined in Eq.~(3.14) of \cite{Russo:2007tc}. For each $i,j = 1, \ldots, (h-1)$, ${\bf D}_{ij}(\veps\,)$ is a spacetime rotation matrix; the $i,j$ indices refer to worldsheet homology cycles. In the complex spacetime coordinates in which the background fields are diagonalized, ${\bf D}_{ij}(\veps\,)$ is diagonal with two non-trivial entries $D(\pm \veps\,)_{ij}$. For $h=2$, the $i,j$ indices can take only one value, so we only have one independent object $D_{11}(\pm\veps\,) \equiv D(\pm\veps\,)$. It is given by
\newcommand{\TwistedOmega}{\widetilde{\Omega}}
\begin{align}
D(\veps\,) & \equiv \frac{1}{2 \pi \ii } \frac{({\cal S}_1^{\veps}\,)^{-1}}{1- {\cal S}_2^{\veps}} \int_{\gamma_{\rm P}} \TwistedOmega^{\veps} \, , \label{eq:Depsdef}
\end{align}
where $\gamma_{\rm P} \equiv a_2 a_1 {a_2}^{-1} {a_1}^{-1}$ is the Pochhammer contour shown in \Fig{fig:pochhammer}, and $\TwistedOmega^{\veps}$ is the Prym differential which has trivial monodromies around the $b_i$ homology cycles and monodromies ${\cal S}_i^{\veps}$ around the $a_i$ homology cycles.
  $\TwistedOmega^{\veps}$ can be expressed with the Schottky group thanks to a relation derived in \cite{Russo:2007tc}, given by Eq.~(3.28) of that reference. In the case $h=2$, it becomes simply
\begin{align}
\TwistedOmega^{\veps}(z) & = \frac{\ex{ 2 \pi \ii \veps \cdot \vec{\Delta}(z)} \, \Omega^{\veps \cdot \tau}(z)}{(\tau_{\veps})_{11} } = \frac{1 - {\cal S}^{\veps \cdot \tau}_2}{1 - {\cal S}_2^{\veps}} \frac{\ex{ 2 \pi \ii \veps \cdot \vec \Delta(z)} \, \Omega^{\veps \cdot \tau}(z)}{\det(\text{Im}\,\tau_{\veps})} \label{eq:epsepst}
\end{align}
  where the second equality is just writing $(\tau_{\veps})_{11}$ in terms of $\det(\text{Im}\,\tau_{\veps})$ with \eq{eq:tepsdef}.

Note that $\gamma_{\rm P}$ crosses each boundary of the worldsheet once in each direction so it starts and ends on the same branch of $\TwistedOmega^{\veps}$ and the integral in \eq{eq:Depsdef} is well-defined.

\begin{figure}
\centering
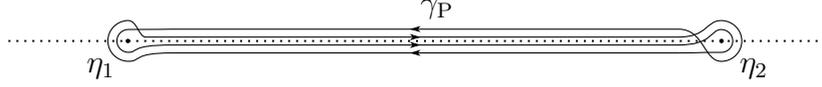
\caption{Our Pochhammer contour (\Fig{fig:pochhammer}) can be deformed arbitrarily close to four copies of the line interval $[\eta_1 , \eta_2 ] \subseteq \mathbf{R}$, with each copy on a different branch of the Prym form $\TwistedOmega^{\veps}$.}
\label{fig:poch2}
\end{figure}

Now we take the formulae \eq{eq:Omegaepsdef}, \eq{eq:zetaepsdef} and \eq{eq:Deltazdef} to get an expression for $\TwistedOmega^{\veps}(z)$ via \eq{eq:epsepst} and interpret them as defining  a one-form not on the worldsheet but on the complex plane, with poles at the Schottky group limit set. If we expand $\TwistedOmega^{\veps}(z)$ as a power series in $k_i$, we see that at leading order it has poles only at the Schottky fixed points, so at leading order we are free to deform the Pochhammer contour through the Schottky circles and arbitrarily close to the line interval $[\eta_1, \eta_2]$ as in \Fig{fig:poch2}.
In this way, we can write the Pochhammer integral as four copies of a real integral taking care to account for the different orientations and branches;
it turns out that we get
\begin{align}
\int_{\gamma_{\rm P}}  \TwistedOmega^{\veps}  & =   (1 - {\cal S}_1^{\veps}\,)(1 - {\cal S}_2^{\veps}\,) \int_{\eta_1}^{\eta_2} \TwistedOmega^{\veps} + {\cal O}(k_i^2) \, . \label{eq:pochdeform}
\end{align}

Then inserting \eq{eq:pochdeform} and \eq{eq:epsepst} into \eq{eq:Depsdef}, we find a relation between $D(\veps\,)$ and $\det(\text{Im}\,\tau_{\veps})$ (Eqs.~(4.13)--(4.15) of \cite{Russo:2007tc}):
\begin{align}
D({\veps}\,) & = - \frac{1}{2 \pi \ii} \frac{1-({\cal S}_1^{\veps})^{-1}}{1 - {\cal S}_2^{\veps}}\frac{1 - {\cal S}_2^{\veps \cdot \tau}}{\det(\text{Im}\,\tau_{\veps})} \int_{0}^{u} \ex{ 2 \pi \ii \veps \cdot \vec \Delta(z)} \Omega^{\veps \cdot \tau}(z) + {\cal O}(k_i^2) \, . \label{eq:Dteps}
\end{align}
Now we can use a relationship between $D(\veps\,)$ and $D(-\veps\,)$ given by Eq.~(3.21) of \cite{Russo:2007tc}; for $h=2$ it can be stated as\footnote{To get this from Eq.~(3.21) of  \cite{Russo:2007tc}, we have to put $\epsilon_h = - \epsilon_2$; the relative sign occurs because our $a_2$ and $b_2$ homology cycles have opposite orientation.}
\begin{align}
D(\veps\,) - D( - \veps) & = - 2 \ii \, \frac{\sin( \pi \epsilon_1)\sin( \pi \epsilon_1 + \pi \epsilon_2)}{\sin(\pi \epsilon_2)} \, . \label{eq:DDeq}
\end{align}
 (this can be found by cutting between the Schottky circles to get a simply-connected fundamental domain for $\overline{\Sigma}_h$, then using Stokes' theorem to integrate $\widetilde{\Omega}^{\veps} \wedge \widetilde{\Omega}^{(-\veps)}$, which vanishes since it's a $(2,0)$ form --- see Appendix A of \cite{Antoniadis:2005sd}).
Inserting \eq{eq:Dteps} in \eq{eq:DDeq}, we can use the fact that $\det(\text{Im}\,\tau_{\veps})$ is even under the substitution $\veps \mapsto - \veps$, as follows from \eq{eq:qkswap}, and solve to find:
\begin{align}
\det( \text{Im}\,\tau_{ \veps} ) & = \frac{1}{4 \pi} \frac{\ex{ - \ii \pi (\epsilon_1 + \epsilon_2)}(1 - {\cal S}_2^{\veps \cdot \tau})}{\sin(\pi (\epsilon_1 + \epsilon_2))} \int_0^u  \ex{ 2 \pi \ii \veps \cdot \vec \Delta(z)}  \Omega^{\veps \cdot \tau}(z) + (\veps \to - \veps) + {\cal O}(k_i^2) \, . \label{eq:bosonicTeps}
\end{align}


\subsection{The twisted determinant on a super Riemann surface}
\label{tepsSRS}

To compute the superstring partition function in our background, we have to find $\det(\text{Im}\,\boldsymbol{\tau}_{\veps})$, appearing in \eq{eq:Reps} and so on. $\det(\text{Im}\,\boldsymbol{\tau}_{\veps})$ is a modified version of $\det(\text{Im}\,\tau_{\veps})$ as computed in the previous subsection; in this subsection we show how to modify the computation for SRSs.
As a starting point, we take the following formula for $\det(\text{Im}\,\tau_{\veps})$ which can be arrived at similarly to \eq{eq:Dteps} but without using \eq{eq:pochdeform} to deform the integration cycle:
\begin{align}
\det( \text{Im}\,\tau_{ \veps} ) & = \frac{1}{4 \pi} \frac{\ex{ - \ii \pi (\epsilon_1 + \epsilon_2)}(1 - {\cal S}_2^{\veps \cdot \tau})}{\sin(\pi (\epsilon_1 + \epsilon_2))(1 - {\cal S}_1^{\veps})(1 - {\cal S}_2^{\veps})} \int_{\gamma_{\rm P}}  \ex{ 2 \pi \ii \veps \cdot \vec \Delta(z)}  \Omega^{\veps \cdot \tau}(z) + (\veps \to - \veps)\, . \label{eq:bostegamP}
\end{align}
To proceed, we replace $\Delta_i(z)$ and $\Omega^\epsilon$ with supersymmetric extensions $\boldsymbol{\Delta}_i({\bf z})$ and $\boldsymbol{\Omega}^{\epsilon}$ respectively, replace the period matrix $\tau$ with the super-period matrix $\boldsymbol{\tau}$ in the phases $\ex{2 \pi \ii (\epsilon \cdot \tau)_i}$, and carry out the integration over a Pochhammer contour $\boldsymbol\gamma_{\rm \bf P} = {\bf a}_2 {\bf a}_1 {\bf a}_2^{-1} {\bf a}_1^{-1}$ on the SRS. The formula, then, is
\begin{align}
\det(\text{Im}\, \boldsymbol{\tau}_{ \veps} ) & = \frac{1}{4 \pi} \frac{\ex{ - \ii \pi (\epsilon_1 + \epsilon_2)}(1 - {\cal S}_2^{\veps \cdot \boldsymbol{\tau}})}{\sin(\pi (\epsilon_1 + \epsilon_2))(1 - {\cal S}_1^{\veps})(1 - {\cal S}_2^{\veps})} \int_{\boldsymbol\gamma_{\rm \bf P}}  \ex{ 2 \pi \ii \veps \cdot \vec{\boldsymbol{\Delta}}({\bf z})}  \boldsymbol \Omega^{\veps \cdot \boldsymbol\tau}({\bf z}) + (\veps \to - \veps)\, . \label{eq:supert}
\end{align}
The integrand in \eq{eq:supert} will be locally of the form $\d {\bf z}\, f(z|\psi)$. We can carry out the $\d \psi$ integral independently and it will just yield the coefficient of $\psi$ in the integrand. The integral is reduced to an ordinary line integral over a Pochhammer contour $\gamma_{\rm P}$ in the reduced space of the SRS. Then as in section \ref{tepsbos}, we are free to expand the integrand as a power series in $k_i$ and deform $\gamma_{\rm P}$ as in \eq{eq:pochdeform} yielding
\begin{align}
\det(\text{Im}\, \boldsymbol{\tau}_{ \veps} ) & = \frac{1}{4 \pi} \frac{\ex{ - \ii \pi (\epsilon_1 + \epsilon_2)}(1 - {\cal S}_2^{\veps \cdot \boldsymbol{\tau}})}{\sin(\pi (\epsilon_1 + \epsilon_2))} \int_0^u \partial_{\psi}\Big( \ex{ 2 \pi \ii \veps \cdot \vec{\boldsymbol{\Delta}}(z|\psi)}  \boldsymbol \Omega^{\veps \cdot \boldsymbol\tau}(z|\psi)\Big) + (\veps \to - \veps) + {\cal O}(k_i)\, . \label{eq:supertred}
\end{align}
Now we need to define the objects appearing in \eq{eq:supertred}. The Prym differentials $\Omega_i^{ \veps}$ we used to compute $\det(\text{Im}\,\tau_{\veps})$ are holomorphic one-forms; the natural analogues on SRSs are holomorphic volume forms: sections of the Berezinian bundle. Just as holomorphic differentials can be written locally as $\d z \, \partial_z f(z)$, sections of the Berezinian can be written locally as $\d {\bf z}\, D_\psi \,  f(z|\psi)$, the combination being invariant under change of superconformal coordinates \cite{Witten:2012ga}. We note that we can write equation \eq{eq:zetaepsdef} for $\zeta_i^{\veps}$  as
\begin{align}
\zeta_i^\epsilon(z) & =   \d z\, \frac{\partial}{\partial z} \Big({\cal S}_i^{\veps} \,  {\sum_\alpha }^{(i)} \ex{ 2 \pi \ii \veps \cdot \vec{N}_\alpha } \log \Big[ \frac{z - T_\alpha(\eta_i)}{z - T_\alpha(\xi_i)} \Big] \\
& \phantom{=====}
+(1 - {\cal S}_i^{\veps} \,  ) \sum_\alpha \ex{2 \pi \ii \veps \cdot \vec{N}_\alpha } \log \Big[ \frac{z - T_\alpha(z_0)}{z - T_\alpha(a_i^\alpha)} \Big] \Big) \, ,
\nonumber
\end{align}
so to find the corresponding SRS volume forms we replace the expressions inside the logarithms with their natural superconformal analogues and replace $\d z \, \partial_z \mapsto \d{\bf z}\, D_\psi$. This yields
\begin{align}
 \boldsymbol{\zeta}^{\veps}_i(z|\psi) & = \d{\bf z} \, D_\psi \Big(
 {\cal S}_i^{\veps} \, {\sum_\alpha }^{(i)} \ex{ 2 \pi \ii \veps \cdot {\vec{N}}_\alpha  } \log \Big[ \frac{\bra{{\bf z}}{\bf  T}_\alpha \ket{ {\bf u}_i} }{\bra{{\bf z}}{\bf  T}_\alpha \ket{ {\bf v}_i} } \Big]  \\
& \hspace{120pt}
+(1 - {\cal S}_i^{\veps}\,) \sum_\alpha \ex{2 \pi \ii \veps \cdot {\vec{N}}_\alpha } \log \Big[ \frac{\bra{{\bf z}}{\bf  T}_\alpha \ket{ {\bf z}_0} }{\bra{{\bf z}}{\bf  T}_\alpha \ket{ {\bf a}_i} }\Big] \Big) \nonumber \\
& =  \d{\bf z} \,\Big({\cal S}_i^{\veps} \, {\sum_\alpha }^{(i)} \ex{ 2 \pi \ii \veps \cdot {\vec{N}}_\alpha }\Big[ \frac{\bra{{\bf z}} \Phi {\bf  T}_\alpha \ket{ {\bf u}_i} }{\bra{{\bf z}}{\bf  T}_\alpha \ket{ {\bf u}_i} } - \frac{\bra{{\bf z}}\Phi{\bf  T}_\alpha \ket{ {\bf v}_i} }{\bra{{\bf z}}{\bf  T}_\alpha \ket{ {\bf v}_i} } \Big]  \label{eq:superzeta} \\
& \hspace{60pt}
+(1 - {\cal S}_i^{\veps}\,) \sum_\alpha \ex{2 \pi \ii \veps \cdot {\vec{N}}_\alpha } \Big[ \frac{\bra{{\bf z}}\Phi {\bf  T}_\alpha \ket{ {\bf z}_0} }{\bra{{\bf z}}{\bf  T}_\alpha \ket{ {\bf z}_0} } - \frac{\bra{{\bf z}}\Phi{\bf  T}_\alpha \ket{ {\bf a}_i} }{\bra{{\bf z}}{\bf  T}_\alpha \ket{ {\bf a}_i} } \Big] \Big)
\nonumber
\end{align}
where we've used $\Phi$ defined in \eq{eq:Phidef}, $\ket{{\bf z}_0}$ is an arbitrary base point, and
\begin{align}
\ket{{\bf a}_i^\alpha}  & =
\begin{cases}
\ket{{\bf u}_i} & \text{ if } {\bf T}_\alpha = {\bf T}_\beta {\bf S}_i^{\ell} \text{ with } \ell \geq 1 \\
\ket{{\bf v}_i} & \text{ otherwise.}
\end{cases}
\end{align}
Then we can write down a basis of $(h-1)$ holomorphic volume forms ${\bf \Omega}^{\veps}_j(z)$  with the expected monodromies along the homology cycles using the analogue of \eq{eq:Omegaepsdef}, noting that the dependence on the base point $\ket{{\bf z}_0}$ cancels out:
\begin{align}
{\bf \Omega}^{\veps}_j(z|\psi) & = \boldsymbol{\zeta}^{\veps}_j(z|\psi) - \frac{1 - {\cal S}_j^{\veps}}{1 - {\cal S}_h^{\veps}} \boldsymbol{\zeta}^{\veps}_h(z|\psi) \, &  j & = 1 , \ldots , (h-1) \, .
\end{align}
We can calculate ${\bf \Omega}^{\veps}_j(z|\psi)$ as a series expansion in $k_i^{1/2}$. Truncating to finite order, we only need to sum \eq{eq:superzeta} over finitely many terms of the super-Schottky group, because if the contribution from ${\bf T}_\alpha$ is ${\cal O}({k_\alpha}^{1/2})$ and the left-most factor of ${\bf T}_\alpha$ is not ${\bf S}_i^{\pm 1}$, then the contribution from ${\bf S}_i^{\pm \ell} {\bf T}_\alpha$ is ${\cal O}({k_i}^{\ell/2} {k_\alpha}^{1/2})$. Restricting ourselves to $h=2$, this means that if we only want to compute to order ${k_i}^{1/2}$ then we only need to sum over the super-Schottky group elements
\begin{align}
{\bf T}_\alpha \in \{ \text{Id}, {\bf S}_1^{\pm 1} , {\bf S}_2^{\pm 1}, ({\bf S}_1 {\bf S}_2)^{\pm 1} ,  ({\bf S}_1^{-1} {\bf S}_2)^{\pm 1}, ({\bf S}_1 {\bf S}_2^{-1})^{\pm 1} ,  ({\bf S}_2 {\bf S}_1)^{\pm 1} \} \, .
\end{align}
Using the fixed points given in \eq{eq:sfpdef}, we obtain the following expression for $\boldsymbol{\Omega}^{\veps}({\bf z}) \equiv \boldsymbol{\Omega}_1^{\veps}({\bf z})$:
\begin{align}{\bf \Omega}^{\veps}(z|\psi) & = \d{\bf z} \,\Big[-\frac{(1-{\cal S}^{\veps}_1) {\cal S}^{\veps}_2 \theta }{(1-{\cal S}^{\veps}_2) (u-z)}+\frac{(1-{\cal S}^{\veps}_1) \phi }{(1-{\cal S}^{\veps}_2) (1-z)} \label{eq:sOmegaSch}
 \\ & \hspace{20pt}
+\frac{\left({\cal S}^{\veps}_1 u(1-{\cal S}^{\veps}_2)-({\cal S}^{\veps}_1-{\cal S}^{\veps}_2-{\cal S}^{\veps}_1 {\cal S}^{\veps}_2 u + u) z+z^2(1-{\cal S}^{\veps}_2)\right) \psi }{(1-{\cal S}^{\veps}_2) (u-z) (1-z) z}
\nonumber \\ & \hspace{20pt}
+ \ex{\ii \pi \varsigma_1}k_1^{{1}/{2}} \Big \{ -\frac{(1-{\cal S}^{\veps}_1) {\cal S}^{\veps}_1  ({\cal S}^{\veps}_2 \theta -\phi )}{(1-{\cal S}^{\veps}_2) z}+\frac{(1-{\cal S}^{\veps}_1) ({\cal S}^{\veps}_2 \theta -u \phi )}{{\cal S}^{\veps}_1 (1-{\cal S}^{\veps}_2) u} \Big \}
\nonumber \\ & \hspace{20pt}
+ \ex{\ii \pi \varsigma_2}k_2^{{1}/{2}}  \Big\{ -\frac{{\cal S}^{\veps}_2 (1-{\cal S}^{\veps}_1)\theta }{u-z}+\frac{{\cal S}^{\veps}_2  (1-{\cal S}^{\veps}_1 u) \phi }{u-z}-\frac{({\cal S}^{\veps}_1 -u) \theta }{{\cal S}^{\veps}_2 u (1-z)}
\nonumber \\ & \hspace{60pt}
-\frac{{\cal S}^{\veps}_2 (1-{\cal S}^{\veps}_1 u) \theta \phi \psi }{(u-z)^2}-\frac{({\cal S}^{\veps}_1 -u) \theta \phi \psi }{{\cal S}^{\veps}_2 u (1-z)^2}-\frac{(1-{\cal S}^{\veps}_1) \phi }{{\cal S}^{\veps}_2 (1-z)}\Big\}
\nonumber \\ & \hspace{20pt}
+\ex{\ii \pi (\varsigma_1 + \varsigma_2)}k_1^{{1}/{2}}  k_2^{{1}/{2}}  \Big\{ \frac{{\cal S}^{\veps}_1 {\cal S}^{\veps}_2  (\phi -{\cal S}^{\veps}_1 u \phi -(1-{\cal S}^{\veps}_1)\theta )}{z}
\nonumber \\ & \hspace{60pt}
+\frac{{\cal S}^{\veps}_2  (\theta (1-{\cal S}^{\veps}_1)-\phi +{\cal S}^{\veps}_1 u \phi )}{{\cal S}^{\veps}_1 u}-\frac{ ((u-{\cal S}^{\veps}_1) \theta -(1-{\cal S}^{\veps}_1)u \phi )}{{\cal S}^{\veps}_1 {\cal S}^{\veps}_2 u}
\nonumber \\ & \hspace{60pt}
+\frac{{\cal S}^{\veps}_1 (u \theta -{\cal S}^{\veps}_1 \theta -u \phi (1-{\cal S}^{\veps}_1))}{{\cal S}^{\veps}_2 u z}+\frac{(1-{\cal S}^{\veps}_1) {\cal S}^{\veps}_1  (1-u) \theta }{(1-{\cal S}^{\veps}_2) u (1-z)}
\nonumber \\ & \hspace{60pt}
-\frac{(1-{\cal S}^{\veps}_1) {\cal S}^{\veps}_1 (1-u) \phi }{(1-{\cal S}^{\veps}_2) {\cal S}^{\veps}_2 u (1-z)}-\frac{(1-{\cal S}^{\veps}_1) {\cal S}^{\veps}_1 {\cal S}^{\veps}_2 (1-u)({\cal S}^{\veps}_2 \theta -\phi )}{(1-{\cal S}^{\veps}_2) (u-z)}
\nonumber \\ & \hspace{60pt}
-\frac{(1-{\cal S}^{\veps}_1) {\cal S}^{\veps}_1 {\cal S}^{\veps}_2  (1-u) \theta \phi \psi }{(1-{\cal S}^{\veps}_2) (u-z)^2}-\frac{(1-{\cal S}^{\veps}_1)^2(1+{\cal S}^{\veps}_1) (1-u) \theta \phi \psi }{{\cal S}^{\veps}_1 (1-{\cal S}^{\veps}_2) u (1-z)^2}
\nonumber \\ & \hspace{60pt}
+\frac{(1-{\cal S}^{\veps}_1) (1-u) (u \phi -{\cal S}^{\veps}_2 \theta )}{{\cal S}^{\veps}_1 (1-{\cal S}^{\veps}_2) {\cal S}^{\veps}_2 (1-z)u }+\frac{(1-{\cal S}^{\veps}_1) ({\cal S}^{\veps}_2)^2 (1-u) \theta }{{\cal S}^{\veps}_1 (1-{\cal S}^{\veps}_2)u (u-z)}
\nonumber \\ & \hspace{60pt}
+\frac{(1-{\cal S}^{\veps}_1) {\cal S}^{\veps}_2 (1-u) \theta \phi \psi }{{\cal S}^{\veps}_1 (1-{\cal S}^{\veps}_2) (u-z)^2}-\frac{(1-{\cal S}^{\veps}_1) {\cal S}^{\veps}_2 (1-u) \phi }{{\cal S}^{\veps}_1 (1-{\cal S}^{\veps}_2) (u-z)} \Big\} \Big] +{\cal O}(k_i) \, , \nonumber
\end{align}
where ${\cal S}^{\veps}_i $ is defined in \eq{eq:periodicity1}.
In our calculation of the twisted super-period matrix, the Prym differential appears not with the monodromies $\veps$ but with $( \veps \cdot \boldsymbol{\tau})$; to find $\Omega^{\veps \cdot \boldsymbol{\tau}}({\bf z})$ we replace ${\cal S}_i^{\veps}$ in \eq{eq:sOmegaSch} with ${\cal S}_i^{\veps \cdot \boldsymbol{\tau}}$. Using $\boldsymbol{\tau}$ from \eq{eq:stauSch}, we find
\begin{align}
{\cal S}_1^{\veps \cdot \boldsymbol{\tau}} & = \ex{2 \pi \ii (\veps \cdot \boldsymbol{\tau})_1}= k_1^{\epsilon_1} u^{\epsilon_2}\left( 1 -  2\epsilon_1 \ex{\ii \pi \varsigma_2}  k_2^{{1}/{2}}\frac{y}{u} \theta \phi \right) + {\cal O}(k_i) \, , \label{superSone}\\
{\cal S}_2^{\veps \cdot \boldsymbol{\tau}} & =  \ex{2 \pi \ii (\veps \cdot \boldsymbol{\tau})_2} = k_2^{\epsilon_2} u^{\epsilon_1}\left( 1-  2\epsilon_2 \ex{\ii \pi \varsigma_1} k_1^{{1}/{2}} \frac{y}{u} \theta \phi \right)  + {\cal O}(k_i) \, . \label{superStwo}
\end{align}
To supersymmetrize the Riemann class $\Delta_i(z)$, we need to replace the cross-ratios in \eq{eq:Deltazdef} with super-projective invariant cross-ratios, and replace the Schottky fixed points $\eta_i, \xi_i$ with the super fixed points ${\bf u}_i = u_i|\theta_i$, ${\bf v}_i = v_i| \phi_i$, and replace the base point $z$ with a super-point ${\bf z} = z|\psi$. The formula becomes, then,
\begin{align}
{\bf \Delta}_i({{\bf z}}) & = \frac{1}{2 \pi \ii } \left\{ - \frac{1}{2} \log k_i + \pi \ii + \sum_{j=1}^h \phantom{}^{(j)} \sum_{\alpha} \phantom{}^{(i)} \log \frac{ \bra{\bf{v}_j} {\bf T}_\alpha \ket{\bf{u}_i}}{\bra{\bf{v}_j} {\bf T}_\alpha \ket{\bf{v}_i} } \frac{ \bra{\bf{z}} {\bf T}_\alpha \ket{\bf{v}_i}}{\bra{\bf{z}} {\bf T}_\alpha \ket{\bf{u}_i} }\right\} \, .
\end{align}
For our purposes, we want to compute ${\bf \Delta}_i({{\bf z}})$ for $h=2$ with the fixed points given in \eq{eq:sfpdef}. At order ${\cal O}(k_i^{1/2})$, we find
\begin{align}
{\bf \Delta}_1({{\bf z}}) & = \frac{1}{2 \pi \ii } \Big\{ - \frac{1}{2} \log k_1 + \pi \ii - \log z + \ex{\ii \pi \varsigma_2}k_2^{1/2} (1-u) \Big( \frac{\psi \theta + \theta \phi}{u(1-z)} + \frac{\theta \phi - \psi \phi}{u-z} \Big)   \\  & \hspace{30pt}
  - \ex{\ii \pi (\varsigma_1 + \varsigma_2)}k_1^{1/2}k_2^{1/2} \frac{1-u}{uz } \left( (1-z) \theta  \psi + (u-z) \psi \phi \right) \Big\} + {\cal O} (k_i)
  \nonumber \\
{\bf \Delta}_2({{\bf z}}) & = \frac{1}{2 \pi \ii } \Big \{ - \frac{1}{2} \log k_2 + \pi \ii + \log \frac{1-z}{u-z} + \frac{1}{u-z}\theta  \psi + \frac{1}{1-z}  \psi \phi
  \\
& \hspace{30pt}
+ \ex{\ii \pi \varsigma_1} k_1^{1/2} \frac{1}{u z} \Big( (u-z) \theta  \psi + z(1-u) \theta \phi   + u (1-z) \psi \phi \Big)  \nonumber
\\
& \hspace{30pt} - \ex{\ii \pi (\varsigma_1 + \varsigma_2)} k_1^{1/2} k_2^{1/2} \frac{(1-u)^2}{u} \Big( \frac{1}{u-z} \theta  \psi + \frac{1}{1-z}  \psi \phi  \Big) \Big\} + {\cal O} (k_i) \, .  \nonumber
\end{align}
Exponentiating these, we get
\begin{align}
\ex{2 \pi \ii \veps \cdot {  \bf \Delta }({\bf z})} & = \ex{\ii \pi (\epsilon_1 + \epsilon_2) } k_1^{-\frac{\epsilon_1}{2}} k_2^{- \frac{\epsilon_2}{2}} z^{- \epsilon_1} \left( \frac{ 1 - z }{u-z} \right)^{\epsilon_2} \times \label{eq:expDelta} \\
& \phantom{===}   \Big[ 1 - \epsilon_1 \Big\{\ex{\ii \pi \varsigma_2}k_2^{1/2} (1-u) \Big( \frac{1}{u(1-z)} \theta  \psi + \frac{1}{u-z}  \psi \phi + \Big(\frac{1}{u-z} -\frac{1}{u(1 - z)}\Big)\theta \phi \Big) \nonumber \\
 & \phantom{===} - \ex{\ii \pi (\varsigma_1 + \varsigma_2)} k_1^{1/2} k_2^{1/2}\frac{1-u}{uz } \big( (1-z) \theta  \psi + (u-z)  \psi \phi \big)  \Big\} \nonumber \\
& \phantom{===} + \epsilon_2 \Big\{ \frac{1}{u-z}\theta  \psi + \frac{1}{1-z}  \psi \phi  -\ex{\ii \pi \varsigma_1} k_1^{1/2}  \frac{1}{u z} \big( (u-z) \theta  \psi + z(1-u) \theta \phi  \nonumber \\
& \phantom{===}   + u (1-z)  \psi \phi \big) - \ex{\ii \pi (\varsigma_1 + \varsigma_2)} k_1^{1/2}k_2^{1/2}\frac{(1-u)^2}{u} \Big( \frac{1}{u-z} \theta  \psi + \frac{1}{1-z}  \psi \phi  \Big) \Big\}\Big] + { \cal O}(k_i) \, .  \nonumber
\end{align}
We have assembled the ingredients to compute $\det(\text{Im}\,\boldsymbol{\tau}_\epsilon)$.
We take $ {\bf \Omega}^{\veps \cdot \tau}_1(z|\psi)$ from \eq{eq:sOmegaSch}, the phases ${\cal S}_i^{\veps \cdot \boldsymbol \tau}$ from \eq{superSone} and \eq{superStwo}, and $ \ex{ 2 \pi \ii \veps \cdot { \boldsymbol \Delta}({z|\psi})}$ from \eq{eq:expDelta} and insert them into \eq{eq:supertred}. We get a sum of integrals of the form
\begin{align}
\det(\text{Im}\,\boldsymbol{\tau}_{\veps}) & =  \sum_{I}  f_I(\boldsymbol{\mu},\veps\,)\frac{\pi}{\sin(\pi (\epsilon_1 + \epsilon_2))} \int_{0}^{u} \d z \, z^{n^I_1 - \epsilon_1} (1-z)^{n^I_2 + \epsilon_2} (u-z)^{n^I_3 - \epsilon_2}  \nonumber \\
& \hspace{160pt} + (\epsilon_i \to - \epsilon_i) + {\cal O}(k_i) \, . \label{eq:twisdetsum}
\end{align}
We can evaluate these integral with the substitution $z = t u$, getting\begin{align}
\frac{\pi}{\sin(\pi(\epsilon_1 + \epsilon_2))} \int_0^u \d z \, \frac{z^{n_1 - \epsilon_1} (1-z)^{n_2 + \epsilon_2} }{(u-z)^{-n_3 + \epsilon_2}}    & = \frac{u^{1+n_1 + n_3 - \epsilon_1 - \epsilon_3}}{(-1)^{1+n_1+ n_3}} \EuScript{I}[n_1,n_2,n_3] \, \, ,
\end{align}
where
\begin{align}
\EuScript{I}[n_1,n_2,n_3]& \equiv \Gamma(1 + n_1 - \epsilon_1) \Gamma(1 + n_3 - \epsilon_2)\Gamma(-1-n_1-n_3+\epsilon_1+\epsilon_2)  \label{eq:intIdef} \\
 & \hspace{40pt} \times  \phantom{}_2F_1 ( -n_2 - \epsilon_2,1+ n_1 - \epsilon_1 ; 2+ n_1 + n_3 - \epsilon_1 - \epsilon_2; u )\, . \nonumber
\end{align}
The hypergeometric function is given by the integral representation
\begin{align}
\phantom{}_2F_1(a,b;c;z) & = \frac{\Gamma(c)}{\Gamma(b)\Gamma(c-b)} \int_0^1 t^{b-1} (1-t)^{c-b-1} (1 - t z)^{-a} \, ,
\end{align}
and we've used the identity $ \Gamma(x)\Gamma(1 - x)\sin(\pi x) \equiv \pi $.
We obtain
\newcommand{\JI}{\EuScript{I}}
\newcommand{\C}{\EuScript{A}}
\begin{align}
\det(\text{Im}\,\boldsymbol{\tau}_{\vec{\epsilon}}) & = \frac{1}{4 \pi^2} \sum_{p,q,r = -2}^{2} \sum_{i,j,n=0}^1 \C_{ijn;pqr} \times (\ex{\ii \pi \varsigma_1} k_1^{1/2})^{i}(\ex{\ii \pi \varsigma_2} k_2^{1/2})^{j}( \theta \phi)^n \,\, \JI[p,q,r] \nonumber \\
 & \hspace{200pt} + {\cal O}(k_i)+ (\epsilon_i \leftrightarrow - \epsilon_i) \, .
\end{align}
At lowest order in $k_i$, $\det(\text{Im}\, \boldsymbol{\tau}_{\vec{\epsilon}} )$ is given as
\newcommand{\nl}{\nonumber \\ & \hspace{10pt} }
\begin{align}\label{eq:b.40}
4 \pi^2 \det(\text{Im}\, \boldsymbol{\tau}_{\vec{\epsilon}} )
  =& k_1^{\frac{\epsilon _1}{2}} k_2^{-\frac{\epsilon _2}{2}} u^{-\epsilon_1} \left(k_2^{\epsilon _2} u^{\epsilon _1}-1\right) \JI[-1,-1,-1] \\ & \hspace{10pt}  +k_1^{-\frac{\epsilon _1}{2}} k_2^{-\frac{\epsilon _2}{2}} u^{\epsilon _3+1} \left(k_2^{\epsilon _2} u^{\epsilon _1}-1\right) \JI[1,-1,-1]
 \nl +k_1^{-\frac{\epsilon _1}{2}} k_2^{-\frac{\epsilon _2}{2}} u^{\epsilon _3} \Big(k_2^{\epsilon _2} u^{\epsilon _1}-k_1^{\epsilon _1} u^{\epsilon _2}+u \big((k_1 u){}^{\epsilon _1} (k_2 u){}^{\epsilon _2}-1\big)\Big) \JI[0,-1,-1]\nl - \theta \phi \,\,
 \epsilon _2 u^{\epsilon_3} k_1^{-\frac{\epsilon _1}{2}} k_2^{-\frac{\epsilon _2}{2}} \left(k_1^{\epsilon _1} u^{\epsilon _2}-1\right) \left(k_2^{\epsilon _2} u^{\epsilon _1}-1\right) \JI[0,-1,-1] \nonumber  + (\veps \leftrightarrow - \veps\,) + {\cal O}(k_i^{1/2}) \, ;
\end{align}
our calculations require the calculation of $\det(\text{Im}\, \boldsymbol{\tau}_{\vec{\epsilon}} )$ to first order in $k_i^{1/2}$; the necessary coefficients $\C_{ijn;pqr} = \C_{ijn;pqr}(k_\ell, \epsilon_m, u)$ for $i,j,n$ ranging from $0$ to $1$ are listed in a Mathematica notebook included as supplemental material on \textsc{ArXiv}.\footnote{
\texttt{Twisted\_determinant\_coefficients.nb}
}

When rewriting $\det(\text{Im}\, \btau)$ in the $p_i$ parametrization (\eq{eq:kToP}) appropriate for the symmetric degeneration (\Fig{fig:stringapple}), one can use
\beqa
\label{eq:hypseries}
  _2F_1 \left( a - \epsilon_1, b - \epsilon_2, c - \epsilon_1 - \epsilon_2, u \right) & = &
  1 + \frac{(a - \epsilon_1)(b - \epsilon_2)}{c - \epsilon_1 - \epsilon_2} \, u + {\cal O}(u^2)
  \nonumber \\ & = & 1 + {\cal O} (p_i) \, ,
\eeqa
so around the symmetric degeneration, all the $_2F_1$ hypergeometric functions can be replaced with 1, and so the factor $\EuScript{I}[n_1,n_2,n_3]$ as defined in \eq{eq:intIdef} depends only on the twists $\veps$ and not on the worldsheet moduli.


\section{Appendix C}
\label{Appc}
\subsection{List of Feynman graphs}
In this appendix we will list all of the two-loop 1-particle-irreducible planar Feynman diagrams we get from the vertices in \eq{matrilagra}. To compare easily with the string theory results, we will order the results by the color indices, \emph{i.e.}, we will list all of the diagrams whose propagators have the three color indices $A$, $B$ and $C$, say; we expect that all of these come from the field theory limit of the worldsheet whose boundaries are on the $A$th, $B$th and $C$th D-branes.

Then we should sum the following diagrams, weighted appropriately, over $A\leq B \leq C$, where we write
\begin{align}
B_{BA} & = B_1 & B_{AC} & = B_2 & B_{CB} & = B_3 & m_{AB}^2 & = m_1^2 & m_{CA}^2 & = m_2^2 & m_{BC}^2 & = m_3^2 \, .
\end{align}
Let us write down the Feynman diagrams with the various sectors of the QFT represented using the following propagators:
\begin{align}
\prop{gluon} & = \text{gluon modes polarized parallel to the background field} \label{ppm} \\
\prop{photon} & = \text{gluon modes polarized perpendicular to the background field} \label{ppn} \\
\prop{ghost} & = \text{Faddeev-Popov ghosts} \label{ppg} \\
\prop{} & = \text{scalars}  \label{pps} \\
\intertext{
Then we get the following Feynman graphs:} \nonumber
\end{align}
\vspace{-100pt}
{\setstretch{2.5}
\begin{align}
\apple{photon}{photon}{photon} & = - (3 - \gamma^2) \frac{(d-2)(d-3)}{2} \frac{g^2}{(4 \pi )^d} \int_0^\infty \bigg(   \frac{\prod_{i=1}^3 \d t_i \,\ex{ - t_i m_i^2}  }{\Delta_0^{d/2 - 1} \Delta_B} \bigg) \frac{t_1 + t_2 + t_3 }{\Delta_0}  \, \label{fnnn}
\\
\apple{photon}{gluon}{photon} & = -(d-2) \frac{g^2}{(4 \pi )^d} \int_0^\infty  \bigg( \frac{\prod_{i=1}^3 \d t_i \,\ex{ - t_i m_i^2}  }{\Delta_0^{d/2 - 1} \Delta_B}  \bigg) \frac{1}{\Delta_B} \bigg[ \frac{\sinh(g B_3 t_3) }{g B_3}  \label{fnmn} \\
& \hspace{30pt} \times \bigg( (1 - \gamma^2) \cosh(g B_2 t_2 -  g B_1 t_1 ) \nonumber \\
& \hspace{60pt} + 2 \cosh( 2 g B_3 t_3 - g B_2 t_2 - g B_1 t_1) \bigg) + \text{cyclic permutations} \bigg] \, \nonumber
\\
\apple{gluon}{photon}{gluon} & = - (d-2) \frac{g^2}{(4 \pi )^d} \int_0^\infty  \bigg( \frac{\prod_{i=1}^3 \d t_i \,\ex{ - t_i m_i^2}  }{\Delta_0^{d/2 - 1} \Delta_B} \bigg) \frac{1}{\Delta_0}  \bigg[ \bigg(2 t_3 + \frac{1-\gamma^2}{2}(t_1 + t_2) \bigg)\nonumber \\
& \hspace{60pt} \times \cosh(2 g B_1 t_1 - 2 g B_2 t_2)  +  \text{cyclic permutations} \bigg] \label{fmnm}
\\
\apple{gluon}{gluon}{gluon} & = - \frac{g^2}{(4 \pi )^d} \int_0^\infty   \bigg( \frac{\prod_{i=1}^3 \d t_i \,\ex{ - t_i m_i^2}  }{\Delta_0^{d/2 - 1} \Delta_B} \bigg) \frac{1}{\Delta_B} \bigg[\frac{\sinh(g B_3 t_3)}{g B_3} \nonumber \\
 & \hspace{30pt} \times \big( 2+ (1 - \gamma^2) \cosh ( 2 g B_1 t_1 - 2  g B_2 t_2) \big)  \nonumber \\ & \hspace{50pt} \times \cosh(2 g B_3 t_3 - g B_2 t_2 - g B_1 t_1 ) + \text{cyclic permutations} \bigg] \label{fmmm} \\
\apple{ghost}{photon}{ghost} & = (1 + \gamma^2) \frac{d-2}{2} \frac{g^2}{(4 \pi )^d} \int_0^\infty \bigg(  \frac{\prod_{i=1}^3 \d t_i \,\ex{ - t_i m_i^2}  }{\Delta_0^{d/2 - 1} \Delta_B} \bigg) \frac{1}{\Delta_0} \big[ t_3\nonumber \\
 & \hspace{200pt}   + \text{ cyclic permutations} \big]  \label{fgng}
\\
\apple{ghost}{gluon}{ghost} & = (1+\gamma^2) \frac{g^2}{(4 \pi )^d} \int_0^\infty  \bigg(  \frac{\prod_{i=1}^3 \d t_i \,\ex{ - t_i m_i^2}  }{\Delta_0^{d/2 - 1} \Delta_B} \bigg) \frac{1}{\Delta_B} \bigg[ \frac{\sinh(g B_3 t_3)}{ g B_3 }\nonumber \\
 & \hspace{60pt}\times   \cosh(2 g B_3 t_3 - g B_1 t_1 - g B_2 t_2)  + \text{ cyclic permutations} \bigg] \label{fgmg} \\
\apple{}{photon}{} & =  - (d-2)  \, n_s \frac{g^2}{(4 \pi )^d} \int_0^\infty   \bigg( \frac{\prod_{i=1}^3 \d t_i \,\ex{ - t_i m_i^2}  }{\Delta_0^{d/2 - 1} \Delta_B} \bigg) \frac{1}{\Delta_0} \bigg[\bigg( t_3 + \frac{1 - \gamma^2}{4} (t_1+t_2)\bigg) \nonumber \\
 & \hspace{200pt} + \text{ cyclic permutations} \bigg] \label{fsns} \\
\apple{}{gluon}{} & =  - 2 \, n_s \frac{g^2}{(4 \pi )^d} \int_0^\infty \bigg(   \frac{\prod_{i=1}^3 \d t_i \,\ex{ - t_i m_i^2}  }{\Delta_0^{d/2 - 1} \Delta_B}  \bigg)  \frac{1}{\Delta_B} \bigg[  \frac{\sinh(g B_3 t_3)}{g B_3} \nonumber \\
 & \hspace{150pt} \times \cosh(2 g B_3 t_3 - g B_1 t_1 - g B_2 t_2)  \nonumber \\
 &  \hspace{20pt} +\frac{1 - \gamma^2}{4} \Big(  \frac{\sinh(g B_1 t_1)}{g B_1} \cosh( g B_3 t_3 - g B_2 t_2) + \big( 1 \leftrightarrow 2) \Big)  \label{fsms} \\
 & \hspace{200pt} + \text{cyclic permutations} \bigg]  \nonumber\\
\apple{photon}{}{photon} & =   \frac{g^2}{(4 \pi )^d} \int_0^\infty   \bigg( \frac{\prod_{i=1}^3 \d t_i \,\ex{ - t_i m_i^2}  }{\Delta_0^{d/2 - 1} \Delta_B} \bigg) \bigg[ \frac{d-2}{2} \big( (1 + \gamma^2)m_{3}^2 - 2 (m_{1}^2 + m_{2}^2 ) \big) \label{fnsn} \\
& \hspace{200pt}  + \text{ cyclic permutations} \bigg] \, . \nonumber  \\
\apple{gluon}{}{gluon} & =  \frac{g^2}{(4 \pi )^d} \int_0^\infty   \bigg( \frac{\prod_{i=1}^3 \d t_i \,\ex{ - t_i m_i^2}  }{\Delta_0^{d/2 - 1} \Delta_B} \bigg) \bigg[ \big( (1 + \gamma^2)m_{3}^2 - 2 (m_{1}^2 + m_{2}^2 ) \big) \label{fmsm}  \\
& \hspace{100pt} \times \cosh(2 g B_1 t_1 - 2 g B_2 t_2)  + \text{ cyclic permutations} \bigg] \, . \nonumber \\
\apple{ghost}{}{ghost} & =   \frac{g^2}{(4 \pi )^d} \int_0^\infty   \bigg( \frac{\prod_{i=1}^3 \d t_i \,\ex{ - t_i m_i^2}  }{\Delta_0^{d/2 - 1} \Delta_B} \bigg) \bigg[ m_{1}^2 + m_{2}^2 -  m_{3}^2   \label{fgsg}  \\
& \hspace{200pt}  + \text{ cyclic permutations} \bigg] \, . \nonumber \\
\apple{}{}{} & =  (1 - n_s)\frac{3 - \gamma^2}{2}   \frac{g^2}{(4 \pi )^d} \int_0^\infty   \bigg( \frac{\prod_{i=1}^3 \d t_i \,\ex{ - t_i m_i^2}  }{\Delta_0^{d/2 - 1} \Delta_B} \bigg)( m_{1}^2 + m_{2}^2 + m_3^2 ) \label{fsss}
\end{align}
\vspace{-75pt}
\begin{align}
\figeightgluons & =  -  \frac{g^2}{(4 \pi )^d} \int_0^\infty \bigg( \prod_{i=1}^2 \frac{\d t_i \,\ex{ - t_i m_i^2} \,  g B_i }{t_i ^{d/2-1}\sinh(g B_i t_i)}  \bigg)  \label{fmm}   \\
 & \hspace{20pt} \times \Big\{2 \cosh(2gB_1 t_1 + 2gB_2 t_2) - \frac{1-\gamma^2}{2} \Big(2 \cosh(2 gB_1 t_1 - 2 gB_2 t_2) \nonumber \\
 & \hspace{60pt} +4 \cosh(2gB_1 t_1)\cosh(2gB_2 t_2)  \Big) \Big\}  + \text{cyclic permutations}\, , \nonumber \\
 \figeight{gluon}{photon}
 & =   \frac{g^2}{(4 \pi )^d} \int_0^\infty \bigg( \prod_{i=1}^2 \frac{\d t_i \,\ex{ - t_i m_i^2} \,  g B_i }{t_i ^{d/2-1}\sinh(g B_i t_i)}  \bigg)  \label{alfmn}   \\
 & \hspace{20pt} \times  (1-\gamma^2) (d-2) 2 \cosh(2gB_1 t_1) + \text{cyclic permutations}\, , \nonumber \\
\figeight{photon}{photon}  & =  -  \frac{g^2}{(4 \pi )^d} \int_0^\infty \bigg( \prod_{i=1}^2 \frac{\d t_i \,\ex{ - t_i m_i^2} \,  g B_i }{t_i ^{d/2-1}\sinh(g B_i t_i)}  \bigg)  \label{fnn}   \\
 & \hspace{20pt} \times \Big\{d-2 -  \frac{1-  \gamma^2}{2} \Big(d-2 +\big(d-2\big)^2 \Big) \Big\} \, . \nonumber \\
 & \hspace{100pt} + \text{cyclic permutations}\, , \nonumber \\
\figeight{}{gluon}  & =   \frac{g^2}{(4 \pi )^d} \int_0^\infty \bigg( \prod_{i=1}^2 \frac{\d t_i \,\ex{ - t_i m_i^2} \,  g B_i }{t_i ^{d/2-1}\sinh(g B_i t_i)}  \bigg)(1-\gamma^2) 2 \cosh(2g B_2 t_2) n_s   \nonumber \\
 & \hspace{100pt} + \text{cyclic permutations}, \label{fms} \\
 \figeight{}{photon}& =     \frac{g^2}{(4 \pi )^d} \int_0^\infty \bigg( \prod_{i=1}^2 \frac{\d t_i \,\ex{ - t_i m_i^2} \,  g B_i }{t_i ^{d/2-1}\sinh(g B_i t_i)}  \bigg)(1  -\gamma^2 ) \big(d-2\big)n_s   \nonumber \\
 & \hspace{100pt} + \text{cyclic permutations}, \label{fns} \\
\figeighttwoNscalars & =  - \frac{g^2}{(4 \pi )^d} \int_0^\infty \bigg( \prod_{i=1}^2 \frac{\d t_i \,\ex{ - t_i m_i^2} \,  g B_i }{t_i ^{d/2-1}\sinh(g B_i t_i)}  \bigg) \big( 1  - \frac{1- \gamma^2 }{2} (1 + n_s ) \big) n_s  \, \nonumber \\
& \hspace{100pt} + \text{cyclic permutations}. \label{fss}
\end{align}
\vspace{-40pt}}
\phantom{}
\newline
\phantom{}
\newline
The diagrams that have `$+$ cyclic permutations' written are to be summed with two additional copies with the replacements $(B_1,B_2,B_3) \mapsto (B_2,B_3,B_1)$ and $(B_3,B_1,B_2)$.

Note that the gauge choice $\gamma^2 = 1$ gives many of these diagrams a much simpler form, for example, the second lines of \eq{fnmn} and \eq{fmmm}, the third line of \eq{fsms} and the third and fourth lines of the \eq{fmm} all vanish in this gauge. In fact, the last example is a special case of the fact that both propagators in the diagrams with quartic vertices must have the same polarization precisely when $\gamma^2 = 1$, which corresponds to the fact that $k_1^{1/2}$ and $k_2^{1/2}$ must be taken from the same CFT in string theory.


\bibliographystyle{utphys}
\bibliography{rs}

\end{document}

%% file: DBranesFields.pdf_tex
\begingroup%
  \makeatletter%
  \providecommand\color[2][]{%
    \errmessage{(Inkscape) Color is used for the text in Inkscape, but the package 'color.sty' is not loaded}%
    \renewcommand\color[2][]{}%
  }%
  \providecommand\transparent[1]{%
    \errmessage{(Inkscape) Transparency is used (non-zero) for the text in Inkscape, but the package 'transparent.sty' is not loaded}%
    \renewcommand\transparent[1]{}%
  }%
  \providecommand\rotatebox[2]{#2}%
  \ifx\svgwidth\undefined%
    \setlength{\unitlength}{255.39621582bp}%
    \ifx\svgscale\undefined%
      \relax%
    \else%
      \setlength{\unitlength}{\unitlength * \real{\svgscale}}%
    \fi%
  \else%
    \setlength{\unitlength}{\svgwidth}%
  \fi%
  \global\let\svgwidth\undefined%
  \global\let\svgscale\undefined%
  \makeatother%
  \begin{picture}(1,0.58465066)%
    \put(0,0){\includegraphics[width=\unitlength]{DBranesFields.pdf}}%
    \put(0.73956036,0.21011125){\color[rgb]{0,0,0}\makebox(0,0)[lb]{\smash{$\vdots$}}}%
    \put(0.73956036,0.35309528){\color[rgb]{0,0,0}\makebox(0,0)[lb]{\smash{$\vdots$}}}%
    \put(-0.04304411,0.42116812){\color[rgb]{0,0,0}\makebox(0,0)[lb]{\smash{$Y_I^1$}}}%
    \put(-0.04304411,0.33783674){\color[rgb]{0,0,0}\makebox(0,0)[lb]{\smash{$Y_I^2$}}}%
    \put(-0.04304411,0.20057873){\color[rgb]{0,0,0}\makebox(0,0)[lb]{\smash{$Y_I^A$}}}%
    \put(-0.0325178,0.545805){\color[rgb]{0,0,0}\makebox(0,0)[lb]{\smash{$x_I$}}}%
    \put(0.72677615,0.28632785){\color[rgb]{0,0,0}\makebox(0,0)[lb]{\smash{$F_{\mu \nu}^A$}}}%
    \put(0.72677615,0.42141704){\color[rgb]{0,0,0}\makebox(0,0)[lb]{\smash{$F_{\mu \nu}^2$}}}%
    \put(0.72677615,0.50831595){\color[rgb]{0,0,0}\makebox(0,0)[lb]{\smash{$F_{\mu \nu}^1$}}}%
    \put(-0.04304411,0.0591502){\color[rgb]{0,0,0}\makebox(0,0)[lb]{\smash{$Y_I^N$}}}%
    \put(0.72677615,0.14629765){\color[rgb]{0,0,0}\makebox(0,0)[lb]{\smash{$F_{\mu \nu}^N$}}}%
    \put(0.66761906,-0.01443798){\color[rgb]{0,0,0}\makebox(0,0)[lb]{\smash{$x_\mu$}}}%
    \put(-0.02399512,0.12813576){\color[rgb]{0,0,0}\makebox(0,0)[lb]{\smash{$\vdots$}}}%
    \put(-0.02399512,0.27111978){\color[rgb]{0,0,0}\makebox(0,0)[lb]{\smash{$\vdots$}}}%
  \end{picture}%
\endgroup%

%% file: DBranePositionsAsym.pdf_tex
\begingroup%
  \makeatletter%
  \providecommand\color[2][]{%
    \errmessage{(Inkscape) Color is used for the text in Inkscape, but the package 'color.sty' is not loaded}%
    \renewcommand\color[2][]{}%
  }%
  \providecommand\transparent[1]{%
    \errmessage{(Inkscape) Transparency is used (non-zero) for the text in Inkscape, but the package 'transparent.sty' is not loaded}%
    \renewcommand\transparent[1]{}%
  }%
  \providecommand\rotatebox[2]{#2}%
  \ifx\svgwidth\undefined%
    \setlength{\unitlength}{277.17648926bp}%
    \ifx\svgscale\undefined%
      \relax%
    \else%
      \setlength{\unitlength}{\unitlength * \real{\svgscale}}%
    \fi%
  \else%
    \setlength{\unitlength}{\svgwidth}%
  \fi%
  \global\let\svgwidth\undefined%
  \global\let\svgscale\undefined%
  \makeatother%
  \begin{picture}(1,0.81792118)%
    \put(0,0){\includegraphics[width=\unitlength]{DBranePositionsAsym.pdf}}%
    \put(0.94478238,0.00972495){\color[rgb]{0,0,0}\makebox(0,0)[lb]{\smash{$x_I$}}}%
    \put(0.7917547,0.00972495){\color[rgb]{0,0,0}\makebox(0,0)[lb]{\smash{$Y_I^2$}}}%
    \put(0.12126742,0.45892183){\color[rgb]{0,0,0}\makebox(0,0)[lb]{\smash{${\rm D}_{(d-1)}^N$}}}%
    \put(-0.00174426,0.39752652){\color[rgb]{0,0,0}\makebox(0,0)[lb]{\smash{$Y_J^N$}}}%
    \put(0.84884759,0.00972495){\color[rgb]{0,0,0}\makebox(0,0)[lb]{\smash{$Y_I^A$}}}%
    \put(0.40809234,0.75465361){\color[rgb]{0,0,0}\makebox(0,0)[lb]{\smash{${\rm D}_{(d-1)}^1$}}}%
    \put(0.85063634,0.60403355){\color[rgb]{0,0,0}\makebox(0,0)[lb]{\smash{${\rm D}_{(d-1)}^2$}}}%
    \put(0.86039046,0.39402767){\color[rgb]{0,0,0}\makebox(0,0)[lb]{\smash{${\rm D}_{(d-1)}^A$}}}%
    \put(0.38731335,0.00972495){\color[rgb]{0,0,0}\makebox(0,0)[lb]{\smash{$Y_I^1$}}}%
    \put(0.22293393,0.00972495){\color[rgb]{0,0,0}\makebox(0,0)[lb]{\smash{$Y_I^N$}}}%
    \put(-0.00174426,0.67432479){\color[rgb]{0,0,0}\makebox(0,0)[lb]{\smash{$Y_J^1$}}}%
    \put(-0.00174426,0.61916609){\color[rgb]{0,0,0}\makebox(0,0)[lb]{\smash{$Y_J^2$}}}%
    \put(-0.00174426,0.30623635){\color[rgb]{0,0,0}\makebox(0,0)[lb]{\smash{$Y_J^A$}}}%
    \put(-0.00174426,0.76811965){\color[rgb]{0,0,0}\makebox(0,0)[lb]{\smash{$x_J$}}}%
  \end{picture}%
\endgroup%

%% file: borderlabels2.pdf_tex
\begingroup%
  \makeatletter%
  \providecommand\color[2][]{%
    \errmessage{(Inkscape) Color is used for the text in Inkscape, but the package 'color.sty' is not loaded}%
    \renewcommand\color[2][]{}%
  }%
  \providecommand\transparent[1]{%
    \errmessage{(Inkscape) Transparency is used (non-zero) for the text in Inkscape, but the package 'transparent.sty' is not loaded}%
    \renewcommand\transparent[1]{}%
  }%
  \providecommand\rotatebox[2]{#2}%
  \ifx\svgwidth\undefined%
    \setlength{\unitlength}{333.95bp}%
    \ifx\svgscale\undefined%
      \relax%
    \else%
      \setlength{\unitlength}{\unitlength * \real{\svgscale}}%
    \fi%
  \else%
    \setlength{\unitlength}{\svgwidth}%
  \fi%
  \global\let\svgwidth\undefined%
  \global\let\svgscale\undefined%
  \makeatother%
  \begin{picture}(1,0.38733343)%
    \put(0,0){\includegraphics[width=\unitlength]{borderlabels2.pdf}}%
    \put(0.05456574,0.32730628){\color[rgb]{0,0,0}\makebox(0,0)[lb]{\smash{$(Y_I^A,B^A)$}}}%
    \put(0.27912171,0.18483913){\color[rgb]{0,0,0}\makebox(0,0)[lb]{\smash{$(Y_I^{B},B^{B})$}}}%
    \put(0.57788324,0.18620798){\color[rgb]{0,0,0}\makebox(0,0)[lb]{\smash{$(Y_I^{C},B^{C})$}}}%
    \put(0.24443941,0.08893225){\color[rgb]{0,0,0}\makebox(0,0)[lb]{\smash{$(\epsilon^1,\dd_I^1)$}}}%
    \put(0.65051249,0.09372339){\color[rgb]{0,0,0}\makebox(0,0)[lb]{\smash{$(\epsilon^2,\dd_I^2)$}}}%
  \end{picture}%
\endgroup%

%% file: homologycycles.pdf_tex
\begingroup%
  \makeatletter%
  \providecommand\color[2][]{%
    \errmessage{(Inkscape) Color is used for the text in Inkscape, but the package 'color.sty' is not loaded}%
    \renewcommand\color[2][]{}%
  }%
  \providecommand\transparent[1]{%
    \errmessage{(Inkscape) Transparency is used (non-zero) for the text in Inkscape, but the package 'transparent.sty' is not loaded}%
    \renewcommand\transparent[1]{}%
  }%
  \providecommand\rotatebox[2]{#2}%
  \ifx\svgwidth\undefined%
    \setlength{\unitlength}{188.30992432bp}%
    \ifx\svgscale\undefined%
      \relax%
    \else%
      \setlength{\unitlength}{\unitlength * \real{\svgscale}}%
    \fi%
  \else%
    \setlength{\unitlength}{\svgwidth}%
  \fi%
  \global\let\svgwidth\undefined%
  \global\let\svgscale\undefined%
  \makeatother%
  \begin{picture}(1,0.47196132)%
    \put(0,0){\includegraphics[width=\unitlength]{homologycycles.pdf}}%
    \put(0.19039035,0.35004867){\color[rgb]{0,0,0}\makebox(0,0)[lb]{\smash{$b_1$}}}%
    \put(0.73029334,0.34978779){\color[rgb]{0,0,0}\makebox(0,0)[lb]{\smash{$b_2$}}}%
    \put(0.39064629,0.20999878){\color[rgb]{0,0,0}\makebox(0,0)[lb]{\smash{$(b_1^{-1} \cdot b_2)$}}}%
  \end{picture}%
\endgroup%

%% file: homologya1a2.pdf_tex
\begingroup%
  \makeatletter%
  \providecommand\color[2][]{%
    \errmessage{(Inkscape) Color is used for the text in Inkscape, but the package 'color.sty' is not loaded}%
    \renewcommand\color[2][]{}%
  }%
  \providecommand\transparent[1]{%
    \errmessage{(Inkscape) Transparency is used (non-zero) for the text in Inkscape, but the package 'transparent.sty' is not loaded}%
    \renewcommand\transparent[1]{}%
  }%
  \providecommand\rotatebox[2]{#2}%
  \ifx\svgwidth\undefined%
    \setlength{\unitlength}{188.30992432bp}%
    \ifx\svgscale\undefined%
      \relax%
    \else%
      \setlength{\unitlength}{\unitlength * \real{\svgscale}}%
    \fi%
  \else%
    \setlength{\unitlength}{\svgwidth}%
  \fi%
  \global\let\svgwidth\undefined%
  \global\let\svgscale\undefined%
  \makeatother%
  \begin{picture}(1,0.47196132)%
    \put(0,0){\includegraphics[width=\unitlength]{homologya1a2.pdf}}%
    \put(0.3999785,0.29196422){\color[rgb]{0,0,0}\makebox(0,0)[lb]{\smash{$(b_1\cdot b_2)$}}}%
  \end{picture}%
\endgroup%

%% file: DoubleSurface.pdf_tex
\begingroup%
  \makeatletter%
  \providecommand\color[2][]{%
    \errmessage{(Inkscape) Color is used for the text in Inkscape, but the package 'color.sty' is not loaded}%
    \renewcommand\color[2][]{}%
  }%
  \providecommand\transparent[1]{%
    \errmessage{(Inkscape) Transparency is used (non-zero) for the text in Inkscape, but the package 'transparent.sty' is not loaded}%
    \renewcommand\transparent[1]{}%
  }%
  \providecommand\rotatebox[2]{#2}%
  \ifx\svgwidth\undefined%
    \setlength{\unitlength}{412.15698242bp}%
    \ifx\svgscale\undefined%
      \relax%
    \else%
      \setlength{\unitlength}{\unitlength * \real{\svgscale}}%
    \fi%
  \else%
    \setlength{\unitlength}{\svgwidth}%
  \fi%
  \global\let\svgwidth\undefined%
  \global\let\svgscale\undefined%
  \makeatother%
  \begin{picture}(1,0.15122202)%
    \put(0,0){\includegraphics[width=\unitlength]{DoubleSurface.pdf}}%
    \put(0.11928181,0.14629829){\color[rgb]{0,0,0}\makebox(0,0)[lb]{\smash{$\Sigma$}}}%
    \put(0.11928181,-0.01624032){\color[rgb]{0,0,0}\makebox(0,0)[lb]{\smash{$\Sigma^*$}}}%
    \put(0.68604697,0.11712769){\color[rgb]{0,0,0}\makebox(0,0)[lb]{\smash{$\overline{\Sigma}$}}}%
  \end{picture}%
\endgroup%

%% file: openclosed.pdf_tex
\begingroup%
  \makeatletter%
  \providecommand\color[2][]{%
    \errmessage{(Inkscape) Color is used for the text in Inkscape, but the package 'color.sty' is not loaded}%
    \renewcommand\color[2][]{}%
  }%
  \providecommand\transparent[1]{%
    \errmessage{(Inkscape) Transparency is used (non-zero) for the text in Inkscape, but the package 'transparent.sty' is not loaded}%
    \renewcommand\transparent[1]{}%
  }%
  \providecommand\rotatebox[2]{#2}%
  \ifx\svgwidth\undefined%
    \setlength{\unitlength}{463.4237793bp}%
    \ifx\svgscale\undefined%
      \relax%
    \else%
      \setlength{\unitlength}{\unitlength * \real{\svgscale}}%
    \fi%
  \else%
    \setlength{\unitlength}{\svgwidth}%
  \fi%
  \global\let\svgwidth\undefined%
  \global\let\svgscale\undefined%
  \makeatother%
  \begin{picture}(1,0.28505027)%
    \put(0,0){\includegraphics[width=\unitlength]{openclosed.pdf}}%
  \end{picture}%
\endgroup%

%% file: pochhammer.pdf_tex
\begingroup%
  \makeatletter%
  \providecommand\color[2][]{%
    \errmessage{(Inkscape) Color is used for the text in Inkscape, but the package 'color.sty' is not loaded}%
    \renewcommand\color[2][]{}%
  }%
  \providecommand\transparent[1]{%
    \errmessage{(Inkscape) Transparency is used (non-zero) for the text in Inkscape, but the package 'transparent.sty' is not loaded}%
    \renewcommand\transparent[1]{}%
  }%
  \providecommand\rotatebox[2]{#2}%
  \ifx\svgwidth\undefined%
    \setlength{\unitlength}{366.75bp}%
    \ifx\svgscale\undefined%
      \relax%
    \else%
      \setlength{\unitlength}{\unitlength * \real{\svgscale}}%
    \fi%
  \else%
    \setlength{\unitlength}{\svgwidth}%
  \fi%
  \global\let\svgwidth\undefined%
  \global\let\svgscale\undefined%
  \makeatother%
  \begin{picture}(1,0.23796864)%
    \put(0,0){\includegraphics[width=\unitlength]{pochhammer.pdf}}%
    \put(0.15232028,0.09151709){\color[rgb]{0,0,0}\makebox(0,0)[lb]{\smash{$\xi_1$}}}%
    \put(0.29357399,0.09151709){\color[rgb]{0,0,0}\makebox(0,0)[lb]{\smash{$\eta_1$}}}%
    \put(0.70222765,0.09151709){\color[rgb]{0,0,0}\makebox(0,0)[lb]{\smash{$\eta_2$}}}%
    \put(0.84348133,0.09151709){\color[rgb]{0,0,0}\makebox(0,0)[lb]{\smash{$\xi_2$}}}%
    \put(0.31668937,0.15380117){\color[rgb]{0,0,0}\makebox(0,0)[lb]{\smash{$a_1$}}}%
    \put(0.66096632,0.15380117){\color[rgb]{0,0,0}\makebox(0,0)[lb]{\smash{$a_2$}}}%
    \put(0.11031958,0.15380117){\color[rgb]{0,0,0}\makebox(0,0)[lb]{\smash{$a_1$}}}%
    \put(0.86456682,0.15380117){\color[rgb]{0,0,0}\makebox(0,0)[lb]{\smash{$a_2$}}}%
    \put(0.45582343,0.20348938){\color[rgb]{0,0,0}\makebox(0,0)[lb]{\smash{$\gamma_{\rm P}$}}}%
  \end{picture}%
\endgroup%

%% file: poch2.pdf_tex
\begingroup%
  \makeatletter%
  \providecommand\color[2][]{%
    \errmessage{(Inkscape) Color is used for the text in Inkscape, but the package 'color.sty' is not loaded}%
    \renewcommand\color[2][]{}%
  }%
  \providecommand\transparent[1]{%
    \errmessage{(Inkscape) Transparency is used (non-zero) for the text in Inkscape, but the package 'transparent.sty' is not loaded}%
    \renewcommand\transparent[1]{}%
  }%
  \providecommand\rotatebox[2]{#2}%
  \ifx\svgwidth\undefined%
    \setlength{\unitlength}{304.775bp}%
    \ifx\svgscale\undefined%
      \relax%
    \else%
      \setlength{\unitlength}{\unitlength * \real{\svgscale}}%
    \fi%
  \else%
    \setlength{\unitlength}{\svgwidth}%
  \fi%
  \global\let\svgwidth\undefined%
  \global\let\svgscale\undefined%
  \makeatother%
  \begin{picture}(1,0.09161564)%
    \put(0,0){\includegraphics[width=\unitlength]{poch2.pdf}}%
    \put(0.50438608,0.07849097){\color[rgb]{0,0,0}\makebox(0,0)[lb]{\smash{$\gamma_{\rm P}$}}}%
    \put(0.09580037,0.00467109){\color[rgb]{0,0,0}\makebox(0,0)[lb]{\smash{$\eta_1$}}}%
    \put(0.89511698,0.00467109){\color[rgb]{0,0,0}\makebox(0,0)[lb]{\smash{$\eta_2$}}}%
  \end{picture}%
\endgroup%